\documentclass[journal, onecolumn, 10pt]{IEEEtran}

\usepackage[utf8]{inputenc}
\usepackage{amsmath}
\usepackage{amsthm}
\usepackage{amsfonts}
\usepackage{amssymb}
\usepackage{bm} 
\usepackage{color}
\usepackage{bbm} 
\usepackage{verbatim} 
\usepackage{breqn} 
\usepackage{enumitem} 
\usepackage[hidelinks]{hyperref} 

\usepackage{lipsum}
\usepackage{caption} 
\usepackage{subcaption} 
\usepackage{comment} 

\usepackage{svg}

\usepackage{float}

\usepackage[sorting=none,backend=bibtex]{biblatex}
\usepackage{tikz}
\usepackage{algorithmic}
\usepackage{graphicx} 
\usepackage{textcomp}
\usepackage{pgfplots}
\usepackage{xcolor}
\usepackage{url}
\usepackage{multirow}
\usepgfplotslibrary{units}
\usetikzlibrary{spy,backgrounds}
\usepackage{pgfplotstable}

\def\*#1{\mathbf{#1}}
\newcommand{\vect}[1]{{\mathbf{#1}}}
\newcommand{\mat}[1]{{\mathbf{#1}}}
\renewcommand{\sf}[1]{{\mathsf{#1}}}
\newtheorem{theorem}{Theorem}
\newtheorem{lemma}{Lemma}
\newtheorem{definition}{Definition}
\newtheorem{remark}{Remark}

\newtheorem{corollary}{Corollary}

\newcommand{\C}{\textnormal{c}}
\newcommand{\NC}{\textnormal{nc}}

\newcommand{\mw}[1]{{\color{black}#1}}
\newcommand{\ab}[1]{{\color{black}#1}}

\newcommand{\ablast}[1]{{\color{black}#1}}
\newcommand{\unsure}[1]{{\color{black}#1}}

\newcommand{\D}{\mathbb{D}}
\newcommand{\I}{\mathbb{I}}

\allowdisplaybreaks[4]
\begin{document}

\title{Whispering Secrets in a Crowd: Leveraging Non-Covert Users for Covert Communications}

\author{Abdelaziz Bounhar, Mireille Sarkiss, Mich\`{e}le Wigger
\thanks{This paper was presented in part at the 2024 IEEE International Conference on Communications \cite{ours_icc}.}}

\maketitle

\begin{abstract}
This paper establishes the fundamental limits of a multi-access system where multiple users communicate to a legitimate receiver in presence of an external warden. Only a specific subset of the users, called covert users, needs their communication to remain undetected to the warden, while the remaining non-covert users have no such constraint.
The fundamental limits show a tradeoff between the different rates that are simultaneously achievable at the various users in function of the secret-key rates that the different users share with the legitimate receiver. 
Interestingly, the presence of the non-covert users can enhance the capacities of the covert users, especially under stringent secret-key budgets. 
Our findings underscore the essential requirement of employing a multiplexing (coded time-sharing) strategy to exhaust the fundamental region of all rates that are simultaneously achievable at the covert and the non-covert users. As a side-product of our results, we also establish the covert-capacity secret-key tradeoff for standard single-user and multi-access covert communication systems (without non-covert users), i.e., the largest covert rates that are achievable under given secret-key rate budgets. Previous works had only established the minimum secret-key rates required at largest covert rates.
\end{abstract}

\section{Introduction}
\label{section:introduction}
\IEEEPARstart{P}{}hysical layer security leverages information-theoretic techniques and the characteristics of wireless channels to establish secure  communication preventing an attacker to intercept or decipher the transmitted data. 
A recent technique within physical layer security is \emph{covert communication}, which requires conveying information without being detected by adversaries (wardens), by other users, or by network monitoring nodes.
Such communication setups are relevant in future IoT and sensor networks, e.g., when adversaries should not be able to detect certain  monitoring activities. 
To maintain  communication undetectable (covert), users  must remain silent for most of the time, which inherently allows them to  transmit only a small number of information bits. 
In the IoT context, such a small number of bits per device suffices for many applications, and as such, covert communication seems an adequate approach to ensure secure IoT communication. Covert communication is also inherently much more energy-efficient than  conventional cryptographic algorithms,  which is particularly beneficial for IoT devices with stringent  battery limitations.\\

The early work of \cite{bash_first} first characterized the fundamental limits of covert communications over AWGN channels. It showed that it is possible to communicate covertly and reliably as long as the   message satisfies  the so-called \emph{square-root law}, i.e., the number of  communicated information bits scales like the square-root of the number of  channel uses. (Recall that without covertness constraints reliable communication is possible when the number of information bits scales linearly in the number of channel uses.)
Similar square-root laws were subsequently identified as the fundamental limits of covert communications for various  other channels and setups \cite{bash_first, bash_p2p, bloch_first, ligong_first}. 
More specifically, \cite{ligong_first} considered communication over arbitrary Discrete Memoryless Channels (DMC) and assumed the existence of an arbitrary large secret-key  shared between the  transmitter and the legitimate receiver. 
In contrast, \cite{bloch_first} assumed the more general setup of rate-limited secret-keys. In particular, it determined the minimum secret-key rate required to communicate at the largest possible covert data-rate. In this work we strengthen this result and characterize the required key-rate for any covert-rate, not only the maximum rate. Or rather,  equivalently,  we characterize the largest covert data-rate that is achievable under any given key-rate budget. 
In all these works, covert \ablast{communication} takes place in the \emph{square-root law regime}. It has been shown that rates beyond this regime are possible when the warden has uncertainty about the channel statistics   \cite{che_uncertainty, ligong_csi_uncertainty_tx, bloch_keyless_csi, bloch_keyless_csi_journal} or in the presence of a jammer \cite{bash_uninformed_jammer, shmuel_multi_antennas_jammer, bloch_coop_jammer}.\\

Network covert communication with either multiple transmitters or multiple legitimate receivers has  also been studied \cite{bloch_journal_embedding_broadcast, ligong_broadcast, sang_multiple_overt_superposition_covert_on_overt, sang_noma_multiple_overt_superposition_covert_on_overt}. 
In particular, \cite{bloch_journal_embedding_broadcast} characterized the limits of covert communication over a single-transmitter two-receiver Broadcast Channel (BC) when the transmitter sends a common  non-covert message to both receivers and a covert message to only one of them. The transmission of this covert message should not be detected by the non-intended receiver. 
The same communication scenario was also studied in  \cite{ligong_broadcast} but  assuming that the code used to send the common message is fixed and given and cannot be optimized to facilitate the embedding of the covert message.
The BC setup with multiple legitimate receivers and an external warden was also studied from a communication-theoretic perspective 
 \cite{sang_multiple_overt_superposition_covert_on_overt, sang_noma_multiple_overt_superposition_covert_on_overt}\ablast{, where it was empirically shown that the detection error probability at the warden vanishes faster in the increasing number of legitimate receivers}.  
The fundamental limits of covert communication over a multi-transmitter single-receiver Discrete Memoryless Multi-Access Channel (\ab{DMMAC}) were established in \cite{bloch_k_users_mac}, assuming that communication from all transmitters has to remain undetected by the external warden. 
Their work characterized the set of all  covert data-rates that are simultaneously achievable from the various transmitters to the \ablast{legitimate} receiver and the secret-key rates that are required to achieve the set of largest possible covert data-rates. 
In this \unsure{manuscript}, we extend these findings and determine the minimum secret-key rates that are required to attain any achievable set of covert data-rates, not only the largest data-rates. 
To this end, we consider a slightly more general model than in \cite{bloch_k_users_mac}, where the different transmitters have access to individual local randomness.\\

\ablast{Additionally, the current work extends} the result in \cite{bloch_k_users_mac} to a scenario that mixes covert and non-covert transmissions.
The covertness constraint thus imposes that the external warden cannot distinguish between the following two hypotheses: 
\begin{IEEEeqnarray}{rCl}
\mathcal{H}=0 \colon &&\quad  \textnormal{only non-covert users transmit}\\
\mathcal{H}=1 \colon && \quad \textnormal{all users transmit}.
\end{IEEEeqnarray}
While the rates of non-covert transmissions are defined in the usual way, i.e., as 
\begin{equation}\label{eq:noncovert_rates}
R_\ell= \frac{\log_2 \mathsf{M}_{\ell}}{n},
\end{equation} for $\log_2 \mathsf{M}_{\ell}$ denoting the number of information bits sent by user $\ell$ and $n$ the blocklength, the rates of the \ablast{covert users} and their secret-key rates are defined according to the square-root law and normalized by the detection capability of the warden: 
\begin{equation}\label{eq:covert_rates}
r_\ell =\frac{ \log_2 \mathsf{M}_{\ell}}{\sqrt{n \delta_n}}
\end{equation}
and 
\begin{equation}\label{eq:key_rates}
k_\ell =\frac{ \log_2 \mathsf{K}_{\ell}}{\sqrt{n \delta_n}}
\end{equation}
for $\log_2 \mathsf{K}_\ell$ denoting the number of \ablast{secret-}key bits shared between the covert user $\ell$ and the \ablast{legitimate} receiver  and  $\delta_n$ (as defined precisely later) denoting an average divergence between the output distributions that the warden observes  under the two hypotheses (i.e., covert users transmitting or not).  Note that in \cite{bloch_first,ligong_first} it was shown that the definitions  in \eqref{eq:covert_rates} and \eqref{eq:key_rates} are meaningful in a setup with only covert users.  \\

In this work, we  determine the set of  all  covert, non-covert, and secret-key rates as defined in \eqref{eq:noncovert_rates}--\eqref{eq:key_rates} that are simultaneously achievable over a given DMMAC with an external warden. In particular, we  identify the rates that are simultaneously achievable without any shared secret-key at all. 
 In our setup, we assume that the external warden has access to the non-covert messages. Our achievability result is thus even robust under such  a  strong assumption regarding  the warden, and trivially remains valid also for less-powerful wardens. \\

Our fundamental \ablast{tradeoff-region} exhibits interesting tradeoffs between the covert and non-covert rates. In fact, for general DMMACs, the largest covert rates are only achievable under reduced non-covert rates and vice versa. Interestingly, this tradeoff even depends on the achievable key rates as we show through our theoretical findings and through numerical examples. The described tradeoffs stem from the inherent tension regarding the choice of the statistics of the inputs at the \emph{non-covert} users: certain statistics induce DMMACs from the covert users to the \ablast{legitimate} receiver that allow for high covert rates, while other statistics allow for high transmission rates for the \emph{non-covert} messages themselves. In contrast, the input statistics of the covert users do not influence the communication rates at the non-covert users, because  to ensure undetectability the number of non-zero symbols has to stay low (sub-linear in the blocklength) and thus the statistics of the covert users asymptotically has no influence on the channel experienced by the non-covert transmissions. \\

To establish the fundamental tradeoff between the set of achievable rates, we propose a random coding scheme and an information-theoretic converse result. The coding scheme multiplexes various  instances of a general scheme over multiple phases, where in each phase a  different set of parameters  is employed. 
Multiplexing allows to attain a somehow limited form of collaboration between the distributed multi-access transmitters, despite the fact that they convey independent messages. As we show, multiplexing is indeed required to exhaust the set of all achievable rates in our setup. The same holds also for non-covert communication over a DMMAC \cite{coverthomas06}. In our mixed covert/non-covert setup, multiplexing is required even when there is only a single covert and a single non-covert  user. This contrasts the findings for the DMMAC when transmission from all  users  needs to be covert. In such case, no multiplexing is required and instead the data stream from each user can be  transmitted using a standalone single-user coding scheme and in the decoding of a given covert message all other transmissions are  ignored \ablast{\cite{tin_paper}}. 
In our  coding scheme that we employ in a given phase, we use a similar coding idea for the covert users, while we use a standard DMMAC coding scheme for the non-covert users. 
In the decoding of the non-covert users, the \ablast{legitimate} receiver can simply ignore the covert users, since they anyways remain silent most of the time. \\

Notice that the study of mixed covert users and non-covert users is novel in this line of work, and so are the results on the fundamental limits. 
Previous works had only considered setups with only non-covert or only covert users. However, even in the works with only covert users, the  minimum secret-key rate required to achieve covertness was only determined in the special case of maximum covert transmission rates. 
The present work  establishes the required secret-key rates for all achievable covert rates, also when they are reduced. As already mentioned, our work also presents new results for the classical single-user and multi-access covert communication scenarios as studied in \cite{bloch_first,bloch_k_users_mac}. 
These new results are crucial to determine the set of  covert rates that are achievable over DMCs and DMMACs under any given secret-key budgets per user, which cannot be obtained from the previous results in \cite{bloch_first,bloch_k_users_mac}.   \\

To simplify notation, in most of this manuscript we restrict to binary input alphabets at the covert users and to two covert and a single non-covert users. 
All results and proofs extend however to arbitrary input alphabets and arbitrary number of covert and non-covert users in a straightforward way. 
For conciseness, we only present the main results for these extended setups in our manuscript. 
As a final extension, we also present the fundamental tradeoff between the message and \ablast{secret-key} rates that are simultaneously achievable in a three-transmitter and two-receiver Discrete Memoryless Interference Channel  (DMIC) with an external warden  where two of the transmitters send individual covert messages to their corresponding legitimate receivers and a third transmitter sends a common non-covert message to both \ablast{legitimate} receivers. 
Since in our mixed covert-/non-covert DMMAC scheme, covert messages \ablast{were} decoded independently, it can also be applied to this DMIC setup with same rates. Following the same arguments as in the converse proof for the DMMAC, it can then be shown that this scheme is also optimal for the proposed DMIC setup, thus establishing its fundamental limits. 
In a similar  spirit, \cite{tin_paper} established the covert capacity of  the DMIC with only covert users based on the capacity-achieving  covert DMMAC scheme and its analysis \cite{bloch_k_users_mac}.\\

To summarize, our main contributions in this manuscript are: 
\begin{itemize}
\item We characterize the fundamental limits of non-covert rates, covert rates, and secret-key rates that are simultaneously achievable over a DMMAC with an external warden (Theorems \ref{main_theorem}, \ref{th:asymp_result}, \ref{th:general_channel_result} and \ref{th:general_channel_non_binary_result}). The related simulation and numerical examples allow us to conclude that the presence of non-covert users can enhance the set of achievable rates of covert users.
\item We establish a connection of our result to the jammer assisted covert communication (Corollary \ref{cor:jammer}).
\item We extend our findings to the DMIC by establishing  the fundamental limits of non-covert, covert, and secret-key rates  over a three-user DMIC with one non-covert and two covert users (Theorem \ref{th:intereference_channel_result}).
\item  For single-user and DMMACs with only covert users, we establish the set of all covert rates that are achievable under a given secret-key rate budget  (Corollaries \ref{cor:two_users} and \ref{cor:SU}). Previous results had only determined the required secret-key rates for maximum covert rates. Our results  determine the required secret-key rates for any achievable set of covert rates. Notice that for the DMMAC (but not for single-user channels), our scheme achieving minimum secret-key rates requires local randomness at a part of the transmitters. 

\end{itemize}. 

\emph{Notation}: In this paper, we follow standard information theory notations. We use calligraphic fonts for sets (e.g. $\mathcal{S}$) and  note by $\left| \mathcal{S} \right|$ the cardinality of a set $\mathcal{S}$. The set $[|1,p|]$ denotes the set of positive integers between $1$ and $p$, i.e. $[|1,p|] = \{1,2, \ldots, p\}$. Random variables are denoted by upper case letters (e.g., $X$), while their realizations are denoted by lowercase letters (e.g. $x$). We write $X^{n}$ and $x^{n}$ for the tuples $(X_1,\ldots, X_n)$ and $(x_1,\ldots, x_n)$, respectively, for any positive integer $n > 0$. For a distribution $P$ on $\mathcal{X}$, we note its product distribution on $\mathcal{X}^{n}$ by $P^{\otimes n}(x^{n}) = \prod_{i=1}^{n} P(x_{i})$. For two distributions $P$ and $Q$ on $\mathcal{X}$, $\D(P\|Q)=\sum_{x \in \mathcal{X}} P(x)\log\left(\frac{P(x)}{Q(x)}\right)$ is the relative entropy. For two distributions $P$ and $Q$ on $\mathcal{X}$, we say that $P$ is absolutely continuous with respect to (w.r.t.) $Q$, noted $P \ll Q$, if for all $x \in \mathcal{X}$ we have $P(x)=0$ if $Q(x)=0$, which is equivalent to the condition that the support of $P$ is included in the support of $Q$.  
The logarithm and exponential functions are in base $e$ and  motivated by continuity of the function $t\log t$ we define $0 \log(0) = 0$. We use Landau notation, i.e., for a function $f(n)$ we write $f(n)=o(g(n))$ if the ratio $f(n)/g(n)$ vanishes as $n\to \infty$, and we write $f(n)=\mathcal{O}(g(n))$ if the cumulation points of the ratio $f(n)/g(n)$ are within a bounded interval. \\

\emph{Paper Outline}: We present our main problem setup, namely a three-user Multiple Access Channel (MAC) with \ablast{two} covert and \ablast{one} non-covert users, in the following Section \ref{sec:problem_statement}. The corresponding results are presented in Section~\ref{sec:main_results}. Section~\ref{sec:extensions} extends these results to arbitrary number of  covert and non-covert users and arbitrary input alphabets, and to the interference channel (IC). Section~\ref{sec:conclusion} concludes the article. The technical proofs are deferred to appendices.

\section{The Mixed Covert/Non-Covert Three-User MAC: Setup}
\label{sec:problem_statement}

\begin{figure}[h]
   \centering
   \includegraphics[scale=0.85]{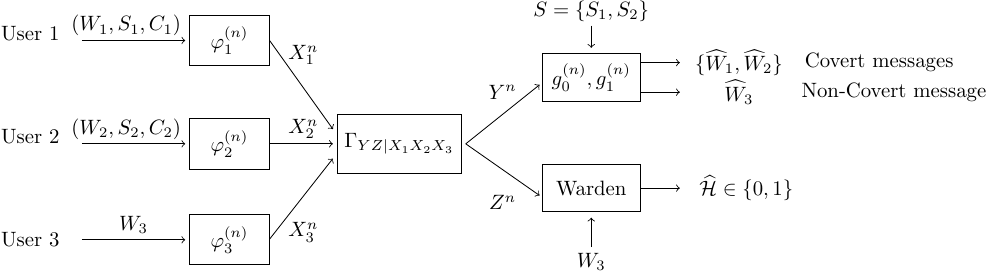}
   \caption{Multi-access communication where communications from Users 1 and  2 have to remain undetectable to the external warden.}
   \label{fig:setup}
\end{figure}
\label{section:problem_setup_and_main_result}

Consider the setup depicted in Figure \ref{fig:setup} where three users  communicate  to a legitimate receiver in the presence of a warden. 
Both Users 1 and  2 wish to communicate covertly, i.e., the warden  should not be able to detect their communication. User 3 does not mind being detected by the warden, and we shall even assume that the warden knows its transmitted message. \\

We thus have two hypotheses  $\mathcal{H}=0$ and $\mathcal{H}=1$, where under $\mathcal{H}=0$ only User 3 sends a message while under $\mathcal{H}=1$ all users send individual messages to the legitimate receiver.
For simplicity, we assume that Users 1 and  2 produce inputs in the binary alphabet  $\mathcal{X}_1=\mathcal{X}_2=\{0,1\}$. User~3's  input alphabet   $\mathcal{X}_3$ is finite but arbitrary otherwise. The legitimate receiver and the warden observe channel outputs in the finite output alphabets $\mathcal{Y}$ and $\mathcal{Z}$. These outputs are produced by a discrete and memoryless multi-access channel (DMMAC) with transition law $\Gamma_{YZ \mid X_1X_2X_3}(\cdot,\cdot| \cdot, \cdot, \cdot)$. This means, if Users 1--3 send input symbols $x_{1,\ab{i}}$, $x_{2,\ab{i}}$, and $x_{3,\ab{i}}$ in time slot $\ab{i}$, then the \ablast{legitimate} receiver and the warden observe symbols $Y_{\ab{i}}$ and $Z_{\ab{i}}$ according to the pmf $\Gamma_{YZ \mid X_1X_2X_3}(\cdot,\cdot| x_{1,\ab{i}},x_{2,\ab{i}},x_{3,\ab{i}})$, irrespective of the previously observed outputs and inputs. 

Define the message,   key sets, and sets of local randomness
\begin{IEEEeqnarray}{rCl}
\mathcal{M}_{1} &\triangleq& \{1, \ldots, \mathsf{M_1} \},\\
\mathcal{M}_{2} &\triangleq& \{1, \ldots, \mathsf{M_2} \},\\
\mathcal{M}_{3} &\triangleq& \{1, \ldots, \mathsf{M_3} \},\\
\mathcal{K}_1 & \triangleq & \{1,\ldots, \mathsf{K_1} \},\\
\mathcal{K}_2 & \triangleq & \{1,\ldots, \mathsf{K_2}\},\\
\mathcal{G}_1& \triangleq & \{1,\ldots, \mathsf{G}_1\}, \\
\mathcal{G}_2& \triangleq & \{1,\ldots, \mathsf{G}_2\},
\end{IEEEeqnarray}
for given positive numbers $\mathsf{M_1}$,  $\mathsf{M_2}$, $\mathsf{M_3}$, $\mathsf{K_1}$,  $\mathsf{K_2}$, $\mathsf{G}_1$, and $\mathsf{G}_2$, and let the messages $W_1$, $W_2$, $W_3$,  the keys $S_1$ and $S_2$, and the local randomness $C_1$ and $C_2$ be uniform over $\mathcal{M}_1, \mathcal{M}_2$, $\mathcal{M}_{3}$, $\mathcal{K}_1$, $\mathcal{K}_2$, $\mathcal{G}_1$, and $\mathcal{G}_2$, respectively. Secret-key $S_1$ is known to User 1 and to the legitimate receiver, and message $W_1$ and local randomness $C_1$ are known  to User 1 only. Similarly, secret-key $S_2$ is known to User 2 and to the legitimate receiver and  message $W_2$ and local randomness $C_2$ are known  to User 2 only. In contrast, message $W_3$ is known to User 3 and the warden. \\

\textit{\underline{Under $\mathcal{H}=0$:}}\\
Users 1 and  2 send the all-zero sequences 
\begin{align}
X_1^n&=0^n,\\
X_2^n&=0^n,
\end{align} 
whereas User 3 applies some  encoding function $\varphi_3^{(n)}\colon \mathcal{M}_3 \to \mathcal{X}_3^n$ to its message $W_3$ and sends the resulting codeword 
\begin{equation}\label{eq:X3}
X_3^n=\varphi_3^{(n)}(W_3)
\end{equation} over the channel.\\

\textit{\underline{Under $\mathcal{H}=1$:}}\\
User 1 applies some  encoding function $\varphi_1^{(n)}\colon \mathcal{M}_1 \times \mathcal{K}_1 \times \mathcal{G}_1\to \mathcal{X}_1^n$ to its message $W_1$, its secret-key $S_1$ and its local randomness $C_1$, and sends the resulting codeword 
\begin{equation}
X_1^n=\varphi_{1}^{(n)}(W_1,S_1, C_1)
\end{equation}
over the channel. Similarly, User 2 applies some encoding function $\varphi_2^{(n)}\colon \mathcal{M}_2 \times \mathcal{K}_2 \times \mathcal{G}_2\to \mathcal{X}_2^n$ to its message $W_2$,  its secret-key $S_2$ and its local randomness $C_2$, and sends the resulting codeword 
\begin{equation}
X_2^n=\varphi_{2}^{(n)}(W_2,S_2,C_2)
\end{equation} over the channel. 

User 3 constructs its channel inputs in the same way as before, see~\eqref{eq:X3}, since it is not necessarily aware of whether $\mathcal{H}=0$ or $\mathcal{H}=1$.

The legitimate receiver, which knows the hypothesis $\mathcal{H}$, decodes the desired messages $W_3$ (under $\mathcal{H}=0$) or  $(W_1,W_2,W_3)$ (under $\mathcal{H}=1$) based on its observed sequence of outputs $Y^n$ and its knowledge of the keys $(S_1,S_2)$.
Thus, under $\mathcal{H}=0$ it uses a decoding function $g_0^{(n)}\colon \mathcal{Y}^n \to \mathcal{M}_3$ to produce the single guess
\begin{equation}
\widehat{W}_3 = g_0^{(n)}( Y^n)
\end{equation}
and  under $\mathcal{H}=1$ it uses a decoding function $g_1^{(n)}\colon \mathcal{Y}^n \times \mathcal{K}_1 \times \mathcal{K}_2 \to \mathcal{M}_1 \times  \mathcal{M}_2 \times  \mathcal{M}_3$ to produce the triple of guesses
\begin{equation}
(\widehat{W}_1,\widehat{W}_2, \widehat{W}_3) = g_1^{(n)}( Y^n, S_1, S_2). 
\end{equation}
Decoding performance of a tuple of encoding and decoding functions $(\varphi_1^{(n)}, \varphi_2^{(n)}, \varphi_3^{(n)}, g_0^{(n)}, g_1^{(n)})$ is measured by the error probabilities under the two hypotheses:
\begin{IEEEeqnarray}{rCl}
    P_{e,0} & \triangleq & \Pr\left(\widehat{W}_3 \neq W_3 \middle|\ \mathcal{H}=0\right), \label{eq:prob2} \\
    P_{e,1} & \triangleq &\Pr\left(\widehat{W}_3 \neq W_3 \text{ or } \widehat{W}_{2} \neq W_{2} \text{ or } \widehat{W}_{1} \neq W_{1} \middle| \mathcal{H}=1\right). \label{eq:prob1}
\end{IEEEeqnarray}

Communication is subject to a covertness constraint at the warden, which observes the channel outputs $Z^n$ as well as the  correct message $W_3$. (Obviously,  covertness assuming that the warden knows $W_3$ implies also covertness in the setup where it does not know $W_3$.) 
For a given codebook $\mathcal{C}$ and for each $w_3\in \mathcal{M}_3$ and $W_3=w_3$, we define the warden's output distribution under $\mathcal{H}=1$ 
\begin{IEEEeqnarray}{rCl}
\label{eq:def_Q_C_w2}
\widehat{Q}_{\mathcal{C}, w_3}^{n}(z^{n}) &\triangleq&  \frac{1}{\mathsf{M}_1 \mathsf{M}_2\mathsf{K}_1\mathsf{K}_2 \mathsf{G}_1\mathsf{G}_2} \sum_{(w_1,s_1,c_1)} \sum_{(w_2,s_2,c_2)} \Gamma^{\otimes n}_{Z \mid X_1X_2X_3} (z^n| x_1^n(w_1,s_1, c_1), x_2^n( w_2,s_2,c_2), x_3^n(w_3)),
\end{IEEEeqnarray}
and  under $\mathcal{H}=0$
\begin{IEEEeqnarray}{rCl}
\Gamma^{\otimes n}_{Z \mid X_1X_2X_3} (z^n| 0^n, 0^n, x_3^n(w_3)),
\end{IEEEeqnarray}
and the divergence between these two distributions: 
\begin{equation}
\label{eq:def_delta_n_w3}
\delta_{n,w_3}\triangleq \mathbb{D}\left(\widehat{Q}_{\mathcal{C}, w_3}^{n} \; \big\| \; \Gamma^{\otimes n}_{Z \mid X_1X_2X_3} ( \cdot | 0^n, 0^n, x_3^n(w_3))\right), \quad \forall w_3 \in \mathcal{M}_3.
\end{equation}
The choice of this measure for covertness is motivated by the fact that any binary hypothesis  test at the warden
satisfies \cite{book_Testing_Statistical_Hypotheses} $\alpha + \beta \geq 1 - \delta_{n,w_3}$, for $\alpha$ and $\beta$  the probabilities of miss-detection and false alarm, respectively.
Ensuring a negligible $\delta_{n,w_3}$ \ablast{for any $w_3 \in \mathcal{M}_3$} is thus sufficient to achieve covertness \ablast{irrespective of the message that is transmitted by the non-covert user}.\\

Our main interest and focus will be on coding schemes $\{(\varphi_1^{(n)}, \varphi_2^{(n)}, \varphi_{3}^{(n)}, g_0^{(n)}, g_1^{(n)})\}_{n=1}^\infty$ that guarantee reliable transmission and undetectability at the warden in the sense of:
\begin{IEEEeqnarray}{rCl}
   \lim_{n \rightarrow \infty} P_{e,\mathcal{H}} & = & 0,\qquad   \mathcal{H}\in\{0,1\},\\
   \lim_{n \rightarrow \infty} \delta_{n,w_3} &=& 0, \qquad \ablast{\forall} w_3 \in \mathcal{M}_3 . 
 \end{IEEEeqnarray}

Our problem is thus multi-objective as we not only wish to have  reliable communication from all the users to the legitimate receiver, i.e. vanishing error probabilities \eqref{eq:prob2} and \eqref{eq:prob1}, but also a vanishing  detectability capability at the warden  \eqref{eq:def_delta_n_w3}.

\begin{remark}[Special Cases of Our Setup]
The setup we introduced in this section includes canonical scenarios as special cases. In fact, when the inputs $X_2$ and $X_3$ do not influence the outputs at the legitimate receiver, Users 2 and 3 can communicate reliably only when $\mathsf{M}_2=\mathsf{M}_3=1$ and  the setup reduces to a covert single-user DMC (from User 1) as studied in \cite{bloch_first, ligong_first}. If only the input of the non-covert user $X_3$ does not influence the outputs, then we fall back to the two-user covert DMMAC studied in \cite{bloch_k_users_mac}. 

If the non-covert user has no data to transmit, i.e., if $\mathsf{M}_3=1$, then the setup specializes to a covert two-user DMMAC with a friendly jammer (User 3) \ablast{whose inputs are revealed to the legitimate receiver and the warden}. 

We shall formally define these special cases in our results Section~\ref{sec:main_result}. 
\end{remark}

\section{The Mixed Covert/Non-Covert Three User MAC: Results}\label{sec:main_results}
We make the following assumptions to avoid degenerate conditions. For any $x_3 \in \mathcal{X}_3$:
\begin{subequations}\label{eq:channel_conditions}
\begin{IEEEeqnarray}{rCl}
\Gamma_{Y \mid X_1X_2X_3}(\cdot | 0, 1, x_3)& \ll & \Gamma_{Y \mid X_1X_2X_3}(\cdot | 0, 0, x_3) , \label{a}\\
\Gamma_{Y \mid X_1X_2X_3}(\cdot | 1, 0, x_3)& \ll & \Gamma_{Y \mid X_1X_2X_3}(\cdot | 0, 0, x_3) ,\label{c}\\
\Gamma_{Z \mid X_1X_2X_3}(\cdot | 0, 1, x_3)& \ll & \Gamma_{Z \mid X_1X_2X_3}(\cdot | 0, 0, x_3) ,\label{e}\\
\Gamma_{Z \mid X_1X_2X_3}(\cdot | 1, 0, x_3)& \ll & \Gamma_{Z \mid X_1X_2X_3}(\cdot | 0, 0, x_3) ,\label{g}\\
\Gamma_{Z \mid X_1X_2X_3}(\cdot | 0, 1, x_3)& \neq & \Gamma_{Z \mid X_1X_2X_3}(\cdot | 0, 0, x_3) ,\label{f} \\
\Gamma_{Z \mid X_1X_2X_3}(\cdot | 1, 0, x_3)& \neq & \Gamma_{Z \mid X_1X_2X_3}(\cdot | 0, 0, x_3).\label{h}
\end{IEEEeqnarray}
\end{subequations}
Conditions \eqref{a} and \eqref{c} exclude the situation in which the legitimate receiver can detect certain input symbols with probability 1, in which case it has been shown that one can communicate $\mathcal{O}(\sqrt{n} \log(n))$ covert \ablast{and reliable} bits in a blocklength $n$, see \cite[Appendix G]{bloch_first}.
Conditions \eqref{e} and \eqref{g} prevent the warden to distinguish the covert users' inputs with probability 1.
Finally, \eqref{f} and \eqref{h}  prevent that  the warden's output distribution does not depend on the covert users' inputs\ablast{, in which case, covertness is trivial}.

\subsection{Useful Definitions}
Before presenting our results we make the following definitions. 
Let $\{\omega_{n}\}_{n=1}^\infty$ be a sequence satisfying 
\begin{subequations}   
    \label{eq:lim_omega_n_def}
    \begin{IEEEeqnarray}{rCl}
        \lim_{n \rightarrow \infty} \omega_{n} &=& 0, \\  
          \lim_{n \rightarrow \infty}\left( \omega_{n}\sqrt{n} -\log n\right) &=& \infty,
    \end{IEEEeqnarray}
\end{subequations}    
and define 
\begin{equation}\label{eq:alphan}
 \alpha_n \triangleq \frac{\omega_n}{\sqrt{n}}.
\end{equation}
We further define for any $x_3 \in \mathcal{X}_3$ the abbreviations
\begin{subequations}\label{eq:D_def}
\begin{IEEEeqnarray}{rCl}
 \mathbb{D}_{Y}^{(1)}(x_3)& \triangleq & \mathbb{D} \left(\Gamma_{Y \mid X_1X_2X_3}(\cdot | 1, 0, x_3) \mid \mid \Gamma_{Y \mid X_1X_2X_3}(\cdot | 0, 0, x_3)  \right), \label{eq:def_divergence_Y_user_1_for_notation}\\
 \mathbb{D}_{Y}^{(2)}(x_3)& \triangleq & \mathbb{D} \left(\Gamma_{Y \mid X_1X_2X_3}(\cdot | 0, 1, x_3) \mid \mid \Gamma_{Y \mid X_1X_2X_3}(\cdot | 0, 0, x_3)  \right), \label{eq:def_divergence_Y_user_2_for_notation}\\
 \mathbb{D}_{Y}^{(1,2)}(x_3)& \triangleq & \mathbb{D} \left(\Gamma_{Y \mid X_1X_2X_3}(\cdot | 1, 1, x_3) \mid \mid \Gamma_{Y \mid X_1X_2X_3}(\cdot | 0, 0, x_3)  \right), \label{eq:def_divergence_Y_user_1_2_for_notation}\\
 \mathbb{D}_{Z}^{(1)}(x_3)& \triangleq & \mathbb{D} \left(\Gamma_{Z \mid X_1X_2X_3}(\cdot | 1, 0, x_3) \mid \mid \Gamma_{Z \mid X_1X_2X_3}(\cdot | 0, 0, x_3)  \right), \label{eq:def_divergence_Z_user_1_for_notation}\\
 \mathbb{D}_{Z}^{(2)}(x_3)& \triangleq & \mathbb{D} \left(\Gamma_{Z \mid X_1X_2X_3}(\cdot | 0, 1, x_3) \mid \mid \Gamma_{Z \mid X_1X_2X_3}(\cdot | 0, 0, x_3)  \right), \label{eq:def_divergence_Z_user_2_for_notation}\\
 \mathbb{D}_{Z}^{(1,2)}(x_3)& \triangleq & \mathbb{D} \left(\Gamma_{Z \mid X_1X_2X_3}(\cdot | 1, 1, x_3) \mid \mid \Gamma_{Z \mid X_1X_2X_3}(\cdot | 0, 0, x_3)  \right)\label{eq:def_divergence_Z_user_1_2_for_notation}. \IEEEeqnarraynumspace
\end{IEEEeqnarray}
\end{subequations}

Define further the function
\begin{IEEEeqnarray}{rCl}
    \chi^2( \rho_1,\rho_2,x_3)
    &\triangleq &\sum_{z \in \mathcal{Z}}  \frac{\left( \frac{\rho_1}{\rho_1+\rho_2}\Gamma_{Z \mid X_1X_2X_3}(z | 1, 0, x_3)  +\frac{\rho_2}{\rho_1+\rho_2} \ \Gamma_{Z \mid X_1X_2X_3}(z | 0, 1, x_3) - \Gamma_{Z \mid X_1X_2X_3}(z | 0, 0, x_3) \right)^2}{\Gamma_{Z \mid X_1X_2X_3}(z | 0, 0, x_3)}. \label{eq:def_xi_distance}
\end{IEEEeqnarray}
We have $\chi^2(\rho_1,\rho_2, x_3)$  the chi-squared distance between the mixture distribution  
\begin{equation}
    \ablast{
        \frac{\rho_1}{\rho_1+\rho_2}\Gamma_{Z \mid X_1X_2X_3}(\cdot  | 1, 0, x_3)  +\frac{\rho_2}{\rho_1+\rho_2} \ \Gamma_{Z \mid X_1X_2X_3}(\cdot | 0, 1, x_3),
    }
\end{equation} and  $\Gamma_{Z \mid X_1X_2X_3}(z | 0, 0, x_3)$. 
Notice that $\chi^2(\rho_1,\rho_2, x_3)=\chi^2\left(\frac{\rho_1}{\rho_1+\rho_2},\frac{\rho_2}{\rho_1+\rho_2}, x_3\right)$, i.e., any normalization of both $\rho_1$ and $\rho_2$ does not change the $\chi^2$ distance.

\subsection{A Basic Coding Scheme}\label{sec:coding}
Our first result in Theorem~\ref{main_theorem} is an achievability result, and we start by explaining the underlying coding scheme. In this subsection we present a special case of the scheme, the general scheme is described in the following Subsection~\ref{sec:gen}. In this special case, we have a deterministic scheme and we simply omit the local randomness in the notation;  the general scheme later can be randomized for one of the users. \\

Fix a finite alphabet $\mathcal{T}$. 
Let $\{\omega_n\}$ be a sequence satisfying \eqref{eq:lim_omega_n_def} and define $\{\alpha_n\}$ as in \eqref{eq:alphan}. Pick a pmf $P_{TX_3}$ over $\mathcal{T} \times \mathcal{X}_3$ and 
  the conditional pmfs
\begin{IEEEeqnarray}{rCl}
P_{X_{1,n} \mid T}(1 \mid t) &= & \rho_{1,t} \alpha_{n}, \quad t\in\mathcal{T},\\ 
P_{X_{2,n} \mid T}(1 \mid t) &= & \rho_{2,t} \alpha_{n}, \quad t \in \mathcal{T}.
\end{IEEEeqnarray}
Define the joint pmf
\begin{equation}\label{eq:Pn}
P_{TX_1X_2X_3Y}^{(n)}\triangleq P_{TX_3}P_{X_{1,n} \mid T}P_{X_{2,n} \mid T}\Gamma_{Y \mid X_1X_2X_3}.
\end{equation}
Let also  $\mu_{n}\triangleq n^{-1/3}$. 

Notice that while both pmfs $P_{X_{1,n} \mid T}$ and $P_{X_{2,n} \mid T}$ used to construct codebooks $\mathcal{C}_1$ and $\mathcal{C}_2$ depend on the blocklength $n$, the pmf $P_{X_3|T}$ used to construct $\mathcal{C}_3$ is independent of $n$.

 Fix a large blocklength $n$ and choose a type-vector   $\boldsymbol{\pi}\in \big\{0, \frac{1}{n}, \frac{2}{n}, \ldots, \frac{n-1}{n} ,1\big\}^{\mathcal{T}}$ with entries summing to $1$, i.e. $\| \boldsymbol{\pi }\|_1=1$, and so that 
\begin{IEEEeqnarray}{rCl}
    \left | \boldsymbol{\pi }(t) - P_T(t) \right | \leq \frac{1}{n}, \quad \forall t \in \mathcal{T},
\end{IEEEeqnarray}
as well as $\boldsymbol{\pi}(t)=0$ whenever $P_T(t)=0$.\\
 
\noindent\underline{\textit{Codebook generation:}}
 Let $t^n=(t_1,\ldots, t_n)$ be any sequence  of type $\boldsymbol{\pi}$, i.e., so that the empirical frequency of symbol $t\in \mathcal{T}$ in $t^n$ equals $\boldsymbol{\pi}(t)$. The $t^n$-sequence acts as a multiplexing sequence that indicates which distribution to use in the   construction of the different entries of the covert and non-covert codewords. As we see in the following, the $i$-th distribution used to construct the $i$-th  entries of all codewords is determined by the value of the symbol $t_i$.
 
\begin{itemize}
    \item For user 1,  generate a codebook
    \begin{equation}\mathcal{C}_{1} =  \left \{  x_{1}^{n}(1,1), \ldots, x_{1}^{n}\big({\mathsf{M}_{1}},{\mathsf{K}_1}\big) \right \}  \end{equation}
    by drawing the $i$-th entry of each codeword $x_1^n(w_1,s_1)$ according to the pmf $P_{X_{1,n} \mid T}(\cdot|t_i)$ independent of all other entries.

    \item For user 2,  generate a codebook
    \begin{equation}\mathcal{C}_{2} =  \left \{  x_{2}^{n}(1,1), \ldots, x_{2}^{n}\big({\mathsf{M}_{2}},{\mathsf{K}_2}\big) \right \}  \end{equation}
    by drawing the $i$-th entry of each codeword $x_2^n(w_2,s_2)$ according to the pmf $P_{X_{2,n} \mid T}(\cdot|t_i)$ independent of all other entries.

    \item For user 3,  generate a codebook
    \begin{equation}\mathcal{C}_3 =  \left \{ \left. x_{3}^{n}(1) , \ldots, x_{3}^{n}\big(\mathsf{M}_3\big) \right \} \right. \end{equation}
        by drawing the $i$-th entry of  each codeword $x_3^n(w_3)$  according to the pmf $P_{X_{3} \mid T}(\cdot|t_i)$  independent of all other entries.
\end{itemize}\bigskip

\noindent\underline{\textit{Encoding and Decoding:}}\\
If $\mathcal{H}=1$, Users 1 and  2 send the codewords $x_1^n(W_1,S_1)$ and $x_2^n(W_2,S_2)$ respectively, and if $\mathcal{H}=0$ they send $x_1^n=0^n$ and $x_2^n=0^n$. User 3  sends the same codeword $x_3^n(W_3)$ under both hypotheses.

\begin{figure}
    \centering
    \includegraphics[scale=0.9]{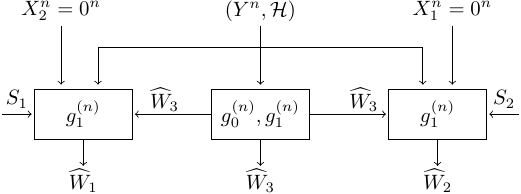}
    \caption{Under $\mathcal{H}=1$, the non-covert message $W_3$ is decoded first, followed by parallel decoding of the two covert messages. Under $\mathcal{H}=0$ only $W_3$ is decoded.}
    \label{fig:decoding_process_2cov_1noncov}
\end{figure}

The legitimate receiver, which observes $Y^n=y^n$ and knows the secret-keys $(S_1,S_2)$ and the true hypothesis $\mathcal{H}$, performs successive decoding starting with message  $W_3$ followed by the messages $W_1$ and $W_2$ in case $\mathcal{H}=1$. (The decoding procedure is also summarized in Figure~\ref{fig:decoding_process_2cov_1noncov}.)

More specifically, under both hypotheses, the legitimate receiver looks for a unique index $w_{3} \in \mathcal{M}_3$ satisfying 
\begin{equation}\label{eq:decoding_user_3}
(t^n,x_3^{n}(w_3), y^{n}) \in \mathcal{T}_{\mu_n}^{n}(P_{TX_3Y }).
\end{equation}
If such a unique index  $w_3$ exists, the \ablast{legitimate} receiver sets $\widehat{W}_3=w_3$. Otherwise it declares an error and stops.

Only under $\mathcal{H}=1$ and after decoding the message $W_3$, the \ablast{legitimate} receiver  decodes the messages of the two covert users. These two decoding steps depend on the set
\begin{equation}
\label{eq:set_a_gamma_n_def_achievability}
    \mathcal{A}_{\eta}^{n} \triangleq \left \{  (x_{1}^{n}, x_{2}^{n} ,x_3^{n}, y^{n}) \in  \mathcal{X}_1^{n} \times \mathcal{X}_2^{n} \times \mathcal{X}_3^{n} \times \mathcal{Y}^{n} : \log \left( \frac{\Gamma_{Y \mid X_1X_2X_3}^{\otimes n}(y^{n} | x_{1}^{n}, x_{2}^{n}, x_3^n) }{\Gamma_{Y \mid X_1X_2X_3}^{\otimes n}(y^{n} |0^n, 0^n, x_3^n) }  \right) \geq \eta \right \}, 
\end{equation}
 where $\eta$ is a given positive constant.

To decode message $W_1$, the legitimate receiver looks for a unique index $w_1$ satisfying 
\begin{equation}\label{eq:decoding_user_1}
\left(x_{1}^{n}(w_1, S_1), 0^n,x_3^{n}(\widehat{W}_3), y^{n}\right)\in \mathcal{A}_{\eta_1}^{n},
\end{equation} 
where $\eta_1$ is a positive constant that needs to be chosen judiciously. (Details on how to choose $\eta_1$ and later $\eta_2$ are presented when we analyze the scheme, see  Equations~\eqref{eq:value_gamma_achievability}
 and \eqref{eq:value_gamma_2_achievability} in Appendix~\ref{sec:proof1}.)
If such a unique index  $w_1$ exists, the \ablast{legitimate} receiver sets $\widehat{W}_1=w_1$. Otherwise it declares an error and stops. 

Similarly, to decode message $W_2$, the \ablast{legitimate} receiver  looks for a unique index $w_2$ satisfying 
\begin{equation}\label{eq:decoding_user_2}
\left(0^n,x_{2}^{n}(w_2, S_2),x_3^{n}(\widehat{W}_3), y^{n} \right) \in \mathcal{A}_{\eta_2}^{n},
\end{equation}
for a well chosen positive constant $\eta_2$.
If such a unique index  $w_2$ exists, it sets $\widehat{W}_2=w_2$, and it declares an error otherwise.
We notice that  the decoding of each covert message uses the previously decoded non-covert message, but assumes that the other covert users send the all-zero sequence.  In fact, the number of non-zero symbols is small in each block, and the all-zero approximation seems not to cause any loss in optimality in terms of achievable rates.

\subsection{Generalization of the Coding Scheme}\label{sec:gen}
We propose a slight generalization of our coding scheme including two new parameters $\phi_1 , \phi_2 \in(0,1]$. In our description, we assume $\phi_1 \geq \phi_2$, otherwise we switch the roles of Users 1 and 2. 

\begin{figure}[h!]
   \centering
   \includegraphics[width=0.45\textwidth]{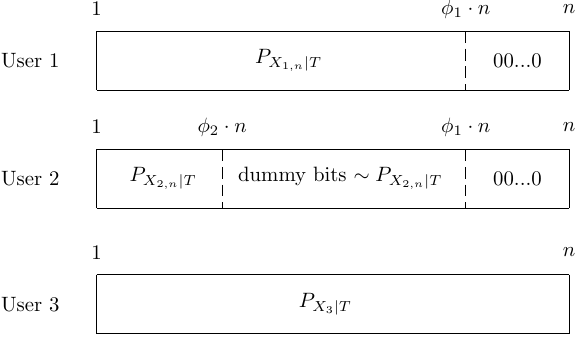}
   \caption{Illustration of the code construction in the generalized coding scheme for the scenario $\phi_1 \geq \phi_2$.}
   \label{fig:generalized_coding_scheme}
\end{figure}

In the generalized scheme, communication at Users 1 and 2 is only over a fraction $\phi_1$ of the time; during the remaining $(1-\phi_1)$ fraction of time both users simply send the zero symbol. User 3 acts as before and transmits over the entire duration of the blocklength $n$. See Figure \ref{fig:generalized_coding_scheme} for an illustration of the scheme. 

For each $t \in \mathcal{T}$ let 
\begin{equation}
\mathcal{L}(t):= \{ j \in [|1,n|] \colon \; t_j=t\},
\end{equation} 
and choose  two disjoint subsets  of time-instances 
$\mathcal{L}_1(t),\mathcal{L}_{1,2}(t)  \subseteq \mathcal{L}(t)$
 of sizes 
 \begin{IEEEeqnarray}{rCl}
 |\mathcal{L}_{1}(t)| &\approx &   n P_T(t)(\phi_1- \phi_2)  \\
 |\mathcal{L}_{1,2}(t)|&  \approx &   n P_T(t) \phi_2  .
 \end{IEEEeqnarray}

 Users 1 and 2 send the following channel inputs depending on whether a time slot $i$ lies in the sets $\mathcal{L}_{1,2}(t)$ or $\mathcal{L}_{1}(t)$ for some $t$ or not. 
\begin{itemize}
\item  In all channel uses $\cup_{t \in \mathcal{T}} \mathcal{L}_{1,2}(t)$:  Users 1 and 2 transmit the corresponding symbols from the  codewords $x_1^n(W_1, S_1)$ and  $x_2^n(W_2,S_2)$ as described in the basic scheme. 
 \item For each $t\in\mathcal{T}$, in  channel uses $\mathcal{L}_{1}(t)$:  User 1  transmits the corresponding symbols from the  codeword $x_1^n(W_1, S_1)$ and  User 2 sends i.i.d. inputs according to $P_{X_{2,n|T=t}}$. Thus, in this scheme only User 2 uses its local randomness $C_2$ to generate the i.i.d. inputs in channel uses $\cup_{t\in \mathcal{T}} \mathcal{L}_1(t)$. 
 \item In the remaining channel uses (i.e., channel uses neither in $\cup_{t \in \mathcal{T}} \mathcal{L}_{1,2}(t)$ nor in $\cup_{t \in \mathcal{T}} \mathcal{L}_{1}(t)$), Users 1 and 2 send   input $0$.\\
 \end{itemize}
 
 The \ablast{legitimate} receiver decodes message $W_3$ as before, see \eqref{eq:decoding_user_3}. To decode message $W_1$, it applies the decoding rule in \eqref{eq:decoding_user_1}, but  focusing only on  channel uses $\cup_{t \in \mathcal{T}} (\mathcal{L}_{1,2}(t) \textsc{ } \cup \textsc{ } \mathcal{L}_{1}(t))$. To decode message $W_2$, it applies the decoding rule in \eqref{eq:decoding_user_2}, but  focusing only on  channel uses  $\cup_{t \in \mathcal{T}} \mathcal{L}_{1,2}(t)$.

\subsection{Main Results}\label{sec:main_result}
Our first result is a finite blocklength achievability result  based on the  coding schemes described in the previous two subsections.

\begin{theorem}
\label{main_theorem}Fix  any pmf $P_{TX_3}$ over 
finite alphabets $\mathcal{T} \times \mathcal{X}_3$ and $(T,X_3)\sim P_{TX_3}$, any sequence $\{\omega_{n}\}_{n=1}^\infty$ as in \eqref{eq:lim_omega_n_def}, any pair $(\phi_1, \phi_2)\in [0,1]^2$, and any nonnegative tuple $\{(\rho_{1, t}, \rho_{2,t})\}_{t\in\mathcal{T}}$. Then,  for any $\epsilon>0$ and  arbitrary small numbers   $\xi_1,\xi_2, \xi_3, \xi_4 , \xi_5, \xi_6 >0$ and for sufficiently large blocklengths  $n$, we can find encoding and decoding functions $\{(\varphi_1^{(n)}, \varphi_2^{(n)}, \varphi_3^{(n)},$ $g_0^{(n)}, g_1^{(n)})\}_n$ 
 with  message sizes $\mathsf{M_1}, \mathsf{M_2}, \mathsf{M_3}$  and secret-key sizes $\mathsf{K_1}, \mathsf{K_2}$ satisfying
 \begin{IEEEeqnarray}{rCl}
 \log(\mathsf{M_1}) &=& (1-\xi_1)\cdot \phi_1 \cdot  \omega_n \sqrt{n} \mathbb{E}_{P_{TX_3}} \left[ \rho_{1,T} \D_{Y}^{(1)}(X_3) \right], \label{eq:th_1_log_m1} \\[1ex]
\log(\mathsf{M_2}) &=& (1-\xi_2) \cdot\phi_2 \cdot\omega_n \sqrt{n} \mathbb{E}_{P_{TX_3}} \left[ \rho_{2,T} \D_{Y}^{(2)}(X_3) \right], \label{eq:th_1_log_m2}\\[1ex]
 \log(\mathsf{M_3}) &=& (1-\xi_3) n \I(X_3;Y \mid X_1=0, X_2=0, T),\label{eq:th_1_log_m3} \\[1ex]
 \log(\mathsf{M_1}) + \log(\mathsf{K_1}) &=& (1+\xi_4) \cdot\phi_1\cdot \omega_{n}\sqrt{n} \mathbb{E}_{P_{TX_3}}  \left[ \rho_{1,T} \D_{Z}^{(1)}(X_3) \right] \label{eq:th_1_log_k1_k2}, \\[1ex]
\log(\mathsf{M_2}) + \log(\mathsf{K_2}) &=& (1+\xi_5)\cdot \phi_2 \cdot\omega_{n}\sqrt{n} \mathbb{E}_{P_{TX_3}}  \left[ \rho_{2,T} \D_{Z}^{(2)}(X_3) \right] \label{eq:th_1_log_k1}\ablast{,}
\end{IEEEeqnarray}
and so that  
\begin{IEEEeqnarray}{rCl}
P_{e, \mathcal{H}}  &\leq & \epsilon, \qquad \forall \mathcal{H} \in \{0,1\} 
 \end{IEEEeqnarray}
and \begin{IEEEeqnarray}{rCl}
\frac{1}{\mathsf{M_3}} \sum_{W_3=1}^{\mathsf{M_3}} \delta_{n,W_3} & = & ( 1+ \xi_6) \cdot \max(\phi_1; \phi_2) \cdot \frac{\omega_n^2 }{2} \mathbb{E}_{P_{TX_3}} \left[ (\rho_{1,T} + \rho_{2,T})^2  \chi^2(\rho_{1,T},\rho_{2,T},X_3) \right ]. \IEEEeqnarraynumspace \label{eq:aux}
\end{IEEEeqnarray}
(Notice that the parameter $\phi_\ell$ influences only the message and key sizes of User $\ell$, $\ell \in \{1, 2\}$, but not of User $3-\ell$. The expected divergence at the warden depends on the term  $\max(\phi_1; \phi_2)$. )
\end{theorem}
\begin{IEEEproof}
Appendix~\ref{sec:proof1} contains the proof in the special case $\phi_1=\phi_2=1$, which is obtained by analyzing the basic coding scheme in Section~\ref{sec:coding}.

The proof in the general case can be obtained by analyzing the generalized coding scheme in Section~\ref{sec:gen}. The analysis is the same as in Appendix~\ref{sec:proof1}, up to some small modifications that allow to introduce the factors $\phi_1$ and $\phi_2$.  We explain the modifications when $\phi_1 \geq \phi_2$, otherwise we switch the indices 1 and 2. 
\begin{itemize}
\item When analyzing $P_{e,1,1}$ in Appendix~\ref{sec:proof1}, restrict to channel uses in $\cup_{t \in \mathcal{T}} \left( \mathcal{L}_{1,2}(t) \cup \mathcal{L}_{1}(t)\right)$ because only these channel uses are used for the decoding of message $W_1$. Since  approximately a fraction $\phi_1$ of the $n$ channel uses are in $\cup_{t \in \mathcal{T}} \left( \mathcal{L}_{1,2}(t) \cup \mathcal{L}_{1}(t)\right)$, we obtain the additional factor $\phi_1$ in \eqref{eq:th_1_log_m1}. 
\item When analyzing $P_{e,1,2}$ , restrict to channel uses in $\cup_{t \in \mathcal{T}}\mathcal{L}_{1,2}(t) $ because only these channel uses are used for the decoding of message $W_2$. Since  approximately a fraction $\phi_2$ of the $n$ channel uses are in $\cup_{t \in \mathcal{T}} \mathcal{L}_{1,2}(t)$, we obtain the additional factor $\phi_2$ in \eqref{eq:th_1_log_m2}. 
\item The details of the resolvability analysis are provided in Appendix~\ref{sec:modified}. The modifications for the generalized scheme allow to introduce the factor $\phi_1$ for bounds \eqref{eq:th_1_log_k1_k2} and \eqref{eq:aux} and the factor $\phi_2$ for \ablast{bounds \eqref{eq:th_1_log_k1} and \eqref{eq:aux}}.
\end{itemize}

\end{IEEEproof}
\medskip
\begin{remark}
To achieve the performance in Theorem~\ref{main_theorem}, the covert user 1 does not require access to local \ablast{randomness} to achieve all tuples for which $\phi_1\geq \phi_2$. Similarly,  covert user 2 does not require access to local randomness to achieve tuples for which $\phi_2\geq \phi_1$. \ablast{(Covert user 2 utilizes common randomness to generate the random inputs in channel uses $\cup_{t} \mathcal{L}_{1}(t)$).}
\end{remark}

\medskip
We observe the difference in the  scalings of the logarithms of the covert-message size  and the  key sizes and the scaling of the logarithm of the  non-covert message size. While the \ablast{formers} grow in the order of $\omega_n \sqrt{n}$, and thus slowlier than $\sqrt{n}$, the logarithm of the non-covert message size scales linearly in $n$. Communication from the non-covert user  thus admits for a positive communication-rate in the traditional sense (ratio between the number of information bits and channel uses), which is not the case for the communications from the  covert users. 

It is further interesting to analyze the influence of the sequence $\{\omega_n\}$.  
The key and covert-message square-root-scalings all depend on the vanishing sequence $\omega_n$. 
Increasing $\omega_n$ proportionally increases the permissible covert-message size but also quadratically increases the average divergence at the warden.
Combined with the observation in the previous paragraph, we conclude that we obtain  meaningful  rates for the covert-messages by dividing the log of their message sizes by the square-roots of the blocklength $n$ and the square-root of the averaged divergences.
This leads to the following definition. 
\begin{definition}\label{def:ach}
A non-negative  tuple $(r_1,r_2,R_3,k_1, k_2)$ is achievable if there exists a sequence (in the blocklength $n$)  of tuples\footnote{Not to overload notation, we did not add a superscript $^{(n)}$ to the parameters $\mathsf{M_1},\mathsf{M_2},\mathsf{M_3},\mathsf{K_1},\mathsf{K_2}$. They however  all depend on the blocklength $n$.} $(\mathsf{M_1},\mathsf{M_2},\mathsf{M_3},\mathsf{K_1},\mathsf{K_2}, \mathsf{G}_1, \mathsf{G}_2)$ and encoding/decoding functions $(\varphi_1^{(n)}, \varphi_2^{(n)}, \varphi_3^{(n)}, g_0^{(n)}, g_1^{(n)})$  satisfying
\begin{IEEEeqnarray}{rCl}
   \lim_{n \rightarrow \infty} P_{e, \mathcal{H}} & = & 0,\qquad  \forall \mathcal{H}\in\{0,1\},\label{eq:Pei}\\ 
   \lim_{n \rightarrow \infty} \delta_{n,W_3} &=& 0, \qquad \forall W_3 \in \mathcal{M}_3 ,  \label{eq:covert_constraint}
 \end{IEEEeqnarray}
and 
\begin{IEEEeqnarray}{rCl}
 r_{\ell} & \ab{=} &  \liminf_{n \rightarrow \infty} \frac{\log(\mathsf{M_{\ell}})}{\sqrt{ n \frac{1}{\mathsf{M_3}} \sum_{W_3 = 1}^{\mathsf{M_3}} \delta_{n,W_3} }}, \qquad  \forall \ell \in \{1,2\},  \label{eq:asymp2}
\\[1ex]
 R_3 &=& \liminf_{n \rightarrow \infty} \frac{\log(\mathsf{M_3})}{n} ,\\[1ex]
 k_{\ell}  &= & \limsup_{n \rightarrow \infty} \frac{\log(\mathsf{K_{\ell}})}{\sqrt{ n \frac{1}{\mathsf{M_3}} \sum_{W_3 = 1}^{\mathsf{M_3}} \delta_{n,W_3} }}, \qquad  \forall \ell \in \{1,2\}.
\end{IEEEeqnarray}
\end{definition}

The following theorem  determines the set of all achievable rate-key tuples $(r_1,r_2,R_3,k_1, k_2)$. 

\begin{theorem}
\label{th:asymp_result}
A nonnegative rate-key tuple $(r_1,r_2,R_3,k_1, k_2)$ is achievable 
 if, and only if, there exists a pmf over $\mathcal{T} \times \mathcal{X}_3$ with $\mathcal{T}=\{1,\ldots, 6\}$ and  $(T,X_3) \sim P_{TX_3}$,  a nonnegative tuple $\{(\rho_{1, t}, \rho_{2,t})\}_{t\in\mathcal{T}}$, and a pair $(\beta_1, \beta_2)\in [0,1]^2$ so that the following inequalities hold:
\begin{IEEEeqnarray}{rCl}
  r_{\ell} & \leq &\beta_\ell  \sqrt{2} \frac{\mathbb{E}_{P_{TX_3}} \left[ \rho_{\ell,T} \D_{Y}^{(\ell)}(X_3) \right]}{\sqrt{\mathbb{E}_{P_{TX_3}} \left[ \left( \rho_{1,T} + \rho_{2,T} \right)^2  \cdot \chi^2(\rho_{1,T}, \rho_{2,T}, X_3) \right]}}, \qquad  \forall \ell \in \{1,2\}, \label{eq:asymp1} \\[1ex]
 R_3 &\leq&  \mathbb{I}(X_3;Y \mid X_1=0,X_2=0,T)\label{eq:asymp3}, \\[1ex]
k_{\ell}  &\geq&\beta_\ell  \sqrt{2} \frac{\mathbb{E}_{P_{TX_3}}  \left[ \rho_{\ell,T} \left( \D_{Z}^{(\ell)}(X_3) - \D_{Y}^{(\ell)}(X_3) \right) \right]} {\sqrt{\mathbb{E}_{P_{TX_3}} \left[ \left( \rho_{1,T} + \rho_{2,T} \right)^2 \cdot \chi^2(\rho_{1,T}, \rho_{2,T}, X_3) \right]}}, \qquad  \forall \ell \in \{1,2\},\label{eq:asympkey}
\end{IEEEeqnarray}
where recall that $\D_Y^{(\ell)}(\cdot)$ and $\D_Z^{(\ell)}(\cdot)$ are defined in \eqref{eq:D_def}.
\end{theorem}
\begin{IEEEproof}
The achievability proof essentially follows   from Theorem~\ref{main_theorem},  by setting $\phi_\ell = \beta_\ell \max(\phi_1\ablast{;} \phi_2)$ and taking $n\to \infty$.
For details, see Appendix~\ref{app:ach}.
For the proof of the converse, see Appendix~\ref{sec:proof2}.
\end{IEEEproof}
\begin{lemma}
    \label{lem:convexity}
    The set of five-dimensional vectors $(r_1, r_2, R_3,k_1, k_2)$ satisfying Inequalities \eqref{eq:asymp1}--\eqref{eq:asympkey} for some choice of pmfs $P_{TX_3}$ and nonnegative pairs $\{(\rho_{1,t}, \rho_{2,t})\}_{t\in \mathcal{T}}$ is a convex set.
\end{lemma}
\begin{IEEEproof}
See Appendix~\ref{app:convexity}. 
\end{IEEEproof}
 \begin{remark}
 Whenever the numerator in \eqref{eq:asympkey} is \ablast{negative}, no secret\mw{-}key is required to establish covert communication in our setup. In particular, whenever $\D_{Z}^{(1)}(x_3) < \D_{Y}^{(1)}(x_3)$ and $ \D_{Z}^{(2)}(x_3) < \D_{Y}^{(2)}(x_3)$ for all $x_3\in\mathcal{X}_3$,  Condition \eqref{eq:asympkey} is always satisfied. 
 \end{remark}
 \smallskip

Our theorem includes various interesting special cases. For example, when User 3 has no message to transmit ($\mathsf{M}_3=1$ and $R_3=0$), its inputs  act as a jamming sequence that shapes the channel and in addition is known to the warden and the \ablast{legitimate} receiver. 
We then obtain the following corollary. 
\begin{corollary}[User 3 acting as a Friendly Jammer]
\label{cor:jammer}
    When User 3 sends no message ($\mathsf{M}_3=1$ and $R_3=0$),  it acts  as a friendly jammer whose inputs are known to the warden and to the \ablast{legitimate} receiver. In this case, a rate-key tuple $(r_1, r_2,k_1,k_2)$ for Users 1 and 2 is achievable if, and only if,  \eqref{eq:asymp1} and  \eqref{eq:asympkey} 
  hold for some choice of $P_{TX_3}$ and pairs $\{(\rho_{1,t}, \rho_{2,t})\}_{t\in\mathcal{T}}$ and $(\beta_1, \beta_2)\in [0,1]^2$. 
  \end{corollary}

Our theorem also includes results for the two-user covert DMMAC and the single-user covert DMC as specials cases. In both cases our results are stronger than the previously known findings in \cite{bloch_k_users_mac} and  \cite{bloch_first}, because we not only determine the required key rate at the largest covert communication rates, but at all rates. The special case of the two-user covert DMMAC is obtained if in our setup we assume that either 
$\mathcal{X}_3 = \{x_3\}$ is \ablast{a} singleton or that the output distributions at the legitimate receiver and the warden  $\Gamma_{Y \mid X_1X_2X_3}$ and $\Gamma_{Z \mid X_1X_2X_3}$ do not depend on the input $X_3$. 
Interestingly, in these cases the cardinality of $\mathcal{T}$ can be set to 1 without loss in optimality. So no multiplexing (coded time-sharing) is needed. Moreover, when $\mathcal{T}=\{t\}$, then the single parameters   $\rho_{1,t}$ and $\rho_{2,t}$ can be chosen to sum to 1 without loss in optimality.
These observations are made precise in the following corollary and its proof.
\begin{corollary}[Only Two Covert Users]\label{cor:two_users}
 Assume that   $\mathcal{X}_3=\{x_3\}$ or  that for any $x_3 \in \mathcal{X}_3$:
\begin{subequations}\label{eq:conditions_nox3}
\begin{IEEEeqnarray}{rCl} 
\Gamma_{Y \mid X_1X_2X_3}(y|x_1,x_2,x_3) &= & \Gamma_{Y|X_1X_2}(y|x_1,x_2), \\
\Gamma_{Z \mid X_1X_2X_3}(y|x_1,x_2,x_3) &= & \Gamma_{Z|X_1X_2}(z|x_1x_2).
\end{IEEEeqnarray}
\end{subequations} Then $R_3=0$ and a message and secret-key rates tuple $(r_1, r_2,  k_1,k_2)$ is achievable if, and only if, there exist nonnegative numbers $\rho_1, \rho_2\geq 0 $ summing to $1$ ($\rho_1+\rho_2=1$) and $\beta_1, \beta_2 \in [0,1]$ so that 
\begin{IEEEeqnarray}{rCl}
  r_{1} &\leq &\beta_1  \sqrt{2} \frac{ \rho_{1}   \mathbb{D} \left(\Gamma_{Y|X_1X_2}(\cdot | 1, 0) \mid \mid \Gamma_{Y|X_1X_2}(\cdot | 0, 0)  \right)}{\sqrt{ \chi^2(\rho_1, \rho_2)}},\label{eq:r1a} \\[1ex] 
    r_{2} &\leq &\beta_2  \sqrt{2} \frac{ \rho_{2}   \mathbb{D} \left(\Gamma_{Y|X_1X_2}(\cdot | 0,1) \mid \mid \Gamma_{Y|X_1X_2}(\cdot | 0, 0)  \right)}{\sqrt{ \chi^2(\rho_1, \rho_2)}} \label{eq:r2a}, \\[1ex]
k_{1}  &\geq&\beta_1  \sqrt{2} \frac{  \rho_{1} \left(\mathbb{D} \left(\Gamma_{Z|X_1X_2}(\cdot | 1, 0) \mid \mid \Gamma_{Z|X_1X_2}(\cdot | 0, 0)  \right) - \mathbb{D} \left(\Gamma_{Y|X_1X_2}(\cdot | 1, 0) \mid \mid \Gamma_{Y|X_1X_2}(\cdot | 0, 0)  \right)\right)} {\sqrt{  \chi^2(\rho_{1}, \rho_{2})  }} , \label{eq:k1a}\\[1ex]
k_{2}  &\geq&\beta_2  \sqrt{2} \frac{  \rho_{2} \left(\mathbb{D} \left(\Gamma_{Z|X_1X_2}(\cdot | 0,1) \mid \mid \Gamma_{Z|X_1X_2}(\cdot | 0, 0)  \right) - \mathbb{D} \left(\Gamma_{Y|X_1X_2}(\cdot | 0,1) \mid \mid \Gamma_{Y|X_1X_2}(\cdot | 0, 0)  \right)\right)} {\sqrt{  \chi^2(\rho_{1}, \rho_{2})  }},
\end{IEEEeqnarray}
where we use the abbreviations
\begin{IEEEeqnarray}{rCl}
\Gamma_{Y|X_1X_2}(y|x_1,x_2) &\triangleq & \Gamma_{Y \mid X_1X_2X_3}(y|x_1,x_2,x_3),\\
\Gamma_{Z|X_1X_2}(y|x_1,x_2)&\triangleq & \Gamma_{Z \mid X_1X_2X_3}(y|x_1,x_2,x_3),\\
 \chi^2(\rho_{1}, \rho_{2}) & \triangleq  &\sum_{z \in \mathcal{Z}}  \frac{\left( \rho_1 \cdot \Gamma_{Z|X_1X_2}(z | 1, 0)  + \rho_2 \cdot \Gamma_{Z|X_1X_2}(z | 0, 1) - \Gamma_{Z|X_1X_2}(z | 0, 0) \right)^2}{\Gamma_{Z|X_1X_2}(z | 0, 0)}. 
\end{IEEEeqnarray}
\end{corollary}
\begin{IEEEproof}
We present the proof assuming that  $\mathcal{X}_3=\{x_3\}$ is a singleton. Under the assumptions \eqref{eq:conditions_nox3} the proof is similar. 

We start by proving that without loss in generality in Theorem~\ref{th:asymp_result} one can restrict to constant  $T$.  To this end, define  $\bar{\rho}_\ell \triangleq \mathbb{E}_{P_{T}}[ \rho_{\ell,T} ]$, and notice that when $X_3$ is a  constant $x_3$, the numerators  of \eqref{eq:asymp1} and \eqref{eq:asympkey} simplify to $\bar{\rho}_\ell \cdot \D_{Y}^{(\ell)}(x_3) $ and $\bar{\rho}_\ell \cdot \left(\D_{Z}^{(\ell)}(x_3) - \D_{Y}^{(\ell)}(x_3) \right)$, respectively. Moreover, in this case, the denominator of \eqref{eq:asymp1} and \eqref{eq:asympkey} can be lower bounded as: 
\begin{IEEEeqnarray}{rCl}
\lefteqn{\mathbb{E}_{P_{T}} \left[ \left( \rho_{1,T} + \rho_{2,T} \right)^2  \cdot \chi^2(\rho_{1,T}, \rho_{2,T}, x_3) \right] }\nonumber\\
& = & \mathbb{E}_{P_{T}} \left( \sum_{z \in \mathcal{Z}}  \frac{\left( \rho_{1,T} \Gamma_{Z \mid X_1X_2X_3}(z | 1, 0, x_3)  +\rho_{2,T}  \Gamma_{Z \mid X_1X_2X_3}(z | 0, 1, x_3) - (\rho_{1,T}+\rho_{2,T}) \Gamma_{Z \mid X_1X_2X_3}(z | 0, 0, x_3) \right)^2}{\Gamma_{Z \mid X_1X_2X_3}(z | 0, 0, x_3)} \right) \nonumber\\
& \geq &  \sum_{z \in \mathcal{Z}}  \frac{\left(\bar{\rho}_{1}\cdot \Gamma_{Z \mid X_1X_2X_3}(z | 1, 0, x_3)  +\bar{\rho}_2\cdot \Gamma_{Z \mid X_1X_2X_3}(z | 0, 1, x_3) -( \bar{\rho}_1+\bar{\rho}_2)\cdot \Gamma_{Z \mid X_1X_2X_3}(z | 0, 0, x_3) \right)^2}{\Gamma_{Z \mid X_1X_2X_3}(z | 0, 0, x_3)} , \label{eq:change}
\end{IEEEeqnarray}
where the inequality holds by  the convexity of the square-function and Jensen's inequality. 

We conclude that \ablast{replacing} $\rho_{1,t}$ and $\rho_{2,t}$ by the respective expectations $\bar{\rho}_1$ and $\bar{\rho}_2$ (and thus $T$ by a constant) does not change the numerator of the right-hand sides of \eqref{eq:asymp1} and \eqref{eq:asympkey}, while it \ablast{divides} all the denominators  by the same factor $\sqrt{\gamma} \geq 1$, for $\gamma$  the ratio between the left- and  right-hand sides of \eqref{eq:change}. Dividing  $\beta_\ell$ by \ablast{$\sqrt{\gamma}$}, we can recover the same constraints on the rates and keys, from which we started.  There is thus no reason to consider non-constant random variables $T$.

Notice  further that for a single $T=t$ the rate and key expressions in Theorem~\ref{th:asymp_result}  only depend on the normalized coefficients $\frac{\rho_{1,t}}{\rho_{1,t}+\rho_{2,t}}$ and $\frac{\rho_{2,t}}{\rho_{1,t}+\rho_{2,t}}$ but not on the absolute values of ${\rho}_{1,t}$ and $\rho_{2,t}$, because $\chi^2(\rho_{1,t}, \rho_{2,t}, x_3)$ also only depends on the ratios $\frac{\rho_{1,t}}{\rho_{1,t}+\rho_{2,t}}$ and $\frac{\rho_{2,t}}{\rho_{1,t}+\rho_{2,t}}$ but not on their absolute values. This  implies that without loss in generality we can restrict to $\rho_{1,t}+\rho_{2,t}=1$, which establishes the above corollary.
\end{IEEEproof}

\begin{remark}
  Corollary~\ref{cor:two_users}  strengthens  the results in \cite{bloch_k_users_mac} for two covert users  because in Corollary~\ref{th:asymp_result} we characterize the required secret-key rates for all achievable covert rates, not only the ones on the boundary of the achievable region.  We recall that we need local randomness only at one of the users.
\end{remark}\bigskip

In a similar way,  we can recover the fundamental limits for a single-user communication system. 
\ablast{We start from Corollary 2 (where $\rho_1+\rho_2 =1$) and assume }that for any $x_1, x_2, x_3$:
\begin{subequations}\label{eq:singleuser}
\begin{IEEEeqnarray}{rCl} 
\Gamma_{Y \mid X_1X_2X_3}(y|x_1,x_2,x_3) &= & \Gamma_{Y|X_1}(y|x_1), \\
\Gamma_{Z \mid X_1X_2X_3}(y|x_1,x_2,x_3) &= & \Gamma_{Z|X_1}(z|x_1),\label{eq:Z}
\end{IEEEeqnarray}
\end{subequations}
for some conditional pmf $\Gamma_{Y|X_1}$. This  immediately implies that Users 2 and 3 cannot communicate reliably, i.e., we can restrict to $r_2=k_2=R_3=0$. Moreover, under Assumption \eqref{eq:Z}, we have 
\begin{IEEEeqnarray}{rCl}
 \chi^2(\rho_{1}, \rho_{2}) & =  &\sum_{z \in \mathcal{Z}}  \frac{\left( \rho_1 \cdot \Gamma_{Z|X_1}(z | 1) + \rho_2  \cdot \Gamma_{Z|X_1}(z | 0)  - \ablast{(\rho_1 + \rho_2) \cdot }\Gamma_{Z|X_1}(z | 0) \right)^2}{\Gamma_{Z|X_1}(z | 0)} \label{eq:first_step_cor_2}\\
&=& \sum_{z \in \mathcal{Z}}  \frac{\left( \rho_1 \cdot \Gamma_{Z|X_1}(z | 1)   - \rho_1\cdot \Gamma_{Z|X_1}(z | 0) \right)^2}{\Gamma_{Z|X_1}(z | 0)}\\
 &=& \rho_1^2  \cdot \sum_{z \in \mathcal{Z}}  \frac{\left(   \Gamma_{Z|X_1}(z | 1)   -  \Gamma_{Z|X_1}(z | 0) \right)^2}{\Gamma_{Z|X_1}(z | 0)},
\end{IEEEeqnarray}
where \ablast{in \eqref{eq:first_step_cor_2}} we used that \ablast{$\rho_1+\rho_2 =1$}. 
Therefore, the $\rho_1$-factor cancels in constraints \eqref{eq:asymp1} and \eqref{eq:asympkey} and the following corollary is obtained. Define 
\begin{IEEEeqnarray}{rCl}
\D_Y& \triangleq&  \mathbb{D} \left(\Gamma_{Y|X_1}(\cdot | 1) \mid \mid \Gamma_{Y|X_1}(\cdot | 0)  \right),\\
\D_Z&\triangleq&  \mathbb{D} \left(\Gamma_{Z|X_1}(\cdot | 1) \mid \mid \Gamma_{Z|X_1}(\cdot | 0)  \right),\\
 \chi^2 & \triangleq  &\sum_{z \in \mathcal{Z}}  \frac{  \left( \Gamma_{Z|X_1}(z | 1)- \Gamma_{Z|X_1}(z | 0 ) \right)^2 }{\Gamma_{Z|X_1}(z | 0)}. 
\end{IEEEeqnarray}

We have the following corollary.
\begin{corollary}[Only a Single Covert User]\label{cor:SU}
Assume \eqref{eq:singleuser} and $r_2=k_2=R_3=0$. Then a message and secret-key rate pair for User 1 $(r_1,k_1)$ is achievable if, and only if, there exists a number $\beta_1 \in [0,1]$ so that 
\begin{IEEEeqnarray}{rCl}
  r_{1} &\leq &\beta_1  \sqrt{2} \frac{\D_Y}{\sqrt{ \chi^2 } }, \\[1ex] 
  k_{1}  &\geq&\beta_1  \sqrt{2} \frac{\D_Z-\D_Y} {\sqrt{\chi^2}} .
\end{IEEEeqnarray}
Defining the \emph{secret-key covert-capacity tradeoff} $r_1^\star(k_1)$ as the largest rate achievable given a key-rate budget $k_1$, 
\begin{equation}
r_1^\star(k_1)= \max \left\{ r_1 : (r_1,k_1) \text{ is achievable} \right\},
\end{equation}
we have: 
\begin{equation}
r_1^\star(k_1)= \min \left\{ k_1 \frac{ \D_Y}{ \max\{\D_Z-\D_Y, 0\}}, \; \sqrt{2}\frac{\D_Y }{\ablast{\sqrt{\chi^2}}} \right\}.
\end{equation} 
\end{corollary}

\begin{figure}[h!]
    \centering
    \includegraphics[width=0.55\textwidth]{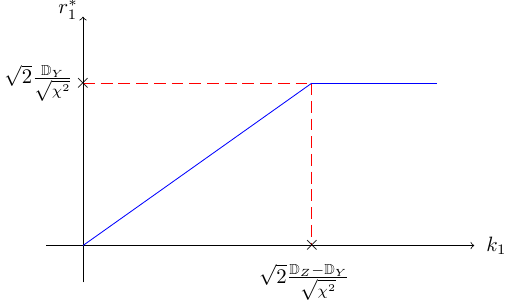}
    \caption{The secret-key covert-capacity tradeoff for the scenario $\mathbb{D}_Z >\mathbb{D}_Y$.}
    \label{fig:r_key_generic_plot}
\end{figure}

For channels where $\D_Z>\D_Y$, the secret-key covert-capacity tradeoff is thus linearly increasing in the secret-key rate $k_1\in \Big[0, \sqrt{2} \frac{\D_Z-\D_Y} {\sqrt{  \chi^2  }} \Big]$, and saturates to the largest covert rate for all larger secret-key rates, see Figure~\ref{fig:r_key_generic_plot}. In particular, for $1$ additional key bit, one can transmit $\frac{\D_Y}{\D_Z-\D_Y}$ covert message bits. For channels where $\D_Z \leq  \D_Y$ the covert capacity is constant and does not require any positive secret-key rate.

Again, Corollary~\ref{cor:SU} not only recovers the result in \cite{bloch_first}  but even strengthens it,  because in \cite{bloch_first}  the required secret-key rate is characterized only for covert capacity and not for achievable covert rates. In particular, based on the results in \cite{bloch_first} it is not possible to characterize the 
\ablast{secret-}key covert capacity tradeoff $r_1^\star(k_1)$. 
\begin{remark}
Corollary~\ref{cor:SU} can be achieved with a deterministic scheme where the single user does not have access to additional local randomness.
\end{remark} 
\medskip 
\subsection{Numerical examples}
Consider binary input alphabets for all three users, i.e. $\mathcal{X}_1=\mathcal{X}_2=\mathcal{X}_3=\{0,1\}$, and output alphabets $\mathcal{Y}=\mathcal{Z}=\{1,\ldots, 6\}$. Assume  the channel transition laws
\begin{subequations}\label{eq:channels}
    \begin{IEEEeqnarray}{rCl}
        \Gamma_{Y \mid X_1X_2X_3}=\begin{bmatrix}
            0.28 & 0.26 & 0.02 & 0.01 & 0.18 &  0.25 \\
            0.12 & 0.36 & 0.29 & 0.06 & 0.11 &  0.06 \\
            0.17 & 0.14 & 0.25 & 0.10 & 0.13 &  0.21 \\
            0.05 & 0.15 & 0.31 & 0.28 & 0.01 &  0.20 \\
            0.08 & 0.18 & 0.02 & 0.25 & 0.39 &  0.08 \\
            0.05 & 0.21 & 0.13 & 0.28 & 0.03 &  0.30 \\
            0.15 & 0.05 & 0.10 & 0.17 & 0.33 &  0.20 \\
            0.05 & 0.25 & 0.10 & 0.20 & 0.10 &  0.30 
        \end{bmatrix}, \nonumber \\
    \end{IEEEeqnarray}
    and
    \begin{IEEEeqnarray}{rCl}
        \Gamma_{Z \mid X_1X_2X_3}= \begin{bmatrix}
            0.15 & 0.11 & 0.57 & 0.01 & 0.06 & 0.10 \\
            0.15 & 0.41 & 0.12 & 0.15 & 0.06 & 0.11 \\
            0.23 & 0.02 & 0.01 & 0.48 & 0.10 & 0.16 \\
            0.14 & 0.17 & 0.21 & 0.12 & 0.24 & 0.12 \\
            0.01 & 0.12 & 0.19 & 0.15 & 0.19 & 0.34 \\
            0.10 & 0.11 & 0.15 & 0.14 & 0.18 & 0.32 \\
            0.05 & 0.15 & 0.15 & 0.20 & 0.10 &  0.35 \\
            0.10 & 0.10 & 0.27 & 0.13 & 0.20 &  0.20
        \end{bmatrix}, \nonumber \\
    \end{IEEEeqnarray}
\end{subequations}
where the six columns correspond to the six output symbols $1,\ldots, 6$ and the eight rows correspond to the eight distinct triples $(x_1,x_2,x_3)$ in increasing alphabetical ordering, i.e., $(0,0,0), (0,0,1), \ldots, (1,1,1)$. 
Notice that above channels satisfy Conditions \eqref{eq:channel_conditions}.\\

\begin{figure}[h!]
    \centering
    \begin{subfigure}{0.45\textwidth}
        \centering
        \includegraphics[width=\textwidth]{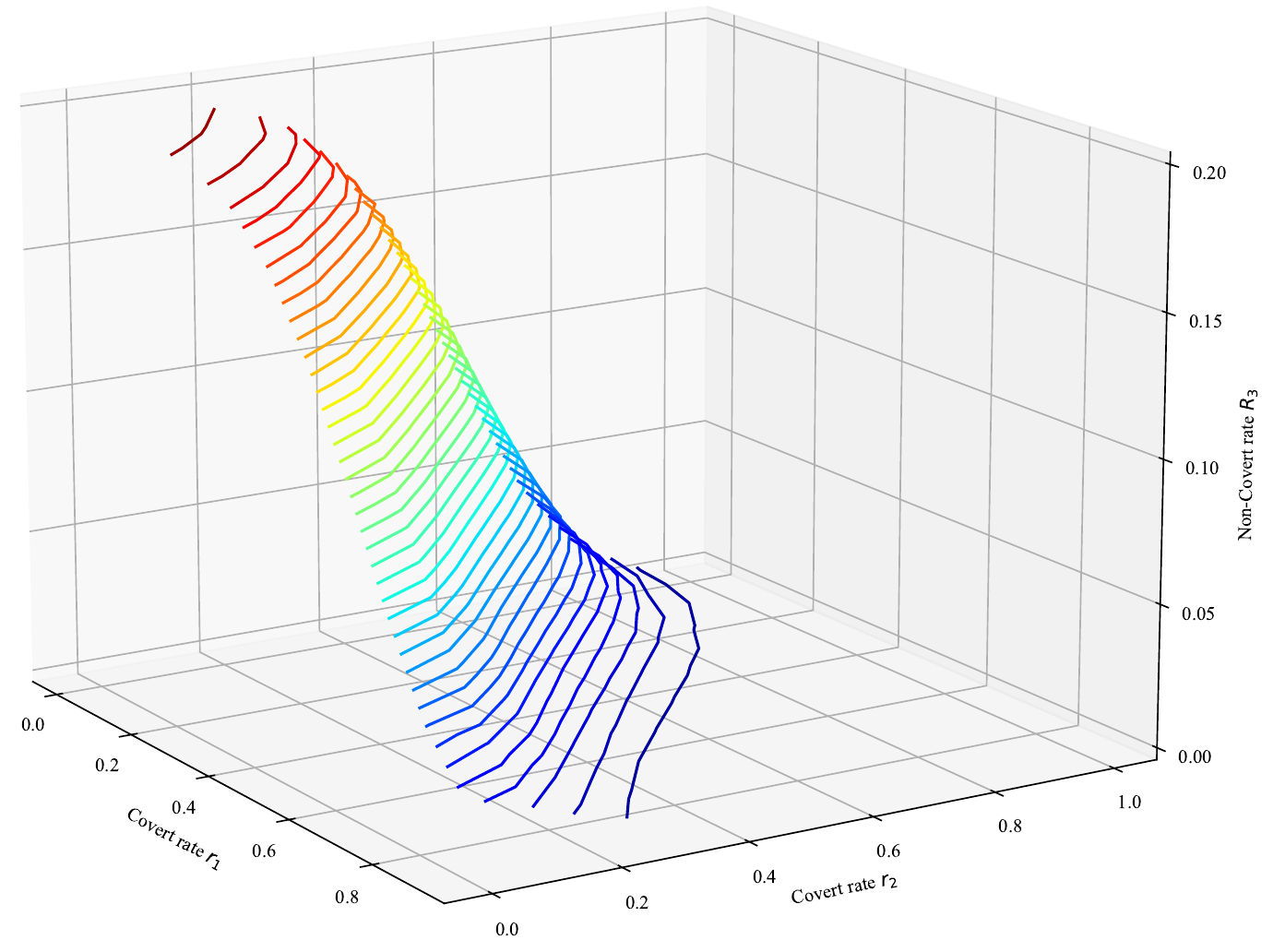}
        \caption{Contour plot of the rate-region $(r_1,r_2,R_3)$}
        \label{fig:contour_plot_3D}
    \end{subfigure}
    \hfill
    \begin{subfigure}{0.45\textwidth}
        \centering
        \includegraphics[width=\textwidth]{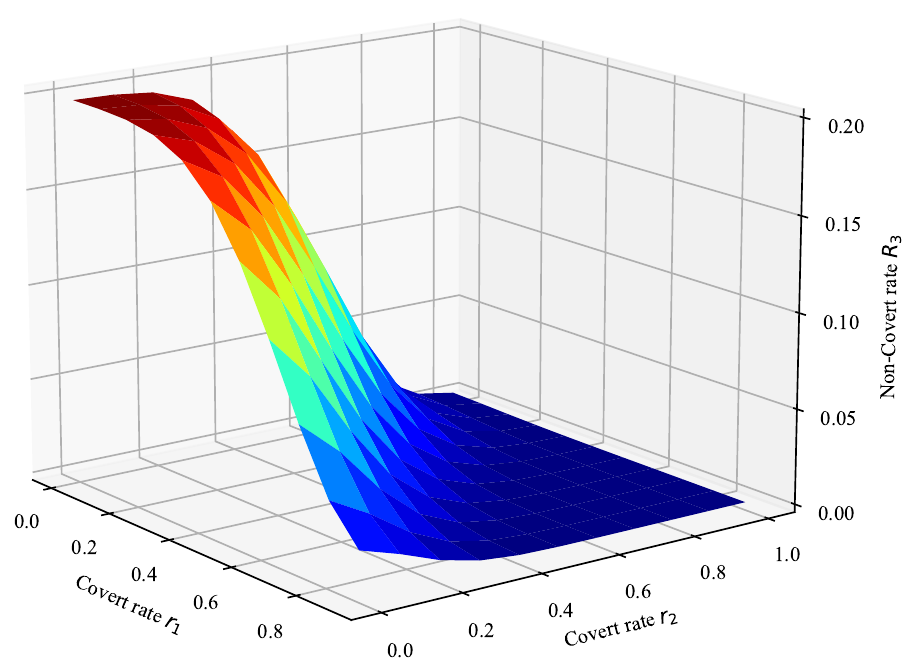}
        \caption{3D-rate region $(r_1,r_2,R_3)$}
        \label{fig:plot_3D}
    \end{subfigure}
    \caption{Rate-region $(r_1,r_2,R_3)$ for the channels in \eqref{eq:channels} and secret-key rate budgets $k_1,k_2\leq 0.8$.}
    \label{fig:simulation_3D}
\end{figure}

 Figure~\ref{fig:simulation_3D} illustrates the set of achievable triples $(r_1,r_2,R_3)$ according to Theorem~\ref{th:asymp_result} for the channels in \eqref{eq:channels} and maximum secret-key rate budgets $k_1, k_2\leq 0.8$.

\begin{figure}[!h]
    \centering
    \begin{tikzpicture}

        \begin{axis}[%
                width=2.919in,
                height=1.172in,
                at={(0in,0in)},
                scale only axis,
                xmin=0,
                xmax=0.9,
                xlabel style={font=\color{white!15!black}},
                xlabel={Covert rate $r_2$},
                ymin=0,
                ymax=0.2,
                ytick={0, 0.1,  0.2},
                ylabel style={font=\color{white!15!black}},
                ylabel={Non-Covert rate $R_3$},
                axis background/.style={fill=white},
                title style={font=\bfseries},
                axis x line*=bottom,
                axis y line*=left,
                legend style={legend cell align=left, align=left,                                   draw=white!15!black,
                                at={(0.75,0.85)}, 
                                anchor=west}
            ]
            \addplot [color=black, dashed]
                  table[row sep=crcr]{%
                1.77972354293735e-07	0\\
                9.24722273239988e-07	0\\
                1.28262630379367e-06	0\\
                1.57923360902797e-06	0\\
                1.60757470770757e-06	0\\
                2.18672888627377e-06	0\\
                3.15583642498565e-06	0\\
                3.83636298066565e-06	0\\
                4.59420092057048e-06	0\\
                7.5713378953655e-06	0\\
                1.40857988992564e-05	0\\
                1.99248310622231e-05	0\\
                2.46620515194424e-05	0\\
                2.55525811790027e-05	0\\
                2.92903849235573e-05	0\\
                3.25239878985047e-05	0\\
                3.58609885302899e-05	0\\
                4.00805258409743e-05	0\\
                4.21970493268816e-05	0\\
                4.53365235073852e-05	0\\
                6.16242574539614e-05	0\\
                6.65426964019677e-05	0\\
                7.6132211504721e-05	0\\
                8.08610547761701e-05	0\\
                8.444004128534e-05	0\\
                8.49350001903799e-05	0\\
                0.000124951506738062	0\\
                0.000128391150556218	0\\
                0.000155430230610769	0\\
                0.000179657514605798	0\\
                0.000191938051779501	0\\
                0.000201248590660078	0\\
                0.000250742995779308	0\\
                0.000254541964520669	0\\
                0.000297255906973181	0\\
                0.000312520736969612	0\\
                0.000320553464335987	0\\
                0.00032362081232753	0\\
                0.000336000888571033	0\\
                0.000397675664332345	0\\
                0.000416549323891534	0\\
                0.000435737439046125	0\\
                0.000445675311085909	0\\
                0.000460841244903013	0\\
                0.000489764080076321	0\\
                0.000501782922922911	0\\
                0.000510895926054142	0\\
                0.00051858060671794	0\\
                0.000534712905862477	0\\
                0.000666379043312699	0\\
                0.000695182186310884	0\\
                0.000717654079503694	0\\
                0.000767938741204489	0\\
                0.000770215563780916	0\\
                0.000789075183164599	0\\
                0.000962149015119825	0\\
                0.00109105093870636	0\\
                0.00114189888547703	0\\
                0.00128112617726145	0\\
                0.0023200975624742	0\\
                0.00263042722092185	0\\
                0.00323838343781641	0\\
                0.00363097000859849	0\\
                0.00385667284300041	0\\
                0.00398281822127545	0\\
                0.00486018653987314	0\\
                0.00537741299713777	0\\
                0.00633287777156006	0\\
                0.00637146015259385	0\\
                0.00639421766952885	0\\
                0.00641071654294489	0\\
                0.0072612319237656	0\\
                0.00802805316291177	0\\
                0.00851102878364846	0\\
                0.0198118015321473	0\\
                0.0342375279487474	0\\
                0.0475080531116297	0\\
                0.0538610694060557	0\\
                0.0612948314141095	0\\
                0.0727582091253656	0\\
                0.0786112146758962	0\\
                0.0786112146758962	0.000350376841645943\\
                0.0565228878692079	0.0642091201256181\\
                0.0535555881184512	0.0723088020210261\\
                0.0516658364032987	0.0771219689519001\\
                0.0472256598296449	0.0847885317389099\\
                0.0409356055296297	0.0902790668071149\\
                0.0121423776174048	0.101923909435917\\
                0.000270120219338757	0.106176748012991\\
                0	0.106176748012991\\
                0	0\\
                3.05863898172023e-10	0\\
                3.23943431192293e-10	0\\
                5.36356586068584e-10	0\\
                7.31143713721128e-10	0\\
                9.05064447941155e-10	0\\
                7.18048434527286e-09	0\\
                2.36782437434929e-08	0\\
                1.66047565735782e-07	0\\
                1.77972354293735e-07	0\\
                };
                 \addlegendentry{$r_1=0.75$} 
                
                \addplot [color=blue, dashdotted]
                  table[row sep=crcr]{%
                0.0209269421595458	0.197534055988625\\
                0	0.197534055988625\\
                0	0\\
                0.366260167743622	0\\
                0.366260167743622	5.92671339086962e-07\\
                0.352467830673336	0.0424563176191503\\
                0.343939259498895	0.0660921094158863\\
                0.339181407707355	0.0787362566994103\\
                0.331392190828389	0.0975435064355721\\
                0.328772634590651	0.103252582194686\\
                0.323773632757054	0.113946412617713\\
                0.321649663783969	0.118235190707539\\
                0.319782634959482	0.121845752369862\\
                0.30705598126345	0.14526089851093\\
                0.302077024896372	0.152408280744081\\
                0.300104990775399	0.15507733345502\\
                0.295096021914815	0.161840713033189\\
                0.289490747688808	0.168581308563562\\
                0.286194502639452	0.17252017700587\\
                0.283322580590531	0.175899359820123\\
                0.277828306137395	0.180667197252443\\
                0.276936739926482	0.181400071218315\\
                0.271389569503115	0.185843088156316\\
                0.268062365695835	0.188216602249852\\
                0.26650468880651	0.18917940579763\\
                0.259215929414188	0.193180954794483\\
                0.25905367335321	0.193251437416159\\
                0.256833409554561	0.194128357322257\\
                0.249730705704971	0.195836584992729\\
                0.241703581766645	0.196886110493879\\
                0.230366610495996	0.197459034796656\\
                0.225782818358957	0.197506105910677\\
                0.203832002692384	0.197526401063909\\
                0.0209269421595458	0.197534055988625\\
                };
                 \addlegendentry{$r_1=0.5$} 
                
                \addplot [color=red, solid]
                  table[row sep=crcr]{%
                0.479011611731705	0.197533997476211\\
                0.410539988152724	0.19753483958795\\
                0	0.19753483958795\\
                0	0\\
                0.653523154298476	0\\
                0.653523154298476	3.12351191284259e-07\\
                0.632458823538284	0.0595218125644257\\
                0.616857973913381	0.1012818747965\\
                0.614658440332804	0.107035517437211\\
                0.602722944584103	0.130441891300352\\
                0.594119614686427	0.144833010129501\\
                0.588539286489944	0.153590532618445\\
                0.579399623769747	0.165505157364048\\
                0.575144712444228	0.170503386973003\\
                0.573578728721614	0.172269156043192\\
                0.566786043546573	0.179332445370669\\
                0.55924302652555	0.18516940617372\\
                0.5525498908793	0.189831822875043\\
                0.547480271942865	0.192594783879944\\
                0.539083280295056	0.195634102869646\\
                0.536143631847814	0.196386133343222\\
                0.53495628444032	0.196651240074962\\
                0.531392617043895	0.197106424759261\\
                0.529525250269319	0.197309363015386\\
                0.528705907572764	0.197390787520215\\
                0.527300550373405	0.197463014950631\\
                0.525123037001012	0.197502586668147\\
                0.522032274031789	0.197513593323215\\
                0.516431893237586	0.197527890018186\\
                0.508952016438069	0.197533488075582\\
                0.479011611731705	0.197533997476211\\
                };
 \addlegendentry{$r_1=0.25$}

        \end{axis}
    \end{tikzpicture}%
    \caption{Rate-region $(r_2,R_3)$  for secret-key rates $k_1,k_2\leq 0.8$ and different rates $r_1$.}
    \label{fig:simulation_r2_vs_r3_for_different_rate_r1}
\end{figure}
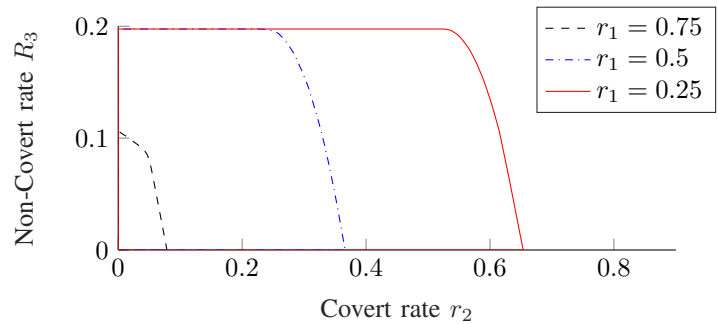

For better visualization, in Figure~\ref{fig:simulation_r2_vs_r3_for_different_rate_r1}, we also show the two-dimensional tradeoff between the rates $(r_2,R_3)$ for $k_1,k_2\leq 0.8$ and  $r_1$ fixed to  $0.25$, $0.5$, or $0.75$.
Furthermore, in Figure \ref{fig:simulation_r1_vs_r2_for_different_rate_r3}, we plot the tradeoff between the two covert users $(r_1,r_2)$ for secret-key rates $k_1,k_2\leq 0.8$ at different values of $R_3\in \{0.1965, 0.15, 0.05\}$.

It is also interesting to study the influence of the non-covert user on the set of achievable $(r_2,k_2)$ rate-key pairs.
To this end, in Figure~\ref{fig:non_covert_improves_covert_capacity} we plot the maximum achievable rate $r_2$ as a function of the secret-key rate $k_2$  without accounting for rates $r_1$ or $r_3$. 
 We compare this maximum rate to the maximum rate achievable for deterministic inputs $X_3=0$  and $X_3=1$. We observe that the gain in optimizing over a \emph{randomized} input $X_3$ 
 achieves larger gains than a simple convex-hull operation.

Finally, in Figure \ref{fig:simulation_time_sharing_beneficial} we show that a multiplexing (coded time-sharing) strategy is better than a single phase transmission.

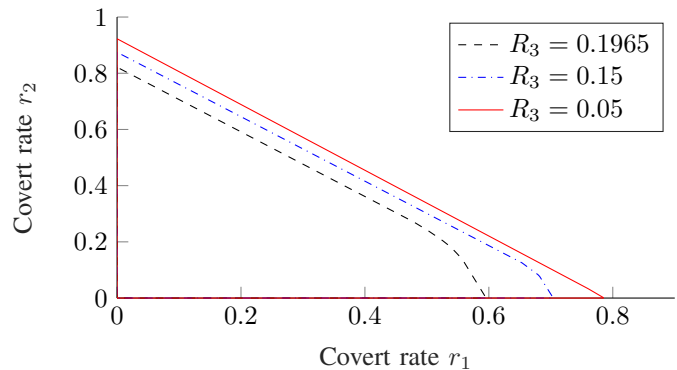
\begin{figure}[!h]
    \centering
    \begin{tikzpicture}

        \begin{axis}[%
                width=2.919in,
                height=1.472in,
                at={(0in,0in)},
                scale only axis,
                xmin=0,
                xmax=0.9,
                xlabel style={font=\color{white!15!black}},
                xlabel={Covert rate $r_1$},
                ymin=0,
                ymax=1,
                ylabel style={font=\color{white!15!black}},
                ylabel={Covert rate $r_2$},
                axis background/.style={fill=white},
                title style={font=\bfseries},
                axis x line*=bottom,
                axis y line*=left,
                legend style={legend cell align=left, align=left, draw=white!15!black}
            ]
            \addplot [color=black, dashed]
                  table[row sep=crcr]{%
                0.00028247093468165	0.821192778396772\\
                0	0.821192778396772\\
                0	0\\
                0.595592789798871	0\\
                0.595592789798871	4.64014149506044e-05\\
                0.558006625995573	0.135090434476761\\
                0.555513157900071	0.141791814663344\\
                0.549866267867281	0.155311166909427\\
                0.548254497569996	0.159157532924254\\
                0.54606638029449	0.164238126967975\\
                0.542968495227319	0.171217391778902\\
                0.540735994725553	0.175846712633924\\
                0.539414965887186	0.178564957415363\\
                0.533510830523509	0.190310300806822\\
                0.526137812795266	0.20314748201889\\
                0.51607007971086	0.21874149685832\\
                0.501269706254916	0.240784042735906\\
                0.478752161958453	0.270812151126057\\
                0.00028247093468165	0.821192778396772\\
                };
        \addlegendentry{$R_3=0.1965$}                
                \addplot [color=blue, dashdotted]
                  table[row sep=crcr]{%
                0.00050276799152913	0.874162793633329\\
                0	0.874162793633329\\
                0	0\\
                0.703830214187137	0\\
                0.703830214187137	0.000489253250488316\\
                0.681981404729577	0.0769953792384557\\
                0.679261163854927	0.0832623849548198\\
                0.653537557906411	0.125141342582735\\
                0.00050276799152913	0.874162793633329\\
                };
                 \addlegendentry{$R_3=0.15$}
                
                \addplot [color=red, solid]
                  table[row sep=crcr]{%
                0.760138346162865	0.0334011911168698\\
                0.000374187680054913	0.922274042076757\\
                0	0.922274042076757\\
                0	0\\
                0.78469230874742	0\\
                0.78469230874742	0.00178500013516695\\
                0.760138346162865	0.0334011911168698\\
                };
                        \addlegendentry{$R_3=0.05$}
        \end{axis}
    \end{tikzpicture}%
    \caption{Rate-region $(r_1,r_2)$ for secret-key rate $k_1,k_2\leq 0.8$ and different rates $R_3$.} 
    \label{fig:simulation_r1_vs_r2_for_different_rate_r3}
\end{figure}
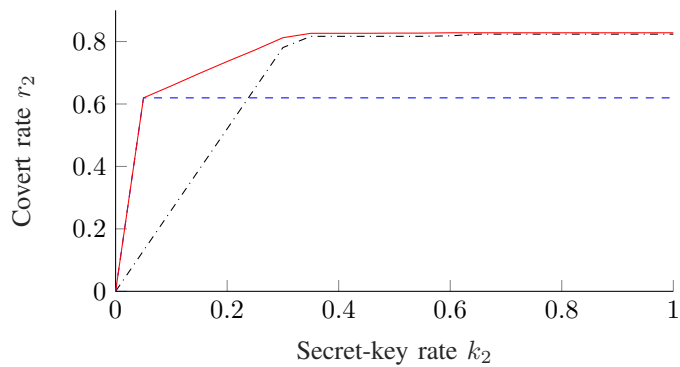
\begin{figure}[!h]
    \centering
    \begin{tikzpicture}
        \begin{axis}[
            width=2.919in,
            height=1.472in,
            at={(0in,0in)},
            scale only axis,
            xmin=0,
            xmax=1,
            xlabel style={font=\color{white!15!black}},
            xlabel={Secret-key rate $k_2$},
            ymin=0,
            ymax=0.9,
            ylabel style={font=\color{white!15!black}},
            ylabel={Covert rate $r_2$},
            axis background/.style={fill=white},
            title style={font=\bfseries},
            axis x line*=bottom,
            axis y line*=left,
            legend style={legend cell align=left, align=left, draw=white!15!black}
        ]
       
            \addplot [color=black, dashdotted]
              table[row sep=crcr]{%
            0	0\\
            0.05	0.130115815930637\\
            0.1	0.260231631861274\\
            0.15	0.39034744779191\\
            0.2	0.520463263722547\\
            0.25	0.650579079653184\\
            0.3	0.780694895583821\\
            0.35	0.816679234764655\\
            0.4	0.816679234764655\\
            0.45	0.816679234764655\\
            0.5	0.816679234764655\\
            0.55	0.816679234764655\\
            0.6	0.818212418241385\\
            0.65	0.823688104809623\\
            0.7	0.823688104809623\\
            0.75	0.823688104809623\\
            0.8	0.823688104809623\\
            0.85	0.823688104809623\\
            0.9	0.823688104809623\\
            0.95	0.823688104809623\\
            1	0.823688104809623\\
            };
            
            \addplot [color=blue, dashed]
              table[row sep=crcr]{%
            0	0\\
            0.05	0.619621086839637\\
            0.1	0.619621086839637\\
            0.15	0.619621086839637\\
            0.2	0.619621086839637\\
            0.25	0.619621086839637\\
            0.3	0.619621086839637\\
            0.35	0.619621086839637\\
            0.4	0.619621086839637\\
            0.45	0.619621086839637\\
            0.5	0.619621086839637\\
            0.55	0.619621086839637\\
            0.6	0.619621086839637\\
            0.65	0.619621086839637\\
            0.7	0.619621086839637\\
            0.75	0.619621086839637\\
            0.8	0.619621086839637\\
            0.85	0.619621086839637\\
            0.9	0.619621086839637\\
            0.95	0.619621086839637\\
            1	0.619621086839637\\
            };
            
            \addplot [color=red, solid]
              table[row sep=crcr]{%
            0	0\\
            0.05	0.619621086839637\\
            0.1	0.658135475139432\\
            0.15	0.697745909281459\\
            0.2	0.7360782728842\\
            0.25	0.772984150029413\\
            0.3	0.812282179138629\\
            0.35	0.826375679242755\\
            0.4	0.82642859609202\\
            0.45	0.826461923639183\\
            0.5	0.827025876391132\\
            0.55	0.827025876391132\\
            0.6	0.82807250538062\\
            0.65	0.82807250538062\\
            0.7	0.82807250538062\\
            0.75	0.82807250538062\\
            0.8	0.82807250538062\\
            0.85	0.82807250538062\\
            0.9	0.82807250538062\\
            0.95	0.82807250538062\\
            1	0.82807250538062\\
            };
            
        \end{axis}
        
        \end{tikzpicture}
        \caption{Covert rate $r_2$ as a function of the  secret-key rate $k_2$ when optimizing over $P_{X_3T}$  (solid line) and when choosing $X_3=0$ or $X_3=1$ deterministically (dashed and dash-dotted lines) for a covert rate $r_1=0.1$ and a secret-key rate $k_1 \leq 0.8$.}
    
        \label{fig:non_covert_improves_covert_capacity}
\end{figure}
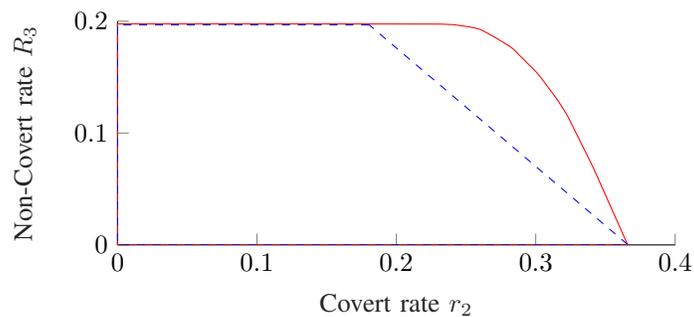
\begin{figure}[!h]
    \centering
    \begin{tikzpicture}
        \begin{axis}[%
            width=2.919in,
            height=1.172in,
            at={(0in,0in)},
            scale only axis,
            xmin=0,
            xmax=0.4,
            xlabel style={font=\color{white!15!black}},
            xlabel={Covert rate $r_2$},
            ymin=0,
            ymax=0.2,
            xtick={0, 0.1, 0.2, 0.3, 0.4},
            ytick={0, 0.1,  0.2},
            ylabel style={font=\color{white!15!black}},
            ylabel={Non-Covert rate $R_3$},
            axis background/.style={fill=white},
            title style={font=\bfseries},
            axis x line*=bottom,
            axis y line*=left,
            legend style={legend cell align=left, align=left, draw=white!15!black}
        ]
        \addplot [color=red]
            table[row sep=crcr]{%
                0.0209393603329447	0.197534050675693\\
                0	0.197534050675693\\
                0	0\\
                0.365874302371556	0\\
                0.365874302371556	0.00020475656488757\\
                0.341260993532471	0.0690075035111324\\
                0.33515527796459	0.0843432013787619\\
                0.323111967569048	0.114542756703675\\
                0.321752065485472	0.117541188433987\\
                0.315786299748673	0.129279386773892\\
                0.313386059962277	0.133311270079798\\
                0.304005523290595	0.149005425754013\\
                0.300323068543084	0.154598560290224\\
                0.283786335605196	0.175179901373978\\
                0.28137989410422	0.177415385091661\\
                0.272150823608734	0.184553522225264\\
                0.260392702464963	0.192644751530112\\
                0.255896812940455	0.193990325298559\\
                0.253087357283375	0.194657732732708\\
                0.251835506846633	0.194944103612356\\
                0.24821124899269	0.195608792433381\\
                0.241039965795286	0.196904759679634\\
                0.230998346134984	0.197390745020062\\
                0.178917199555822	0.197488598853102\\
                0.0209393603329447	0.197534050675693\\
                };
        
        \addplot [color=blue, dashed]
            table[row sep=crcr]{%
                0.18053796335818	0.19666811197991\\
                0	0.19666811197991\\
                0	0\\
                0.0159641442318079	0\\
                0.365564895078077	0\\
                0.366261291736838	0\\
                0.366261291736838	3.11518313179177e-09\\
                0.18053796335818	0.19666811197991\\
                };
        
        \end{axis}
        
    \end{tikzpicture}%
    
    \caption{Rate-region $(r_2,R_3)$ for secret-key rates $k_1, k_2\leq 0.8$ and $r_1=0.5$ in function of the allowed cardinality $|\mathcal{T}|$: we have $|\mathcal{T}|=6$ for the solid line and a degenerate region with  $|\mathcal{T}|=1$  for the dashed line.}

    \label{fig:simulation_time_sharing_beneficial}
\end{figure}

\section{Extensions}
\label{sec:extensions}
In this section, we broaden the scope of our findings from the multi-access scenario with 2 covert users and 1 non-covert user and with binary covert input alphabets $\{0,1\}$ to: 
\begin{itemize}
\item arbitrary numbers of covert and non-covert users; 
\item  arbitrary finite input alphabets at the covert users; and
\item the interference channel with two covert users and one non-covert user.
\end{itemize}

\subsection{Arbitrary Number of  Covert and Non-Covert Users}
\label{sec:general_setup}
\begin{figure}[h]
   \centering
   \includegraphics[width=0.99\textwidth]{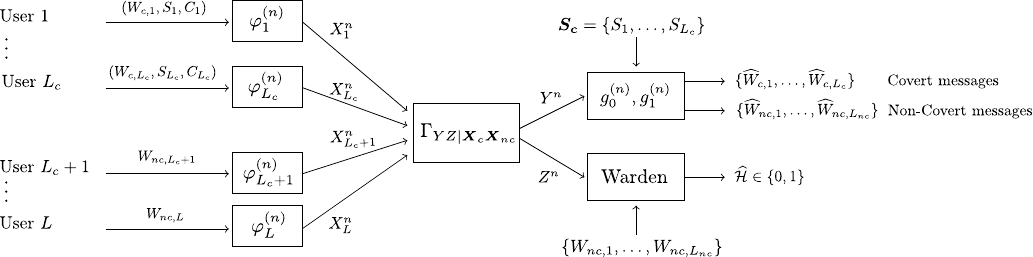}
   \caption{Multi-access communication model with $L_{\C}\geq 0$ covert users and $L_{\NC} \geq 0$  non-covert users.}
   \label{fig:general_setup}
\end{figure}

Consider the setup depicted in Figure \ref{fig:general_setup} where $L_{\C}$ covert  users  and $L_{\NC}$ non-covert users communicate  individual messages $W_1, \ldots, W_{L}$ for $L \triangleq L_{\C}+L_{\NC}$, to a legitimate receiver in the presence of a warden. Each message $W_{\ell}$ is uniformly distributed over the set $\mathcal{M}_{\ell} \triangleq \{1,\ldots, \mathsf{M}_{\ell}\}$, $ \ell \in \{1, \ldots, L\}$,  and independent of all other messages. Covert users can secure their transmissions at hand of secret-keys  $S_{\ell}$ which are uniform over the sets $ \{1,\ldots, \mathsf{K}_{\ell}\}$ and independent of each other and the messages, and also with independent local randomness $C_\ell$, which is uniform over $\{1,\ldots, \mathsf{G}_\ell\}$. The legitimate receiver and the warden observe channel outputs  produced by a memoryless interference channel with  finite output alphabets $\mathcal{Y}$ and $\mathcal{Z}$ and transition law $\Gamma_{YZ|X_1 \cdots X_{L}}$.
Similarly to Section~\ref{sec:problem_statement},  it is assumed that  the messages $W_{L_{\C}+1},\ldots, W_{L}$ transmitted by the non-covert users are known to the warden.

Each covert user $\ell=1, \ldots, L_{\C}$ produces its channel inputs $X^n_\ell \triangleq (X_{\ell,1},\ldots, X_{\ell,n})$ in the binary input alphabet $\mathcal{X}_{\ell}=\{0,1\}$, while each  non-covert user $\ell=L_{\C}+1, \ldots, L$ produces  inputs $X^n_\ell$ pertaining to an arbitrary finite alphabet $\mathcal{X}_{\ell}$.
Under $\mathcal{H}=0$ only the non-covert users transmit their messages:
\begin{IEEEeqnarray}{rCl}\label{eq:no_cover}
X_{\ell}^n(W_\ell) =\varphi_\ell^{(n)}(W_\ell), \quad \ell \in \{ L_{\C}+1, \ldots, L\},
\end{IEEEeqnarray}
for some encoding function $\varphi_{\ell}^{(n)}(\cdot)$ on appropriate domains, while the covert users send the all-zero sequences:
\begin{IEEEeqnarray}{rCl}\label{eq:cover}
X_{\ell}^n=0^n \quad \ell \in \{1,\ldots, L_{\C}\}.
\end{IEEEeqnarray}
In this case, the output distribution induced at the warden is  given by $\Gamma_{Z|\boldsymbol{X}_\C \boldsymbol{X}_{\NC}}^{\otimes n} (\cdot \mid \bm{0}^n , \bm{X}^n_{\NC}(\bm{W_{\NC}}) )$, 
where we define 
\begin{equation}
    \bm{W}_{\NC}\triangleq (W_{L_{\C}+1}, \ldots, W_L)
\end{equation}
and where
 \begin{equation}
    \bm{X}^n_{\NC}(\bm{W_{\NC}}) \triangleq (X_{L_{\C}+1}^n(W_{L_{\C}+1}), \ldots, X_L^n(W_{L})). 
\end{equation}

Under $\mathcal{H}=1$ all users communicate. That means,  \eqref{eq:no_cover} continues to hold, but \eqref{eq:cover} has to be replaced by 
\begin{IEEEeqnarray}{rCl}\label{eq:covert}
X_{\ell}^n(W_\ell, S_\ell)=\varphi_\ell^{(n)}(W_\ell, S_\ell, C_\ell), \quad \ell \in \{1, \ldots,  L_{\C}\},
\end{IEEEeqnarray}
for appropriate encoding functions. 
Thus, under $\mathcal{H}=1$, the output distribution induced at the warden is  given by 
\begin{IEEEeqnarray}{rCl}
\label{eq:def_Q_C_w_general}
\widehat{Q}_{\mathcal{C},\bm{w_{\NC}}}^{n}(z^{n}) &\triangleq&  \frac{1}{\prod_{\ell=1}^{L_{\C}} \mathsf{M}_{\ell} \mathsf{K}_{\ell} \mathsf{C}_\ell} \sum_{\bm{w_{\C}}, \bm{s_{\C}}, \bm{c}_{\C} }\Gamma_{Z|\boldsymbol{X}_\C \boldsymbol{X}_{\NC}}^{\otimes n} \left(z^n|  \bm{X}_{\C}^n(\bm{w_{\C}}, \bm{s_{\C}}), \bm{X}_{\NC}^n(\bm{w_{\NC}})  \right).
\end{IEEEeqnarray}
where we define $\bm{w}_{\C}\triangleq (w_{1}, \ldots, w_{L_{\C}})$,  $\bm{s}_{\C}\triangleq (s_{1}, \ldots, s_{L_{\C}})$, and  $\bm{c}_{\C}\triangleq (c_{1}, \ldots, c_{L_{\C}})$,  and 
 \begin{equation}
    \bm{X}^n_{\C}(\bm{w_{\C}}, \bm{s}_{\C},  \bm{c}_{\C}) \triangleq (X_{1}^n(w_{1},s_1, c_1), \ldots, X_{L_{\C}}^n(w_{L_{\C}}, s_{L_{\C}}, c_{{L_{\C}}})). 
\end{equation}

The error probabilities under the two hypotheses are defined as:
\begin{IEEEeqnarray}{rCl}
    P_{e,0} & \triangleq & \Pr\left(\bigcup_{\ell = L_{\C}+1, \ldots, L}  \widehat{W}_{\ell} \neq W_{\ell} \textsc{ } \Bigg| \textsc{ } \mathcal{H}=0\right), \label{eq:general_error_proba_H_0} \\
    P_{e,1} & \triangleq &\Pr\left(\bigcup_{ \ell = 1, \ldots, L}  \widehat{W}_{\ell} \neq W_{\ell} \textsc{ } \Bigg| \textsc{ } \mathcal{H}=1\right). \label{eq:general_error_proba_H_1}
\end{IEEEeqnarray}
For a given non-covert messages vector $\bm{w_{\NC}}$,   covertness at the warden is measured by the divergence
   \begin{equation}
\label{eq:def_delta_n_w2}
\delta_{n,\bm{w_{\NC}}}\triangleq \mathbb{D}\left(\widehat{Q}_{\mathcal{C},\bm{w_{\NC}}}^{n}\;  \Big\| \; \Gamma_{Z|\boldsymbol{X}_\C \boldsymbol{X}_{\NC}}^{\otimes n} (z^n \mid \bm{0}^n , \bm{X}^n_{\NC}(\bm{w_{\NC}}) ) \right).
\end{equation}

\begin{definition}
A tuple $(r_1, \ldots, r_{L_{\C}}, R_{L_{\C}+1}, \ldots, R_{L}, k_1, \ldots, k_{L_{\C}})$ is achievable if there exists a sequence (in the blocklength $n$)  of tuples $(\mathsf{M}_1, \ldots, \mathsf{M}_{L}, \mathsf{K}_1, \ldots, \mathsf{K}_{L_{\C}},\mathsf{G}_1, \ldots, \mathsf{G}_{L_{\C}})$ and encoding/decoding functions $\{(\varphi_1^{(n)}, \ldots, \varphi_{L}^{(n)}, g_0^{(n)}, g_1^{(n)})\}$  satisfying
\begin{IEEEeqnarray}{rCl}
    \label{eq:covert_constraint} 
    \lim_{n \rightarrow \infty} P_{e,\mathcal{H}} & = & 0,\qquad  \forall \mathcal{H}\in\{0,1\}\\
       \lim_{n \rightarrow \infty} \delta_{n,\bm{W_{\NC}}} &=& 0, \qquad \forall\bm{W_{\NC}} \in \mathcal{M}_{L_{\C}+1} \times \ldots \times \mathcal{M}_{L}\label{eq:Pei}
\end{IEEEeqnarray}
and 
\begin{IEEEeqnarray}{rCl}
r_{\ell} &=&   \liminf_{n \rightarrow \infty} \frac{\log(\mathsf{M}_{\ell})}{\sqrt{n \mathbb{E}_{\bm{W_{\NC}}} \left[ \delta_{n,\bm{W_{\NC}}} \right]}}, \qquad \ell\in\{1,\ldots, L_{\C}\},\\
\\[1ex]
R_{\ell} &=& \liminf_{n \rightarrow \infty} \frac{\log(\mathsf{M}_{\ell})}{n} ,\qquad \ell\in\{ L_{\C}+1,\ldots, L\},\\[1ex]
k_\ell &=&\limsup_{n \rightarrow \infty} \frac{\log(\mathsf{K}_{\ell})}{\sqrt{ n \mathbb{E}_{\bm{W_{\NC}}} \left[ \delta_{n,\bm{W_{\NC}}} \right]}}, \qquad \ell\in\{1,\ldots, L_{\C}\},
\end{IEEEeqnarray}
where $\bm{W_{\NC}}$ is uniformly distributed over the set of all possible vectors $\bm{w}_{\NC}$.
\end{definition}
 
For conciseness, and similarly to \eqref{eq:def_divergence_Y_user_1_for_notation}, \eqref{eq:def_divergence_Y_user_2_for_notation}, \eqref{eq:def_divergence_Z_user_1_for_notation}, \eqref{eq:def_divergence_Z_user_2_for_notation}, for any $\ell \in \{1, \ldots, L_\C\}$, and  tuple $\bm{x}_{\NC}\in\mathcal{X}_{L_\C + 1} \times \ldots \times \mathcal{X}_L$, we define the abbreviations
\begin{IEEEeqnarray}{rCl}
   \mathbb{D}_Y^{(\ell)}(\bm{x}_{\NC}) = \mathbb{D}\left( \Gamma_{Y|\bm{X}_{\C}\bm{X}_{\NC}}(\cdot \mid \bm{e}_{\ell}, \bm{x}_{\NC}) \| \Gamma_{Y|\bm{X}_{\C}\bm{X}_{\NC}}(\cdot \mid \bm{0}, \bm{x}_{\NC}) \right), \label{eq:def_divergence_Y_user_ell_for_notation} \\
   \mathbb{D}_Z^{(\ell)}(\bm{x}_{\NC}) = \mathbb{D}\left( \Gamma_{Z|\bm{X}_{\C}\bm{X}_{\NC}}(\cdot \mid \bm{e}_{\ell}, \bm{x}_{\NC}) \| \Gamma_{Z|\bm{X}_{\C}\bm{X}_{\NC}}(\cdot \mid \bm{0}, \bm{x}_{\NC}) \right),\label{eq:def_divergence_Z_user_ell_for_notation}
\end{IEEEeqnarray}
where  $\bm{e}_\ell$ denotes the $\ell$-th canonical basis vector $\bm{e}_{\ell}=(0,\ldots, 0, 1,0,\ldots, 0)$ of dimension $L_{\C}$.
Redefine also the set  $\mathcal{T}\triangleq \{1, \ldots, L + L_{\C} + 1\}$, and for any nonnegative tuple $\bm{\rho}=(\rho_1, \ldots, \rho_L)$ and   $\bm{x}_{\NC}\in \mathcal{X}_{L_{\C}+1} \times \cdots \mathcal{X}_{L}$, define
(similarly to \eqref{eq:def_xi_distance}):
    \begin{IEEEeqnarray}{rCl}
        \chi^2(\bm{\rho}, \bm{x}_{\NC}) &\triangleq& \sum_{z \in \mathcal{Z}} \frac{\left(\sum_{\ell=1}^{L_{\C}} \frac{\rho_{\ell}}{\| \bm{\rho}\|_1 } \Gamma_{Z|\bm{X}_{\C}\bm{X}_{\NC}}(z \mid \bm{e}_{\ell}, \bm{x}_{\NC}) - \Gamma_{Z|\bm{X}_{\C} \bm{X}_{\NC}} (z \mid \bm{0} , \bm{x}_{\NC} )\right)^2}{\Gamma_{Z|\bm{X}_{\C} \bm{X}_{\NC}}(z \mid \bm{0} , \bm{x}_{\NC} )}. \label{eq:def_xi_2_distance_general_setup_binary}
\end{IEEEeqnarray}

\begin{theorem}
    \label{th:general_channel_result}
A message and secret-key rate tuple $(r_1, \ldots, r_{L_{\C}}, R_{L_{\C}+1}, \ldots, R_{L}, k_1, \ldots, k_{L_{\C}})$ is achievable if, and only if, 
there exists a random variable $T$ over $\mathcal{T}$ and a random tuple $\bm{X}_{\NC}\triangleq (X_{L_{\C}+1}, \ldots, X_L)$ over $ \mathcal{X}_{L_{\C}+1} \times \cdots \times \mathcal{X}_{L}$  of joint pmf $P_{T\bm{X}_{\NC}}$, as well as a set of  nonnegative tuples  $\{\boldsymbol{\rho}_t \triangleq (\rho_{1,t}, \ldots, \rho_{L_{\C},t})\}_{t\in \mathcal{T}}$ and numbers $\beta_1, \ldots, \beta_{L_{\C}}\in [0,1]$,  so that the following inequalities hold:    
    \begin{IEEEeqnarray}{rCl}
    r_{\ell} & \mw{\leq} & \beta_\ell  \sqrt{2} \frac{\mathbb{E}_{P_{T\bm{X}_{\NC}}} \left[ \rho_{\ell,T} \mathbb{D}_Y^{(\ell)}(\bm{X}_{\NC}) \right]}{\sqrt{\mathbb{E}_{P_{T\bm{X}_{\NC}}} \left[ \|\bm{\rho}_T\|_1^2 \cdot  \chi^2(\bm{\rho}_T, \bm{X}_{\NC}) \right]}}\label{eq:rate_user_uc}, \quad \forall \ell \in \{1, \ldots, L_{\C}\}, \\[1ex]
     \sum_{j \in \mathcal{J}} R_{j} &\leq& I(\bm{X}_{\NC,\mathcal{J}};Y \mid \bm{X}_\C=0, \bm{X}_{nc,\mathcal{J}^{c}}, T), \quad \forall \mathcal{J} \subseteq \{L_{\C+1}, \ldots, L\}, \label{eq:rate_user_unc} \\[1ex]
     k_{\ell} &\geq& \beta_\ell \sqrt{2} \frac{\mathbb{E}_{P_{T\bm{X}_{\NC}}}  \left[ \rho_{\ell,T} \left( \mathbb{D}_Z^{(\ell)}(\bm{X}_{\NC}) - \mathbb{D}_Y^{(\ell)}(\bm{X}_{\NC}) \right) \right]} {\sqrt{\mathbb{E}_{P_{T\bm{X}_{\NC}}} \left[\|\bm{\rho}_T\|_1^2\cdot  \chi^2(\bm{\rho}_T, \bm{X}_{nc})  \right]}} \label{eq:general_key_rate}, \quad \forall \ell \in \{1, \ldots, L_{\C}\}.
    \end{IEEEeqnarray}
where for any subset of non-covert users $\mathcal{J} \subseteq \{L_{\C+1}, \ldots, L\}$:
\begin{equation}
    \bm{X}_{\NC,\mathcal{J}} = \{X_{\ell} : \ell \in \mathcal{J} \}.
\end{equation}

\end{theorem}
\begin{IEEEproof}
 A straightforward extension of the proof of  Theorem~\ref{th:asymp_result} and  omitted.   
\end{IEEEproof}

\begin{lemma}
    \label{lem:convexity}
    The set of $(L+L_{\C})$-dimensional vectors $(r_1, \ldots, r_{L_{\C}}, R_{L_{\C}+1}, \ldots, R_{L}, k_1, \ldots, k_{L_{\C}})$ satisfying Inequalities $\eqref{eq:rate_user_uc}-\eqref{eq:general_key_rate}$ for some choice of pmfs $P_{T\bm{X}_{nc}}$, tuples $\{\bm{\rho}_t\}_{t\in\mathcal{T}}$, and numbers $\beta_1, \ldots, \beta_{L_{\C}} \in \ablast{[}0,1]$ is a convex set. \end{lemma}
\begin{IEEEproof}
Similar to the proof of Lemma~\ref{lem:convexity} in Appendix~\ref{app:convexity} and   omitted. 
\end{IEEEproof}
\medskip

\subsection{Arbitrary Input Alphabets at the Covert Users}
\label{sec:general_setup_non_binary}
We extend the result in the previous subsection to  arbitrary input alphabets $\mathcal{X}_{\ell}$ at the covert user $\ell \in \{1, \ldots, L_{\C}\}$. The only restriction is that each $\mathcal{X}_\ell$ contains the 0 symbol, which we still consider to be the ``off-symbol" sent under $\mathcal{H}=0$. 

To state our main result for arbitrary covert input alphabets, we extend Definition  \eqref{eq:def_xi_2_distance_general_setup_binary} as follows. Given a tuple $\bm{\rho}\triangleq (\rho_1, \ldots, \rho_{L_{\C}})$, a set of  pmfs $\{ \psi_{\ell}(\cdot)\}_{\ell=1}^{L_{\C}}$ over $\mathcal{X}_\ell \backslash \{0\}$, and a  tuple $\bm{x}_{\NC}$, define: 
\begin{IEEEeqnarray}{rCl}
        \chi^2(\bm{\rho}, \{ \psi_\ell \},\bm{x}_{\NC}) &\triangleq& \sum_{z \in \mathcal{Z}} \frac{\left(\sum_{\ell=1}^{L_{\C}}  \frac{\rho_\ell}{\| \bm{\rho}\|_1} \sum_{x_\ell \in \mathcal{X}_{\ell}} \psi_{\ell}(x_\ell)\Gamma_{Z|\bm{X}_\C\bm{X}_{\NC}}(z \mid x_{\ell}\cdot \bm{e}_{\ell}, \bm{x}_{\NC}) - \Gamma_{Z|\bm{X}_\C \bm{X}_{\NC}} (z \mid \bm{0} , \bm{x}_{\NC} )\right)^2}{\Gamma_{Z|\bm{X}_\C \bm{X}_{\NC}}(z \mid \bm{0} , \bm{x}_{\NC} )}. \label{eq:def_xi_2_distance_general_setup_non_binary}
\end{IEEEeqnarray}
In a similar way, we extend the definitions of \eqref{eq:def_divergence_Y_user_ell_for_notation} and \eqref{eq:def_divergence_Z_user_ell_for_notation}, which now depend on the non-zero symbol $x_\ell \in \mathcal{X}_{\ell}$ used by  the covert user $\ell$. For given $x_\ell \in \mathcal{X}_\ell \backslash \{0\}$ and $\bm{x}_{\NC}$, define:
\begin{IEEEeqnarray}{rCl}
   \mathbb{D}_Y^{(\ell)}(x_\ell, \bm{x}_{\NC}) = \mathbb{D}\left( \Gamma_{Y|\bm{X}_{\C} \bm{X}_{\NC}}(\cdot \mid x_\ell \cdot \bm{e}_{\ell}, \bm{x}_{\NC}) \| \Gamma_{Y|\bm{X}_{\C}\bm{X}_{\NC}}(\cdot \mid \bm{0}, \bm{x}_{\NC}) \right), \label{eq:def_divergence_Y_user_ell_for_notation_non_binary} \\
   \mathbb{D}_Z^{(\ell)}(x_\ell, \bm{x}_{\NC}) = \mathbb{D}\left( \Gamma_{Z|\bm{X}_{\C}\bm{X}_{\NC}}(\cdot \mid x_\ell\cdot  \bm{e}_{\ell}, \bm{x}_{\NC}) \| \Gamma_{Z|\bm{X}_{\C}\bm{X}_{\NC}}(\cdot \mid \bm{0}, \bm{x}_{\NC}) \right). \label{eq:def_divergence_Y_user_ell_for_notation_non_binary}
\end{IEEEeqnarray}

\begin{theorem}[Arbitrary Covert Input Alphabets]
    \label{th:general_channel_non_binary_result}
  A message and secret-key rate tuple $(r_1,\ldots, r_{L_{\C}}, R_{L_{\C}+1}, \ldots, R_L, k_{1}, \ldots, k_{L_{\C}})$ is achievable   if, and only if, there exists a tuple of random variables $(T, \bm{X}_{\NC})$ over $\mathcal{T} \times \mathcal{X}_{L_{\C}+1}\times \cdots \times \mathcal{X}_{L}$ distributed according to a joint pmf $P_{T\bm{X}_{\NC}}$, nonnegative tuples $\{\bm{\rho}_{t}\triangleq (\rho_{1,t}, \ldots, \rho_{L_{\C},t})\}_{t\in\mathcal{T}}$ and numbers $\beta_1, \ldots, \beta_{L_\C}\in \ablast{[}0,1]$, and sets of marginal pmfs $\{ \bm{\psi}_T \triangleq \{ \psi_{1,t}, \ldots, \psi_{L_{\C},t}\}\}_{t \in \mathcal{T}}$  over $\mathcal{X}_{1}\backslash 0, \ldots, \mathcal{X}_{L_{\C}}\backslash \{0\}$ so that the following inequalities hold:    
    \begin{IEEEeqnarray}{rCl}
    r_{\ell} &\mw{\leq}&\beta_\ell  \sqrt{2} \frac{\mathbb{E}_{P_{T\bm{X}_{\NC}}} \left[ \rho_{\ell,T} \sum_{x_\ell \in \mathcal{X}_{\ell}} \psi_{\ell,T}(x_\ell) \cdot \mathbb{D}_Y^{(\ell)}(x_\ell, \bm{X}_{\NC}) \right]}{\sqrt{\mathbb{E}_{P_{T\bm{X}_{\NC}}} \left[ \| \bm{\rho}_{T} \|_1^2 \cdot \chi^2(\bm{\rho}_T, \bm{\psi}_T, \bm{X}_{\NC}) \right]}}\label{eq:rate_user_uc_non_binary}, \quad \forall \ell \in \{1, \ldots, L_{\C}\}, \\[1ex]
    \sum_{j \in \mathcal{J}} R_{j} &\leq& I(\bm{X}_{\NC,\mathcal{J}};Y \mid \bm{X}_{\C}=\bm{0}, \bm{X}_{\NC,\mathcal{J}^{c}}, T), \quad \forall \mathcal{J} \subseteq \{ L_{\C}+1, \ldots, L\}, \label{eq:rate_user_unc_non_binary} \\[1ex]
     k_{\ell} &\geq&\beta_\ell \sqrt{2} \frac{\mathbb{E}_{P_{T\bm{X}_{\NC}}}  \left[ \rho_{\ell,T} \sum_{x_\ell \in \mathcal{X}_{\ell}} \psi_{\ell,T}(x_\ell) \cdot \left( \mathbb{D}_Z^{(\ell)}(x_\ell, \bm{X}_{\NC}) - \mathbb{D}_Y^{(\ell)}(x_\ell, \bm{X}_{\NC}) \right) \right]} {\sqrt{\mathbb{E}_{P_{T\bm{X}_{\NC}}} \left[  \| \bm{\rho}_{T} \|_1^2 \cdot  \chi^2(\bm{\rho}_T, \bm{\psi}_T, \bm{X}_{\NC})  \right]}}, \quad \forall \ell \in \{1, \ldots, L_{\C}\}. \label{eq:general_key_rate_non_binary}
      \end{IEEEeqnarray}   
\end{theorem}
\begin{IEEEproof}
Achievability can be proved by analyzing an extension of the coding schemes proposed in Sections~\ref{sec:coding}  and \ref{sec:gen}, where in the code construction the entries of the codewords at a covert User $\ell=1, \ldots, L_{\C}$ are drawn conditionally i.i.d. given the $t^n$ sequence according to the conditional pmf 
\begin{equation}
P_{X_{\ell,n}|T}(x_\ell|t) = \psi_{\ell,t}(x_\ell) \cdot \rho_{\ell,t} \cdot\alpha_n, \quad x_\ell \in \mathcal{X}_{\ell} \backslash \{0\}
\end{equation}
and 
\begin{equation}
P_{X_{\ell,n}|T}(0|t) =1- \rho_{\ell,t} \cdot\alpha_n.
\end{equation}
The converse proof is obtained by generalizing the converse in Appendix~\ref{sec:proof2} to non-binary input alphabets at the covert users, similarly to \ablast{\cite[Appendix G]{bloch_k_users_mac_arxiv}}. \\
Details of both proofs are omitted. 
\end{IEEEproof}

\subsection{Interference channels}
\label{sec:interference_channels}
\begin{figure}[!h]
   \centering
   \includegraphics[scale=0.85]{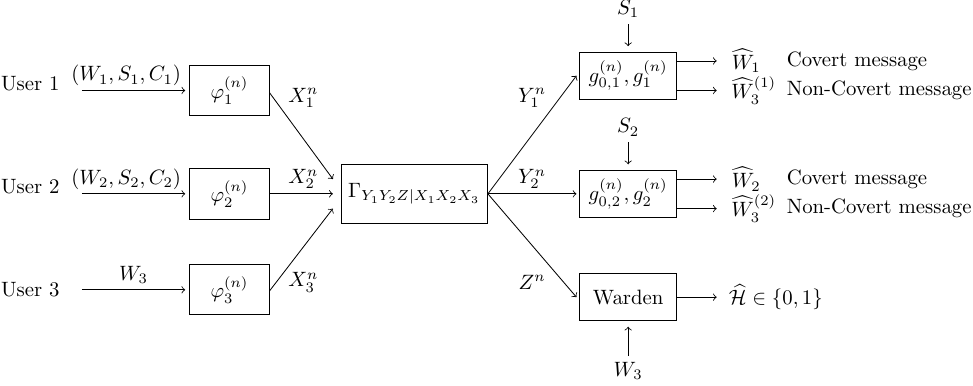}
   \caption{Interference channel with two covert users and a non-covert user sending a common message to both \ablast{legitimate} receivers.}
   \label{fig:setup_interference_channel}
\end{figure}

In this section, we consider the two-receiver discrete memoryless interference channel (DMIC)  in Figure~\ref{fig:setup_interference_channel}. We have two covert users, each sending a message to their respective \ablast{legitimate} receiver and a non-covert user sending a common message to both \ablast{legitimate} receivers. The transition law of the DMIC is denoted \ablast{$\Gamma_{{Y_1Y_2Z \mid X_1X_2X_3}}$}.

Encodings are as defined in Section~\ref{sec:problem_statement} for the multi-access channel.  Decoding now takes place at two different legitimate receivers $\ell\in \{1,2\}$,  which both know the hypothesis $\mathcal{H}$. 

Under $\mathcal{H}=0$ \ablast{each of the two legitimate receivers} $\ell$ decodes  the common message $W_3$ by producing the guess
\begin{equation}
\widehat{W}_{3}^{(\ell)} = g_{0,\ell}^{(n)}( Y_\ell^n)
\end{equation}
using a decoding function $g_{0,\ell}^{(n)}(\cdot)$.

Under $\mathcal{H}=1$, it 
uses a decoding function $g_\ell^{(n)}\colon \mathcal{Y}^n \times \mathcal{K}_\ell \to \mathcal{M}_\ell \times  \mathcal{M}_3$ to produce the pair of guesses
\begin{equation}
(\widehat{W}_\ell, \widehat{W}_3^{(\ell)}) = g_\ell^{(n)}(Y_\ell^n, S_\ell).
\end{equation}
Achievability is defined analogously to the DMMAC, see Definition~\ref{def:ach}, but where the definitions of the probabilities of error  \eqref{eq:prob2} and \eqref{eq:prob1} need to be replaced by 
\begin{IEEEeqnarray}{rCl}
    P_{e,0} & \triangleq & \Pr\left(\widehat{W}_3^{(1)} \neq W_3  \text{ or } \widehat{W}_3^{(2)} \neq W_3 \Big| \mathcal{H}=0\right), \label{eq:prob2a} \\
    P_{e,1} & \triangleq &\Pr\left(\widehat{W}_3^{(1)}  \neq W_3 \text{ or } \widehat{W}_3^{(2)} \neq W_3 \text{ or } \widehat{W}_{2} \neq W_{2} \text{ or } \widehat{W}_{1} \neq W_{1} \Big| \mathcal{H}=1\right). \label{eq:prob1a}
\end{IEEEeqnarray} 
Let $\mathcal{T}\triangleq \{1,\ldots, 7\}$. We have the following theorem for the DMIC.
\begin{theorem}
    \label{th:intereference_channel_result}
A message and secret-key rates tuple $(r_1,r_2,R_3,k_1,k_2)$ is achievable over the DMIC  
$\Gamma_{{Y_1Y_2Z \mid X_1X_2X_3}}$  
 if, and only if, there exists a pair of random variables $(T, X_3)$ over $\mathcal{T} \times \mathcal{X}_3$ 
distributed according to a pmf $P_{TX_3}$ over $\mathcal{T}$ and $\mathcal{X}_3$, nonnegative tuples  $\{  (\rho_{1,t}, \rho_{2,t})\}_{t\in \mathcal{T}}$, and $\beta_1, \beta_2 \in [0,1]$, so that for all $\ell \in \{1,2\}$, the following three inequalities hold:    
    \begin{IEEEeqnarray}{rCl}
      r_{\ell} &\leq&\mw{\beta_\ell} \sqrt{2} \frac{\mathbb{E}_{P_{TX_3}} \left[ \rho_{\ell,T} \D_{Y_{\ell}}^{(\ell)}(X_3) \right]}{\sqrt{\mathbb{E}_{P_{TX_3}} \left[ \left( \rho_{1,T} + \rho_{2,T} \right)^2 \chi^2(\rho_{1,T},\rho_{2,T}, X_3) \right]}}\label{eq:intereference_channel_r1}, \\[1ex]
     R_3 &\ab{\leq} & \min \left( \mathbb{I}(X_3;Y_1 \mid X_1=0,X_2=0,T) ; \mathbb{I}(X_3;Y_2 \mid X_1=0,X_2=0,T) \right) \label{eq:intereference_channel_r3}, \\[1ex]
     k_{\ell}  &\geq&\beta_\ell  \sqrt{2} \frac{\mathbb{E}_{P_{TX_3}}  \left[ \rho_{\ell,T} \left( \D_{Z}^{(\ell)}(X_3) - \D_{Y_{\ell}}^{(\ell)}(X_3) \right) \right]} {\sqrt{\mathbb{E}_{P_{TX_3}} \left[ \left( \rho_{1,T} + \rho_{2,T} \right)^2 \chi^2(\rho_{1,T},\rho_{2,T}, X_3) \right]}}.\label{eq:intereference_channel_k}
    \end{IEEEeqnarray}
\end{theorem}
Notice that here in Theorem~\ref{th:intereference_channel_result}, we have a different output signal $Y_\ell$ for each \ablast{legitimate r}eceiver $\ell \in \{
1,2\}$.
\begin{IEEEproof}
Analogous to the DMMAC proof, but with the following modifications for the achievability and converse proofs.
\subsubsection{Achievability}
In the decoding, \eqref{eq:decoding_user_3} needs to be replaced by the following two conditions (one for each \ablast{legitimate} receiver)
\begin{align}
(t^n,x_3^{n}(w_3), y_1^{n}) &\in \mathcal{T}_{\mu_n}^{n}(P_{TX_3Y_1}), \label{eq:decoding_user_3_interf_1} \\
(t^n,x_3^{n}(w_3), y_2^{n}) &\in \mathcal{T}_{\mu_n}^{n}(P_{TX_3Y_2}), \label{eq:decoding_user_3_interf_2}
\end{align}
\ablast{and} \eqref{eq:decoding_user_1} and \eqref{eq:decoding_user_2}  need to be replaced by 
\begin{equation}
(x_{1}^{n}(w_1, S_1), 0^n,x_3^{n}(\widehat{W}_3^{(1)}), y_1^{n})\in \mathcal{A}_{\eta_1}^{n},
\end{equation} 
and
\begin{equation}
(0^n,x_{2}^{n}(w_2, S_2),x_3^{n}(\widehat{W}_3^{(2)}), y_2^{n}) \in \mathcal{A}_{\eta_2}^{n}.
\end{equation}
Accordingly, in the analysis,  the DMMAC output $Y$ also need to be replaced by either of  the two DMIC outputs $Y_1$ or $Y_2$.
\subsubsection{Converse}
Replace the DMMAC output $Y$ by the two DMIC outputs $Y_\ell$  when deriving the bounds on  $r_\ell$ and on $k_\ell$, and perform the steps to bound $R_3$ twice: once using output $Y_1$ instead of $Y$ and once using output $Y_2$.
\end{IEEEproof}

\medskip

    Notice that now the rate of the non-covert message is the minimum rate at which the legitimate receivers can decode the non-covert message. This is because \eqref{eq:decoding_user_3_interf_1} requires  condition
    \begin{equation}
        R_3 \leq \mathbb{I}(X_3;Y_1 \mid X_1=0,X_2=0,T)
    \end{equation}
    for reliable decoding, 
    whereas \eqref{eq:decoding_user_3_interf_2} requires
    \begin{equation}
        R_3 \leq \mathbb{I}(X_3;Y_2 \mid X_1=0,X_2=0,T).
    \end{equation}
   
\section{Conclusion}
\label{sec:conclusion}
This paper characterizes the fundamental limits of a multi-access communication setup with covert users (sharing common secret-keys of fixed rates with the \ablast{legitimate} receiver) and non-covert users, all communicating with the same \ablast{legitimate} receiver in presence of a warden.
Our findings illustrate an intricate interplay among three pivotal quantities: the covert users' rates, the non-covert users' rate and the secret-key rates.
Similarly  to multi-access scenarios without covert constraints, our results emphasize the necessity of a multiplexing (coded time-sharing)  strategy in the code constructions,  so as to exhaust the entire tradeoff of achievable covert and non-covert rates. This holds even with only a single non-covert user. 
Our results also prove  that the presence of the non-covert users can increase the covert-capacity under a stringent \ablast{secret-key} rate constraint.

The scenario considered in this work contains as interesting special cases   covert communication over a single-user DMC or over a DMMAC (without non-covert users). Our results also imply new findings for these previously studied special cases. In fact, with our new results\ablast{,} we can characterize the minimum \ablast{secret-key} rates that are required to achieve any set over covert data-rates, while previous works only characterized the \ablast{secret-key} rates required to transmit at largest covert rates. With our new results we can thus determine the set of covert rates that are achievable over a DMC or a DMMAC under a given stringent secret-key budget, which was not possible with the previous \ablast{findings}.
Notice that while for the single-user DMC our findings apply both to setups with and without local randomness at the transmitter, for the DMMAC our scheme requires local randomness at least at some of the transmitters when we do not communicate at largest possible covert data-rates.

We further showed that our findings on the mixed covet and non-covert DMMAC naturally also extend to related setups including the interference channels or channels with a friendly jammer. 
Further interesting research directions include studies of fading channels, or non-synchronized transmissions.

\begin{appendices}

\section{Proof of Theorem~\ref{main_theorem}}\label{sec:proof1}

\subsection{Analysis of the Decoding Error Probability  of  the  Scheme in Section~\ref{sec:coding}}
\label{sec:error_proba_analysis}
In this section, we analyze the expected error probabilities $\mathbbm{E}_{\mathcal{C}}[  P_{e,0} ]$ and $\mathbb{E}_{\mathcal{C}}[  P_{e,1}]$, where  expectations are with respect to the random codebooks.\\
\subsubsection{Analysis of $\mathbb{E}_{\mathcal{C}}[P_{e,0}]$}
By standard arguments and because for all $t \in \mathcal{T}$, $P_{X_{1,n}}(0 \mid t)$ and $P_{X_{2,n}}(0 \mid t)$ tend to $1$ as $n\to \infty$:
\begin{equation}\label{eq:Pe0}
\lim_{n\to\infty} \mathbb{E}_{\mathcal{C}}[  P_{e,0}] =0
\end{equation}
whenever
\begin{equation}\label{eq:R2}
\varlimsup_{n\to\infty} \frac{1}{n} \log(\mathsf{M_3})\leq  I(X_3;Y \mid X_1=0,X_2=0,T) .
\end{equation}
\subsubsection{Analysis of $\mathbb{E}_{\mathcal{C}}[P_{e,1}]$} 
Define the   probabilities 
\begin{IEEEeqnarray}{rCl}
    P_{e,1,1} & \triangleq &\Pr\left( \widehat{W}_1 \neq W_1 \textsc{ }\Big| \mathcal{H}=1 \right) \label{eq:prob11},\\
    P_{e,1,2} & \triangleq &\Pr\left( \widehat{W}_2 \neq W_2 \textsc{ }\Big| \mathcal{H}=1 \right) \label{eq:prob12},\\
     P_{e,1,3} & \triangleq & \Pr\left(\widehat{W}_3 \neq W_3 \textsc{ }\Big| \mathcal{H}=1, \widehat{W}_1 = W_1, \widehat{W}_2 = W_2\right), \label{eq:prob3}
   \end{IEEEeqnarray}  
   and notice that 
   by the union bound we have
\begin{equation}
 P_{e,1} \leq P_{e,1,1} + P_{e,1,2} + P_{e,1,3}.
\end{equation}
In the following, we analyze  each of the three summands separately. 

\textit{\underline{Analyzing $\mathbb{E}_{\mathcal{C}}[P_{e,1,1}]$:}}
\label{subsection:achievability_err_w1_h1}

By the symmetry of the code construction and the uniformity of the messages and the key, we can assume that $W_{1}=1$, $S_1=1$ and $W_3=w_3$. Then, we have:
     \begin{IEEEeqnarray}{rCl}
            \displaystyle \mathbb{E}_{\mathcal{C}}[P_{e,1,1}] &\leq  &\Pr [(X_1^n(1,1), 0^n, X_3^n(w_3), Y^n) \notin \mathcal{A}_{\eta_1}^{n} ]  + \sum_{w_1=2}^{\mathsf{M_1}}  \quad  \Pr [(X_1^n(w_1,1), 0^n, X_3^n(w_3), Y^n) \in \mathcal{A}_{\eta_1}^{n} ]  \label{eq:pe_m1m2} 
    \end{IEEEeqnarray}

We first analyze  a specific term in the summation of  (\ref{eq:pe_m1m2}). For any $w_1 \in [|2, \mathsf{\mathsf{M_1}}|]$ we have:
     \begin{IEEEeqnarray}{rCl}
         \lefteqn{ \Pr [( X_1^n(w_1,1), 0^n, X_3^n(w_3), Y^n) \in \mathcal{A}_{\eta_1}^{n} ] } \\
        & = &\displaystyle \mathop{\mathbb{E}}_{X_{1}^{n}(w_1,1), X_3^n(w_3), Y^n} \left [\left \{ \mathbbm{1} (X_{1}^{n}(w_1,1), 0^n, X_3^n(w_3), Y^{n})\in \mathcal{A}_{\eta_1}^{n} \right \}  \right] \\
        & \overset{(a)}{\leq} &e^{-\eta_1} \displaystyle \mathop{\mathbb{E}}_{X_{1}^{n}(w_1,1), X_3^n(w_3), Y^n} \left [
        \frac{\Gamma_{Y \mid X_1X_2X_3}^{\otimes n}(Y^{n} | X_1^{n}(w_1,1), 0^n, X_3^n(w_3))}{\Gamma_{Y \mid X_1X_2X_3}^{\otimes n}(Y^{n} | 0^{n}, 0^n, X_3^n(w_3))} 
        \underbrace{\mathbbm{1}
        \left \{ (X_{1}^{n}(w_1,1), 0^n, X_3^n(w_3), Y^{n}) \in \mathcal{A}_{\eta_1}^{n} \right \}}_{\leq 1} \right]\\
        &   \overset{(b)}{\leq} & e^{-\eta_1} \displaystyle \mathop{\mathbb{E}}_{X_{1}^{n}(w_1,1), X_3^n(w_3), Y^n} \left [
        \frac{\Gamma_{Y \mid X_1X_2X_3}^{\otimes n}(Y^{n} | X_1^{n}(w_1,1), 0^n,  X_3^n(w_3))}{\Gamma_{Y \mid X_1X_2X_3}^{\otimes n}(Y^{n} | 0^{n}, 0^n, X_3^n(w_3))} 
   \right]\\
        &   = &    e^{-\eta_1}\prod_{i=1}^n \quad  \mathop{\mathbb{E}}_{X_{1,i}(w_1,1), X_{3,i}(w_3), Y_i} \left[     \frac{\Gamma_{Y \mid X_1X_2X_3}(Y_i| X_{1,i}(w_1,1), 0, X_{3,i}(w_3))}{\Gamma_{Y \mid X_1X_2X_3}(Y_i| 0, 0, X_{3,i}(w_3))} \right]  \\
        &   = &    e^{-\eta_1} \prod_{t \in \mathcal{T}} \left( \mathop{\mathbb{E}}_{P_{X_3Y|T=t} P_{X_{1,n}|T=t}} \left[     \frac{\Gamma_{Y \mid X_1X_2X_3}(Y | X_1, 0, X_3)}{\Gamma_{Y \mid X_1X_2X_3}(Y| 0, 0, X_3)} \right] \right)^{n \boldsymbol{\pi}(t)}  \\
        &   \overset{(c)}{=}  &    e^{-\eta_1} \prod_{t \in \mathcal{T}} \left( \mathop{\mathbb{E}}_{P_{X_3Y|T=t}} \left[     \frac{\Gamma_{Y|X_2X_3}^{(t)}(Y | 0, X_3)}{\Gamma_{Y \mid X_1X_2X_3}(Y| 0, 0, X_3)} \right] \right)^{n \boldsymbol{\pi}(t)}\ablast{.} \label{eq:pe_exp_agamma_first_part}
    \end{IEEEeqnarray}
Here,  $(a)$ holds by the definition of the set $\mathcal{A}_{\eta_1}^n$ in \eqref{eq:set_a_gamma_n_def_achievability}; $(b)$ by replacing the indicator function by the all-one \ablast{function} and $(c)$ upon defining 
    \begin{equation}
    \Gamma_{Y|X_2X_3}^{(t)}(y|x_2,x_3) \triangleq \sum_{x_1 \in \{0,1\}} P_{X_{1,n} \mid T}(x_1 \mid t) \Gamma_{Y \mid X_1X_2X_3}(y|x_1,x_2,x_3).
    \end{equation}
    
For any $t\in\mathcal{T}$ we have:
\begin{IEEEeqnarray}{rCl}
   \mathbb{E}_{P_{X_3Y|T=t}} \left[ \frac{\Gamma_{Y|X_2X_3}^{(t)}(Y | 0, X_3)}{\Gamma_{Y \mid X_1X_2X_3}(Y| 0, 0, X_3)} \right] = 1- \rho_{1,t} \alpha_n + \rho_{1,t}\alpha_n \cdot \mathbb{E}_{P_{X_3Y|T=t}}\left[  \frac{ \Gamma_{Y \mid X_1X_2X_3}(Y| 1, 0, X_3)}{\Gamma_{Y \mid X_1X_2X_3}(Y| 0, 0, X_3)} \right] \label{eq:rewriting_bound_achievability_1}
    \end{IEEEeqnarray}
 Under our assumption \eqref{a} that for any $x_3$ we have $\Gamma_{Y \mid X_1=1, X_2=0, X_3=x_3} \ll \Gamma_{Y \mid X_1=0, X_2=0, X_3=x_3}$, we can conclude that $\frac{ \Gamma_{Y \mid X_1X_2X_3}(y| 1, 0, x_3)}{\Gamma_{Y \mid X_1X_2X_3}(y| 0, 0, x_3)}$ is uniformly upper-bounded for all realizations of $y$ and $x_3$, i.e., 
 \begin{equation}
 \frac{ \Gamma_{Y \mid X_1X_2X_3}(y| 1, 0, x_3)}{\Gamma_{Y \mid X_1X_2X_3}(y| 0, 0, x_3)} \leq \Delta_Y,
 \end{equation} 
 for some finite $\Delta_Y > 0$. We  continue with this upper bound to deduce
\begin{IEEEeqnarray}{rCl}
    \lefteqn{\mathbb{E}_{P_{X_3Y|T=t}}\left[  \frac{\Gamma_{Y \mid X_1X_2X_3}(Y| 1, 0, X_3)}{\Gamma_{Y \mid X_1X_2X_3}(Y| 0, 0, X_3)} \right]} \nonumber \\
    &= &  \mathbb{E}_{P_{X_3\mid T=t} }\Bigg[(1-\rho_{1,t} \alpha_n) (1-\rho_{2,t} \alpha_n)\sum_{y}  \Gamma_{Y \mid X_1X_2X_3}(y| 1, 0, X_3)+  \rho_{1,t}\alpha_n(1-\rho_{2,t}\alpha_n )  \sum_{y}   \frac{\Gamma_{Y \mid X_1X_2X_3}(y| 1, 0, X_3)^2}{\Gamma_{Y \mid X_1X_2X_3}(y| 0, 0, X_3)}  \nonumber \\
    && \hspace{1.6cm} + (1- \rho_{1,t}\alpha_n)\rho_{2,t}\alpha_n   \sum_{y}   \frac{\Gamma_{Y \mid X_1X_2X_3}(y| 1, 0, X_3)  }{\Gamma_{Y \mid X_1X_2X_3}(y| 0, 0, X_3)}\cdot \Gamma_{Y \mid X_1X_2X_3}(y| 0,1, X_3)  \nonumber \\
    && \hspace{1.6cm} +    \rho_{1,t}\rho_{2,t}\alpha_n^2 \sum_{y}   \frac{\Gamma_{Y \mid X_1X_2X_3}(y| 1, 0, X_3) }{\Gamma_{Y \mid X_1X_2X_3}(y| 0, 0, X_3)} \cdot \Gamma_{Y \mid X_1X_2X_3}(y| 1,1, X_3)  \Bigg] \IEEEeqnarraynumspace \\
      & \leq & \mathbb{E}_{P_{X_3\mid T=t} }\left[(1-\rho_{2,t} \alpha_n)(1-\rho_{1,t} \alpha_n ) +\left((\rho_{1,t}+\rho_{2,t}) \alpha_n - \rho_{1,t}\rho_{2,t}\alpha_n^2\right) \Delta_Y\right] \\[1.3ex]
    &=&(1-\rho_{2,t} \alpha_n)(1-\rho_{1,t} \alpha_n ) + \left((\rho_{1,t}+\rho_{2,t}) \alpha_n + \rho_{1,t}\rho_{2,t}\alpha_n^2\right) \Delta_Y  \label{eq:dd}     
\end{IEEEeqnarray}
which with \eqref{eq:rewriting_bound_achievability_1} yields: 
\begin{IEEEeqnarray}{rCl}
      \mathbb{E}_{P_{X_3 Y\mid T=t} }\left[ \frac{\Gamma_{Y|X_2X_3}^{(t)}(Y | 0, X_3)}{\Gamma_{Y \mid X_1X_2X_3}(Y| 0, 0, X_3)} \right] 
    &\leq & 1 - \rho_{1,t}(\rho_{1,t}+\rho_{2,t})\alpha_n^2 (1-\Delta_Y) + \mathcal{O}(\alpha_n^3).  \label{eq:dd}
\end{IEEEeqnarray}

Combining  \eqref{eq:pe_exp_agamma_first_part} with \eqref{eq:dd},
we obtain: 
\begin{IEEEeqnarray}{rCl}
\lefteqn{\sum_{w_1=2}^{\mathsf{M_1}}  \quad  \displaystyle \mathop{\mathbb{E}}_{X_1^{n}(w_1,1),X_3^n(w_3), Y^n} \left [\mathbbm{1}  \left \{ ( X_{1}^{n}(w_1,1), 0^n, X_3^n(w_3), y^{n})\in \mathcal{A}_{\eta_1}^{n} \right \}  \right]} \nonumber \\ 
  & \leq & \mathsf{M_1} e^{-\eta_1}   \prod_{t \in \mathcal{T}} \quad \left(1 -  \rho_{1,t}(\rho_{1,t}+\rho_{2,t})\alpha_n^2 (1-\Delta_Y) + \mathcal{O}(\alpha_n^3)\right)^{n \boldsymbol{\pi}(t)}\\
& = &  \mathsf{M_1} e^{-\eta_1} e^{n  \sum_{t\in\mathcal{T}} \boldsymbol{\pi}(t) \log \left(1 -  \rho_{1,t}(\rho_{1,t}+\rho_{2,t})\alpha_n^2 (1-\Delta_Y) + \mathcal{O}(\alpha_n^3) \right)} \\
& \leq &  \mathsf{M_1} e^{-\eta_1} e^{- n  \sum_{t\in\mathcal{T}} \boldsymbol{\pi}(t)  \left(  \rho_{1,t}(\rho_{1,t}+\rho_{2,t})\alpha_n^2 (1-\Delta_Y) + \mathcal{O}(\alpha_n^3) \right)} \\
& \leq & \mathsf{M_1} e^{-\eta_1}e^{  -\omega_n^2 \sum_{t\in\mathcal{T}} (P_T(t)+\mu_n) \left[  \rho_{1,t}(\rho_{1,t}+\rho_{2,t}) (1-\Delta_Y) + \mathcal{O}(\alpha_n) \right] }
\label{eq:pe_exp_agamma_first_part_left}
    \end{IEEEeqnarray}
    Notice that the term in the last exponent tends to 0 as $n\to \infty$ because $\omega_n \to 0$ and the sum over $t$ is bounded. 
 We conclude that the  term in \eqref{eq:pe_exp_agamma_first_part_left}, and thus the  sum in \eqref{eq:pe_m1m2},   tend to 0  as $n\to \infty$, if 
\begin{equation}\label{eq:limit_M1}
    \lim_{n\to\infty}\left( \log \mathsf{M_1} - \eta_1  \right)=-\infty.
\end{equation}

We next bound the first summand on the right-hand side of \eqref{eq:pe_m1m2}. 
To this end, start by noticing the following:
\begin{equation}
    \Pr \left[\log \left(\frac{\Gamma_{Y \mid X_1X_2X_3}^{\otimes n}(Y^{n}| X_{1}^{n}(1,1), 0^n, X_3^n(w_3)) } {\Gamma_{Y \mid X_1X_2X_3}^{\otimes n}(Y^{n} | 0^n, 0^n, X_3^n(w_3)) }\right) \leq \eta_1 \right] = \Pr \left[ \sum_{i=1}^{n} \log \left(\frac{\Gamma_{Y \mid X_1X_2X_3}(Y_i| X_{1,i}, 0, X_{3,i})}{\Gamma_{Y \mid X_1X_2X_3}(Y_i| 0, 0, X_{3,i})} \right) \leq \eta_1 \right] \label{eqn:error_proba_analysis_decomp_t_1_0}
\end{equation}
where $X_{1,i}, X_{3,i}$ and $Y_i$ denote the $i$-th entries of $X_{1}^{n}(1,1), X_3^n(w_3)$, and $Y^n$, and probabilities are with respect to the randomness in the code construction and the channel. For each $ i\in\{1,\ldots, n\}$, consider the random variable
\begin{equation}\label{eq:Xi_j}
\Xi_i\triangleq  \log \left(\frac{\Gamma_{Y \mid X_1X_2X_3}(Y_i| X_{1,i}, 0, X_{3,i})}{\Gamma_{Y \mid X_1X_2X_3}(Y_i| 0, 0, X_{3,i})} \right) ,
\end{equation}
where as above, the tuple $(X_{1,i}, X_{3,i}, Y_i)$ follows the joint pmf $P_{X_{1,t_i}}(x_1)P_{X_3|T=t_i}(x_3)\Gamma_{Y \mid X_1X_2X_3}(y|x_1,0,x_3)$. Let 
\begin{equation}
\Lambda_Y \triangleq \min_{ (x_1, x_3,y)}  \log \left(\frac{\Gamma_{Y \mid X_1X_2X_3}(y| x_1, 0,x_3)}{\Gamma_{Y \mid X_1X_2X_3}(y| 0, 0, x_3 )}\right)  ,
\end{equation}
where the minimum is only over triples $(x_1,x_3,y)$ for which the ratio $\frac{\Gamma_{Y \mid X_1X_2X_3}(y| x_1, 0,x_3)}{\Gamma_{Y \mid X_1X_2X_3}(y| 0, 0, x_3 )}$ is non-zero. (Since we prevent the ratio inside the log from being 0 and the sets $\mathcal{X}_1, \mathcal{X}_3, \mathcal{Y}$ are all finite, the minimum must exist.) By above definition and Assumption~\eqref{a}, we have  $\Pr[ |\Xi_i| \leq c]=1$ for $c \triangleq \max\{ | \log \Lambda_Y|, | \log \Delta_Y|\}$. Moreover, the first and second moments of the random variable $\Xi_i$, for $i\in\{1,\ldots, n\}$, satisfy
\begin{IEEEeqnarray}{rCl}
\mathbb{E} \left[\Xi_i\right] 
       & \overset{(a)}{=}& \rho_{1,t_i} \alpha_n   \mathbb{E}_{P_{X_3 \mid T=t_i}}\cdot  \left[\D_{Y}^{(1)}(X_3) \right ] \\
          \mathop{\mathbb{E}} \left[\Xi_i^2\right] &\overset{(b)} \leq & \rho_{1,t_i} \alpha_n \cdot  (\log \Delta_Y)^2, \label{eq:variance_log_squared_bernstein_step}
\end{IEEEeqnarray}
where $(a)$ \ablast{and $(b)$} \ablast{hold} because when $x_1=0$ the log term in the definition of $\Xi_i$, \eqref{eq:Xi_j}, is zero.
We can thus apply a large deviation argument to bound the probability 
\begin{equation}
\Pr\left[ \frac{1}{n} \sum_{i=1}^n  \Xi_i < \alpha_n   \mathbb{E}_{\boldsymbol{\pi}P_{X_3|T}} \left[ \rho_{1,T} \D_{Y}^{(1)}(X_3) \right] - a \right],
\end{equation}
for any $a>0$.  Since above probability coincides with the probabilities in \eqref{eqn:error_proba_analysis_decomp_t_1_0} for $\eta_1= n\left( \alpha_n   \mathbb{E}_{\boldsymbol{\pi}P_{X_3|T}} \left[ \rho_{1,T} \D_{Y}^{(1)}(X_3) \right] - a \right)$, we obtain  
 by Bernstein's inequality:
 \begin{IEEEeqnarray}{rCl}
    \Pr \left[\log \left(\frac{\Gamma_{Y \mid X_1X_2X_3}^{\otimes n}(Y^{n}| X_{1}^{n}(1,1), 0^n, X_3^n(w_3)) } {\Gamma_{Y \mid X_1X_2X_3}^{\otimes n}(Y^{n} | 0^n, 0^n, X_3^n(w_3)) }\right) \leq \eta_1 \right] \leq 2 e^{- \frac{n a^2}{  \alpha_n \mathbb{E}_{ \boldsymbol{\pi}}[\rho_{1,T}] (\log \Delta_Y)^2 +2/3 a c}},\label{eq:last}
\end{IEEEeqnarray}
for any $\eta_1 \geq  n\left( \alpha_n   \mathbb{E}_{\boldsymbol{\pi}P_{X_3|T}} \left[ \rho_{1,T} \D_{Y}^{(1)}(X_3) \right] - a \right)$. 
Noting that $\lim_{n \to \infty} \boldsymbol{\pi}(t) \to P_T(t)$, we  specialize above bound to the choice
\begin{IEEEeqnarray}{rCl}
    \label{eq:value_gamma_achievability}
    \eta_1 &=& (1-\mu_1)n \alpha_n \mathbb{E}_{P_{TX_3}} \left[ \rho_{1,T} \D_{Y}^{(1)}(X_3) \right]\label{eq:M1t}
    \end{IEEEeqnarray}
    for an arbitrary $\mu_1 \in(0,1)$, in which case $a$ scales as $\alpha_n$ and the exponent scales as $- n \alpha_n= -\ablast{\omega_n} \sqrt{n}$ and thus tends to $-\infty$ in the limit as $n\to \infty$. As a consequence,  the probability \eqref{eq:last}, and thus  \eqref{eqn:error_proba_analysis_decomp_t_1_0}, 
 vanish exponentially fast in the blocklength $n$ as  $n\to \infty$. 
 
 Choosing further 
    \begin{IEEEeqnarray}{rCl}
    \log(\mathsf{\mathsf{M_1}}) &=& (1-\xi_1) \alpha_n {n}  \mathbb{E}_{P_{TX_3}} \left[ \rho_{1,T} \D_{Y}^{(1)}(X_3) \right],
\end{IEEEeqnarray}
for any small number $\xi_1 > \mu_1$
by~\eqref{eq:limit_M1}, we finally  conclude that the entire probability $\mathbb{E}_{\mathcal{C}}[P_{e,1,1}]$ on the right-hand side of \eqref{eq:pe_m1m2}  vanishes exponentially fast in the blocklength.

\textit{\underline{Analyzing $\mathbb{E}_{\mathcal{C}}[P_{e,1,2}]$:}}\\
\label{subsection:achievability_err_u2}
By symmetry, the same steps allow one to conclude also that for any small numbers $\mu_2 > 0$ and $\xi_2 > \mu_2$ and   under the choices 
\begin{IEEEeqnarray}{rCl}
    \label{eq:value_gamma_2_achievability}
    \eta_2 &\triangleq& (1-\mu_2) \alpha_n n  \mathbb{E}_{P_{TX_3}} \left[ \rho_{2,T} \mathbb{D}_{Y}^{(2)}(X_3) \right],\\
\label{eq:M2t}
    \log(\mathsf{\mathsf{M_2}}) &=& (1-\xi_2) \alpha_n n  \mathbb{E}_{P_{TX_3}} \left[ \rho_{2,T} \mathbb{D}_{Y}^{(2)}(X_3) \right],
\end{IEEEeqnarray}
the probability of decoding error  $\mathbb{E}_{\mathcal{C}}[P_{e,1,2}]$ of Message $W_2$ vanishes exponentially fast in the blocklength $n$.

\textit{\underline{Analyzing $\mathbb{E}_{\mathcal{C}}[P_{e,1,3}]$:}}\\
As in the analysis of $P_{e,0}$, we deduce that under condition~\eqref{eq:R2}:
\begin{equation}\label{eq:Pe12}
\lim_{n\to\infty} \mathbb{E}_{\mathcal{C}}[P_{e,1,3}] =0
\end{equation}

\subsection{Channel Resolvability Analysis}\label{sec:channel_resolvability_analysis}

\subsubsection{Auxiliary Lemma and Definitions}

We will need the following lemma, which  
is an immediate consequence of  \cite[Lemma~1]{bloch_k_users_mac}. 
\begin{lemma}
\label{lemma_1_asymptotics}
For each blocklength $n$, consider two  pmfs $P_{X_{1,n}}$ and $P_{X_{2,n}}$ over the binary alphabets $\mathcal{X}_1=\mathcal{X}_2=\{0,1\}$ respectively, such  that
\begin{equation}
\lim_{n\to \infty} P_{X_{\ell,n}}(1) =0 , \quad \ell \in\{1,2\}. 
\end{equation}
Let $\mathcal{X}_3$, $\mathcal{Z}$, and $\Gamma_{Z \mid X_1X_2X_3}$ be as defined earlier. Then, for  all sufficiently large values of $n$,
  the conditional pmfs 
\begin{IEEEeqnarray}{rCl}
    \Gamma_{Z \mid X_3} (z \mid x_3) \triangleq \sum_{(x_1,x_2) \in \mathcal{X}_1 \times \mathcal{X}_2} P_{X_{1,n}}(x_1) P_{X_{2,n}}(x_2) \Gamma_{Z \mid X_1X_2X_3}(z| x_1, x_2, x_3), \quad x_3 \in \mathcal{X}_3,\;  z \in \mathcal{Z}, 
\end{IEEEeqnarray}
satisfy 
\begin{align}
  \lefteqn{  \mathbb{D}( \Gamma_{Z \mid X_3}(\cdot \mid x_3)\,  \|\,  \Gamma_{Z\mid X_1X_2X_3}(\cdot \mid 0, 0, x_3) } \quad \\&= (1+o(1)) \cdot \frac{\left( P_{X_{1,n}}(1) +P_{X_{2,n}}(1) \right)^2}{2} \chi^2\left( \frac{P_{X_{1,n}}(1)}{P_{X_{1,n}}(1) +P_{X_{2,n}}(1)}, \frac{P_{X_{2,n}}(1)}{P_{X_{1,n}}(1) +P_{X_{2,n}}(1)}, x_3 \right). 
\end{align}
\end{lemma}

The following definitions will be useful in our resolvability analysis. 
For any $t \in \mathcal{T}$, define the averaged channels
 \begin{IEEEeqnarray}{rCl}
    \label{eqn:def_Wz_X2}
    \Gamma_{Z|X_3}^{(t)}(z|x_3) &\triangleq& \sum_{(x_1,x_2) \in \mathcal{X}_1 \times \mathcal{X}_2} P_{X_{1,n} \mid T}(x_1 \mid t)P_{X_{2,n} \mid T}(x_2 \mid t)\Gamma_{Z \mid X_1X_2X_3}(z|x_1, x_2, x_3)\ablast{,}\\
     \Gamma_{Z \mid X_2X_3}^{(t)}( z|x_2,x_3) &\triangleq&  \sum_{x_1} P_{X_{1,n} \mid T}(x_1 \mid t) \Gamma_{Z \mid X_1X_2X_3}(z|x_1,x_2,x_3)\ablast{,}\\
     \Gamma_{Z \mid X_1X_3}^{(t)}( z|x_1,x_3)&\triangleq& \sum_{x_2} P_{X_{2,n} \mid T}(x_2 \mid t) \Gamma_{Z \mid X_1X_2X_3}(z|x_1,x_2,x_3) ,
 \end{IEEEeqnarray}
and the corresponding product channels
 \begin{IEEEeqnarray}{rCl}
 \label{eqn:def_Q_tilde_n}
   \tilde{\Gamma}_{\ab{Z \mid X_3}}^{\ab{n}}(z^n \mid x_3^n) &\triangleq&
 \prod_{i=1}^n \Gamma_{Z \mid X_3}^{(t_i)}(z_i \mid x_{3,i})\ablast{,}\\
 \tilde{\Gamma}_{\ab{Z \mid X_2X_3}}^{\ab{n}}(z^n \mid x_2^n, x_3^n) &\triangleq& \prod_{i=1}^n  \Gamma_{Z \mid X_2X_3}^{(t_i)}( z_i \mid x_{2,i},x_{3,i})\ablast{,} \\
\tilde{\Gamma}_{\ab{Z \mid X_1X_3}}^{\ab{n}}(z^n \mid x_1^n, x_3^n) &\triangleq& \prod_{i=1}^n \Gamma_{Z \mid X_1X_3}^{(t_i)}( z_i \mid x_{1,i},x_{3,i}).\label{eq:tG}
 \end{IEEEeqnarray}

\subsubsection{The Proof}
Recall that the warden's output distribution under $\mathcal{H}=1$ for a given codebook $\mathcal{C}$ and  message $w_3 \in \mathcal{M}_3$:
\begin{IEEEeqnarray}{rCl}
\widehat{Q}_{\mathcal{C}, w_3}^{n}(z^{n})
& = & \frac{1}{\mathsf{M_1}\mathsf{M_2}} \frac{1}{\mathsf{K_1}\mathsf{K_2}} \sum_{w_1 =1}^{\mathsf{M_1}} \sum_{s_1=1}^{\mathsf{K_1}} \sum_{w_2 =1}^{\mathsf{M_2}} \sum_{s_2=1}^{\mathsf{K_2}} \Gamma^{\otimes n}_{Z \mid X_1X_2X_3} (z^n| x_1^n(w_1,s_1), x_2^n(w_1,s_2), x_3^n(w_3)).
\end{IEEEeqnarray}

In this section, we show the limit
\begin{equation}
\mathbb{E}_{\mathcal{C}}\left[\mathbb{D}\left( \widehat{Q}_{\mathcal{C}, w_3}^n \Big\| \Gamma^{\otimes n}_{Z \mid X_1X_2X_3} ( \cdot | 0^n, 0^n, x_3^n(w_3))\right) \right] \to 0 \qquad \forall w_3 \in \mathcal{M}_3,
\end{equation}
where expectation is with respect to the random code construction. 
Fix a message $w_3 \in \mathcal{W}_3$ and a codeword $x_3^n(w_3)$. 
We start by  expanding the  divergence of interest as follows:
\begin{align}
    \label{eqn:resolvability_first_steps}
    \mathbb{D}\left( \widehat{Q}_{\mathcal{C}, w_3}^n \Big\| \Gamma^{\otimes n}_{Z \mid X_1X_2X_3} ( \cdot | 0^n, 0^n, x_3^n(w_3))\right) &= \mathbb{D}\left( \widehat{Q}_{\mathcal{C}, w_3}^n \Big\| \tilde{\Gamma}_{\ab{Z \mid X_3}}^{\ab{n}}\right) + \mathbb{D}\left( \tilde{\Gamma}_{\ab{Z \mid X_3}}^{\ab{n}} \Big\| \Gamma^{\otimes n}_{Z \mid X_1X_2X_3} ( \cdot | 0^n, 0^n, x_3^n(w_3))\right)\\
    &+ \sum_{z^n} \left(\widehat{Q}_{\mathcal{C}, w_3}^n(z^n) - \tilde{\Gamma}_{\ab{Z \mid X_3}}^{\ab{n}}(z^n \mid x_3^n(w_3)) \right ) \log \left( \frac{\tilde{\Gamma}_{\ab{Z \mid X_3}}^{\ab{n}}(z^n \mid x_3^n(w_3))}{\Gamma^{\otimes n}_{Z \mid X_1X_2X_3} ( \cdot | 0^n, 0^n, x_3^n(w_3))} \right ).
\end{align}
Defining $\ab{\nabla_0}$ as the minimum probability in the support of $\Gamma_{Z \mid X_1X_2X_3}(z|0,0,x_3(w_3))$:
\begin{equation}\ab{\nabla_0} = \min_{z,x_3 \in \text{supp}(\Gamma_{Z \mid X_1X_2X_3}(z|0,0,x_3(w_3)))} \Gamma_{Z \mid X_1X_2X_3}(z|0,0,x_3(w_3)),
\end{equation} 
 by Pinsker's inequality\footnote{For any two distributions $P$ and $Q$ on the same alphabet $\mathcal{X}$ we have $\mathbb{V}(P,Q) \leq  \sqrt{\frac{\mathbb{D}(P,Q)}{2}}$.}, we can conclude the following:
\begin{align}
    \label{eq:resolvability_original}
    &\left | \mathbb{D}\left( \widehat{Q}_{\mathcal{C}, w_3}^n \Big\| \Gamma^{\otimes n}_{Z \mid X_1X_2X_3} ( \cdot | 0^n, 0^n, x_3^n(w_3))\right) -  \mathbb{D}\left( \tilde{\Gamma}_{\ab{Z \mid X_3}}^{\ab{n}} \Big\| \Gamma^{\otimes n}_{Z \mid X_1X_2X_3} ( \cdot | 0^n, 0^n, x_3^n(w_3))\right) \right | \\
    &\leq \mathbb{D}\left( \widehat{Q}_{\mathcal{C}, w_3}^n \Big\| \tilde{\Gamma}_{\ab{Z \mid X_3}}^{\ab{n}}\right) + n \log\left( \frac{1}{\ab{\nabla_0}} \right)  \sqrt{\frac{1}{2}\mathbb{D}\left( \widehat{Q}_{\mathcal{C}, w_3}^n \Big\| \tilde{\Gamma}_{\ab{Z \mid X_3}}^{\ab{n}}\right)}. \label{eq:abs_diff}
\end{align}
We shall separately analyze the divergences $\mathbb{D}( \tilde{\Gamma}_{\ab{Z \mid X_3}}^{\ab{n}}(\cdot \mid x_3^n(w_3)) \| \Gamma^{\otimes n}_{Z \mid X_1X_2X_3} ( \cdot | 0^n, 0^n, x_3^n(w_3)))$  and $\mathbb{D}\left( \widehat{Q}_{\mathcal{C}, w_3}^n \Big\| \tilde{\Gamma}_{\ab{Z \mid X_3}}^n\right)$, or more precisely, their expectations over the choices of the codebooks.

\textit{Analysis of the expected divergence $\mathbb{D}\left( \widehat{Q}_{\mathcal{C}, w_3}^n \Big\| \tilde{\Gamma}_{\ab{Z \mid X_3}}^{\ab{n}}(\ab{\cdot} \mid x_3^n)\right)$:}
Let $Z^n$ be the output sequence observed at the warden under $\mathcal{H}=1$ and $W_3$ the message of the non-covert user. Given that $W_3=w_3$ and for given codebooks $\mathcal{C}$,  we then have $Z^n\sim  \widehat{Q}_{\mathcal{C}, w_3}^n$.
Consider the average (over the codebooks) expected divergence
\begin{IEEEeqnarray}{rCl}  
 \lefteqn{  \mathop{\mathbb{E}}_{ \mathcal{C}} 
 \left[\mathbb{D}\left( \widehat{Q}_{\mathcal{C}, w_3}^n \Big\| \tilde{\Gamma}_{\ab{Z \mid X_3}}^{\ab{n}}(\cdot \mid X_3^n(w_3))\right) \right]}\nonumber\\
&=& \mathop{\mathbb{E}}_{ \{X_1^{n}(\ab{w_1,s_1})\}, \{X_2^{n}(\ab{w_2,s_2})\}, X_3^n(w_3)} \left [\sum_{z^n}  \widehat{Q}_{\mathcal{C}, w_3}^{n}(z^{n})\log \left( \frac{ \widehat{Q}_{\mathcal{C}, w_3}^{n}(z^{n} )}{\tilde{\Gamma}_{\ab{Z \mid X_3}}^{\ab{n}}(z^n \mid X_3^n(w_3)) } \right) \right] \label{eq:firsta}\\[1.2ex]
&  \stackrel{(a)}{=} & \mathop{\mathbb{E}}_{ \{X_1^{n}(\ab{w_1,s_1})\}, \{X_2^{n}(\ab{w_2,s_2})\}, X_3^n(w_3)}\Bigg[  \nonumber \\
&& \hspace{3cm} \mathop{\mathbb{E}}_{ \ab{Z^n}}\Bigg[
\log \left( \frac{\sum_{(w_1',w_2',s_1',s_2')} \Gamma_{Z \mid X_1X_2X_3}^{\otimes n}(Z^n| X_1^n(w_1',s_1'), X_2^n(w_2',s_2'), X_3^n(w_3))}{\mathsf{M_1} \mathsf{M_2} \mathsf{K_1} \mathsf{K_2}  \cdot \tilde{\Gamma}_{\ab{Z \mid X_3}}^{\ab{n}}(Z^n \mid X_3^n(w_3))} \right)  \Bigg] \Bigg] \nonumber \\ \\
&   \overset{(b)}{\leq} & \mathop{\mathbb{E}}_{\substack{X_1^{n}(1,1), \\ X_2^{n}(1,1), \\ X_3^n(w_3), Z^n}} \Bigg[    \log \Bigg( \Bigg. \mathop{\mathbb{E}}_{\substack{\{ X_1^{n}(\ab{w_1,s_1}) \} \setminus  X_1^{n}(1,1), \\ \{ X_2^{n}(\ab{w_2,s_2}) \} \setminus  X_2^{n}(1,1)}} \Bigg[  \frac{\sum_{(w_1',w_2',s_1',s_2')} \Gamma_{Z \mid X_1X_2X_3}^{\otimes n}(Z^{n}  | X_{1}^{n}(w_1',s_1'), X_{2}^{n}(w_2',s_2'), X_3^n(w_3))}{\mathsf{M_1} \mathsf{M_2} \mathsf{K_1} \mathsf{K_2}  \cdot \tilde{\Gamma}_{\ab{Z \mid X_3}}^{\ab{n}}(Z^n \mid X_3^n(w_3))} 
\Bigg) \Bigg] \\[1.2ex]
&  = & \mathbb{E}\left[ 
\log \left(\sum_{(w_1,w_2,s_1,s_2)} \mathop{\mathbb{E}}_{\substack{\{ X_1^{n}(\ab{w_1,s_1}) \} \setminus  X_1^{n}(1,1), \\ \{ X_2^{n}(\ab{w_2,s_2}) \} \setminus  X_2^{n}(1,1)}}  \left[ \frac{\Gamma_{Z \mid X_1X_2X_3}^{\otimes n}(Z^{n} | X_1^{n}(w_1',s_1'), X_2^{n}(w_2',s_2'), X_3^n(w_3))
}{\mathsf{M_1} \mathsf{M_2} \mathsf{K_1} \mathsf{K_2}  \cdot \tilde{\Gamma}_{\ab{Z \mid X_3}}^{\ab{n}}(Z^n \mid X_3^n(w_3))} 
\right ]
\right ) \right] 
\\[1.2ex]
& \overset{(c)}{=} &
\mathbb{E}\left[ \log \left( \frac{(\mathsf{M_1} \mathsf{K_1}-1)(\mathsf{M_2}  \mathsf{K_2} -1)}{\mathsf{M_1} \mathsf{M_2} \mathsf{K_1} \mathsf{K_2} } +  \frac{  \Gamma_{Z \mid X_1X_2X_3}^{\otimes n}(Z^{n} | X_{1}^{n}(1,1), X_{2}^{n}(1,1), X_3^n(w_3))}{\mathsf{M_1} \mathsf{M_2} \mathsf{K_1} \mathsf{K_2}  \cdot \tilde{\Gamma}_{\ab{Z \mid X_3}}^{\ab{n}}(Z^n \mid X_3^n(w_3))}  \right. \right. \nonumber \\[1.2ex]
&& \hspace{1.9cm} + \sum_{(w_2',s_2')\neq (1,1)}  \mathop{\mathbb{E}}_{ X_2^{n}(w_2', s_2')} \left[ \frac{  \Gamma_{Z \mid X_1X_2X_3}^{\otimes n}(Z^{n} | X_{1}^{n}(1,1), X_{2}^{n}(w_2',s_2'), X_3^n(w_3))}{\mathsf{M_1} \mathsf{M_2} \mathsf{K_1} \mathsf{K_2}  \cdot \tilde{\Gamma}_{\ab{Z \mid X_3}}^{\ab{n}}(Z^n \mid X_3^n(w_3))} \right] \nonumber \\
& &  \hspace{2.6cm}  + \left. \left.  \sum_{(w_1',s_1')\neq (1,1)}  \mathop{\mathbb{E}}_{ X_1^{n}(w_1', s_1')}  \left[ \frac{  \Gamma_{Z \mid X_1X_2X_3}^{\otimes n}(Z^{n} | X_{1}^{n}(w_1',s_1'), X_{2}^{n}(1,1), X_3^n(w_3))}{\mathsf{M_1} \mathsf{M_2} \mathsf{K_1} \mathsf{K_2}  \cdot \tilde{\Gamma}_{\ab{Z \mid X_3}}^{\ab{n}}(Z^n \mid X_3^n(w_3))}  \right] \right) \right]  \IEEEeqnarraynumspace\\[1.2ex]
& \overset{(d)}{=}  &
\mathbb{E}\left[ \log \left( \frac{(\mathsf{M_1} \mathsf{K_1}-1)(\mathsf{M_2}  \mathsf{K_2} -1)}{\mathsf{M_1} \mathsf{M_2} \mathsf{K_1} \mathsf{K_2} }  +  \frac{  \Gamma_{Z \mid X_1X_2X_3}^{\otimes n}(Z^{n} | X_{1}^{n}(1,1), X_{2}^{n}(1,1), X_3^n(w_3))}{\mathsf{M_1} \mathsf{M_2} \mathsf{K_1} \mathsf{K_2}  \cdot \tilde{\Gamma}_{\ab{Z \mid X_3}}^{\ab{n}}(Z^n \mid X_3^n(w_3))}  \right. \right. \nonumber \\
&& \hspace{1.5cm}  \left. \left. + \frac{ (\mathsf{M_2}\mathsf{K_2}-1) \tilde{\Gamma}_{\ab{Z \mid X_1X_3}}^{\ab{n}}(Z^{n} | X_{1}^{n}(1,1), X_3^n(w_3))}{\mathsf{M_1} \mathsf{M_2} \mathsf{K_1} \mathsf{K_2}  \cdot \tilde{\Gamma}_{\ab{Z \mid X_3}}^{\ab{n}}(Z^n \mid X_3^n(w_3))} +  \frac{ (\mathsf{M_1}\mathsf{K_1}-1) \tilde{\Gamma}_{\ab{Z \mid X_2X_3}}^{\ab{n}}(Z^{n} | X_{2}^{n}(1,1), X_3^n(w_3))}{\mathsf{M_1} \mathsf{M_2} \mathsf{K_1} \mathsf{K_2}  \cdot \tilde{\Gamma}_{\ab{Z \mid X_3}}(Z^n \mid X_3^n(w_3))}   \right ) \right]
\end{IEEEeqnarray}
\begin{IEEEeqnarray}{rCl}
&\leq &
\mathbb{E}\left[ \log \left( 1 +  \frac{  \Gamma_{Z \mid X_1X_2X_3}^{\otimes n}(Z^{n} | X_{1}^{n}(1,1), X_{2}^{n}(1,1), X_3^n(w_3))}{\mathsf{M_1} \mathsf{M_2} \mathsf{K_1} \mathsf{K_2}  \cdot \tilde{\Gamma}_{\ab{Z \mid X_3}}^{\ab{n}}(Z^n \mid X_3^n(w_3))}  \right. \right. \nonumber  \\
&& \hspace{1.5cm}  \left. \left. + \frac{ \tilde{\Gamma}_{\ab{Z \mid X_1X_3}}^{\ab{n}}(Z^{n} | X_{1}^{n}(1,1), X_3^n(w_3))}{\mathsf{M_1}  \mathsf{K_1}  \cdot \tilde{\Gamma}_{\ab{Z \mid X_3}}^{\ab{n}}(Z^n \mid X_3^n(w_3))} +  \frac{ \tilde{\Gamma}_{\ab{Z \mid X_2X_3}}^{\ab{n}}(Z^{n} | X_{2}^{n}(1,1), X_3^n(w_3))}{ \mathsf{M_2}  \mathsf{K_2}  \cdot \tilde{\Gamma}_{\ab{Z \mid X_3}}^{\ab{n}}(Z^n \mid X_3^n(w_3))}   \right ) \right]  ,  \label{eq:cr_1}
\end{IEEEeqnarray}
where in before equation $(c)$ the warden's output sequence $Z^n$   is generated from $X_1^n(W_1,S_1)$, $X_2^n(W_2, S_2)$, and $X_3^n(w_3)$ according to the memoryless channel law $\Gamma_{Z \mid X_1X_2X_3}^{\otimes n}$, and starting with $(c)$ it is generated according to the same channel law but based on the random codewords $X_1^n(1,1)$, $X_2^n(1, 1)$, and $X_3^n(w_3)$.

 Above sequence of (in)equalities are justified as follows: 
 \begin{itemize}
 \item[$(a)$] holds by rewriting the summation as an expectation  over  $Z^n$;
 \item[$(b)$] holds by applying Jensen's inequality over all expectations except the expectations over $X_1^n(W_1,S_1), X_2^n(W_2,S_2), X_3^n(w_3), Z^n$  and by assuming that $W_1=W_2=S_1=S_2=1$ and thus $Z^n$ is generated from $X_1^n(1,1)$, $X_2^n(1, 1)$, and $X_3^n(w_3)$ according to $\Gamma_{Z \mid X_1X_2X_3}^{\otimes n}$\ablast{.} This  assumption is without loss of optimality by the symmetry of the code construction; 
 \item[$(c)$]\hspace{-1.5mm},\hspace{-0.mm}$(d)$ \;  hold by the linearity of expectation and because for $(w_1', s_1')\neq (1,1)$ and $(w_2',s_2' )\neq 1$
\begin{IEEEeqnarray}{rCl}  
  \mathop{\mathbb{E}}_{X_1^{n}(w_1',s_1'), X_2^{n}(w_2',s_2')} \left[  \Gamma_{Z|X_{1} X_{2}X_3}^{\otimes n}(Z^{n} | X_1^{n}(w_1',s_1'), X_2^{n}(w_2',s_2'), X_3^n(w_3))\right]
  = \tilde{\Gamma}_{\ab{Z \mid X_3}}^{\ab{n}}(Z^n \mid X_3^n(w_3))
\end{IEEEeqnarray}
and 
\begin{IEEEeqnarray}{rCl}  
  \mathop{\mathbb{E}}_{X_1^{n}(w_1',s_1')} \left[  \Gamma_{Z|X_{1}X_2X_3}^{\otimes n}(Z^{n} | X_1^{n}(w_1',s_1'), X_2^{n}(1,1), X_3^n(w_3))\right]
  = \tilde{\Gamma}_{\ab{Z \mid X_2X_3}}^{\ab{n}}(Z^n \mid X_2^n(1,1),X_3^n(w_3))\\
    \mathop{\mathbb{E}}_{X_2^{n}(w_2',s_2')} \left[  \Gamma_{Z|X_{1} X_{2}X_3}^{\otimes n}(Z^{n} | X_1^{n}(1,1), X_2^{n}(w_2',s_2'), X_3^n(w_3))\right]
  =\tilde{ \Gamma}_{\ab{Z \mid X_1X_3}}^{\ab{n}}(Z^n \mid X_1^n(1,1),X_3^n(w_3)).
\end{IEEEeqnarray}
\end{itemize}

Define for any triple $\boldsymbol{\theta}= (\theta_0,\theta_1, \theta_2)$ the set
\begin{IEEEeqnarray}{rCl}  
\label{eq:def_b_tau_set_resolvability}
    \mathcal{B}_{\boldsymbol{\theta}}^{n} \triangleq \Bigg \{  (x_1^n, x_2^n, x_3^n, z^n) \in  \mathcal{X}_1^n \times \mathcal{X}_2^n \times \mathcal{X}_3^n \times \mathcal{Z}^{n} &:&  \log \left( \frac{\Gamma_{Z \mid X_1X_2X_3}^{\otimes n}(z^n \mid x_1^n, x_2^n, x_3^n) }{\Gamma_{Z \mid X_1X_2X_3}^{\otimes n}(z^n \mid 0^n,0^n, x_3^n)}  \right) \leq \theta_0, \nonumber \\
    && \log \left( \frac{\tilde{\Gamma}_{\ab{Z \mid X_1X_3}}^{\ab{n}}(z^n \mid x_1^n, x_3^n) }{\Gamma_{Z \mid X_1X_2X_3}^{\otimes n}(z^n \mid 0^n, 0^n, x_3^n)}  \right) \leq \theta_1, \nonumber \\
    && \log \left( \frac{\tilde{\Gamma}_{\ab{Z \mid X_2X_3}}^{\ab{n}}(z^n \mid x_2^n, x_3^n) }{\Gamma_{Z \mid X_1X_2X_3}^{\otimes n}(z^n \mid 0^n, 0^n, x_3^n)}  \right) \leq \theta_2 \Bigg \},\label{eq:186}
\end{IEEEeqnarray}
and denote by $A$ and $B$ the events $\{(X_1^n(1,1), X_2^n(1,1), X_3^n(w_3), Z^n) \in \mathcal{B}_{\boldsymbol{\theta}}^{n}\}$ and $\{(X_1^n(1,1), X_2^n(1,1), X_3^n(w_3), Z^n) \notin \mathcal{B}_{\boldsymbol{\theta}}^{n}\}$ respectively.
Using the total law of expectation we rewrite  \eqref{eq:cr_1} as: 
\begin{IEEEeqnarray}{rCl}  
   \lefteqn{  \mathop{\mathbb{E}}_{ \{X_1^{n}(w_1',s_1')\}_{(w_1',s_1')}, \{X_2^{n}(w_2',s_2')\}_{(w_2',s_2')}, X_3^n(w_3)} \left[\mathbb{D}\left( \widehat{Q}_{\mathcal{C}, w_3}^n \Big\| \tilde{\Gamma}_{\ab{Z \mid X_3}}^{\ab{n}}(\cdot \mid X_3^n(w_3))\right) \right]  } \quad \nonumber \\
&\leq& \mathbb{E}\Bigg[ \log \Bigg( 1 +  \frac{  \Gamma_{Z \mid X_1X_2X_3}^{\otimes n}(Z^{n} | X_{1}^{n}(1,1), X_{2}^{n}(1,1), X_3^n(w_3))}{\mathsf{M_1} \mathsf{M_2} \mathsf{K_1} \mathsf{K_2}  \cdot \tilde{\Gamma}_{\ab{Z \mid X_3}}^{\ab{n}}(Z^n \mid X_3^n(w_3))} + \frac{\tilde{\Gamma}_{\ab{Z \mid X_1X_3}}^{\ab{n}} \ab{(Z}^n \mid \ablast{X_1^n, X_3^n})}{\mathsf{M_1}  \mathsf{K_1}  \cdot \tilde{\Gamma}_{\ab{Z \mid X_3}}^{\ab{n}}(Z^n \mid X_3^n(w_3))} \nonumber \\
&& \hspace{1.65cm} +  \frac{  \tilde{\Gamma}_{\ab{Z \mid X_2X_3}}^{\ab{n}}(\ablast{Z^n} \mid \ablast{X_2^n, X_3^n})}{ \mathsf{M_2}  \mathsf{K_2}  \cdot \tilde{\Gamma}_{\ab{Z \mid X_3}}^{\ab{n}}(Z^n \mid X_3^n(w_3))} \Bigg ) \Bigg| (X_1^n(1,1), X_2^n(1,1), X_3^n(w_3), Z^n) \in \mathcal{B}_{\boldsymbol{\theta}}^{n} \Bigg] \cdot \Pr[A] \nonumber \\
&&+\mathbb{E}\Bigg[ \log \Bigg( 1 +  \frac{  \Gamma_{Z \mid X_1X_2X_3}^{\otimes n}(Z^{n} | X_{1}^{n}(1,1), X_{2}^{n}(1,1), X_3^n(w_3))}{\mathsf{M_1} \mathsf{M_2} \mathsf{K_1} \mathsf{K_2}  \cdot \tilde{\Gamma}_{\ab{Z \mid X_3}}^{\ab{n}}(Z^n \mid X_3^n(w_3))} + \frac{\tilde{\Gamma}_{\ab{Z \mid X_1X_3}}^{\ab{n}}\ab{(Z}^n \mid \ablast{X_1^n, X_3^n})}{\mathsf{M_1}  \mathsf{K_1}  \cdot \tilde{\Gamma}_{\ab{Z \mid X_3}}^{\ab{n}}(Z^n \mid X_3^n(w_3))} \nonumber \\
&&\hspace{1.8cm}+  \frac{ \tilde{\Gamma}_{\ab{Z \mid X_2X_3}}^{\ab{n}}(z^n \mid \ablast{X_2^n, X_3^n})}{ \mathsf{M_2}  \mathsf{K_2}  \cdot \tilde{\Gamma}_{\ab{Z \mid X_3}}^{\ab{n}}(Z^n \mid X_3^n(w_3))} \Bigg ) \Bigg| (X_1^n(1,1), X_2^n(1,1), X_3^n(w_3), Z^n) \notin \mathcal{B}_{\boldsymbol{\theta}}^{n} \Bigg] \cdot \Pr[B].
\label{eq:dde}
\end{IEEEeqnarray}
To bound the first summand, we observe:
\begin{IEEEeqnarray}{rCl}
\lefteqn{  \mathbb{E}\Bigg[ \log \Bigg( 1 +  \frac{  \Gamma_{Z \mid X_1X_2X_3}^{\otimes n}(Z^{n} | X_{1}^{n}(1,1), X_{2}^{n}(1,1), X_3^n(w_3))}{\mathsf{M_1} \mathsf{M_2} \mathsf{K_1} \mathsf{K_2}  \cdot \tilde{\Gamma}_{\ab{Z \mid X_3}}^{\ab{n}}(Z^n \mid X_3^n(w_3))} + \frac{\tilde{\Gamma}_{\ab{Z \mid X_1X_3}}^{\ab{n}}(\ab{Z}^n \mid \ablast{X_1^n, X_3^n})}{\mathsf{M_1}  \mathsf{K_1}  \cdot \tilde{\Gamma}_{\ab{Z \mid X_3}}^{\ab{n}}(Z^n \mid X_3^n(w_3)) }}\nonumber \\
&& \hspace{1.65cm} +  \frac{  \tilde{\Gamma}_{\ab{Z \mid X_2X_3}}^{\ab{n}}(\ab{Z}^n \mid \ablast{X_2^n, X_3^n})}{ \mathsf{M_2}  \mathsf{K_2}  \cdot \tilde{\Gamma}_{\ab{Z \mid X_3}}^{\ab{n}}(Z^n \mid X_3^n(w_3))} \Bigg ) \Bigg| (X_1^n(1,1), X_2^n(1,1), X_3^n(w_3), Z^n) \in \mathcal{B}_{\boldsymbol{\theta}}^{n} \Bigg] \cdot \Pr[A] \nonumber \\
& \stackrel{(a)}{\leq} & \mathbb{E}\Bigg[ \log \Bigg( 1 +  \frac{ e^{\theta_0}  \Gamma_{Z \mid X_1X_2X_3}^{\otimes n}(Z^{n} | 0^n, 0^n, X_3^n(w_3))}{\mathsf{M_1} \mathsf{M_2} \mathsf{K_1} \mathsf{K_2}  \cdot \tilde{\Gamma}_{\ab{Z \mid X_3}}^{\ab{n}}(Z^n \mid X_3^n(w_3))} + \frac{e^{\theta_1}  \Gamma_{Z \mid X_1X_2X_3}^{\otimes n}(Z^{n} | 0^n, 0^n, X_3^n(w_3))}{\mathsf{M_1}  \mathsf{K_1}  \cdot \tilde{\Gamma}_{\ab{Z \mid X_3}}^{\ab{n}}(Z^n \mid X_3^n(w_3))}  \nonumber \\
&& \hspace{1.8cm} +  \frac{e^{\theta_2}  \Gamma_{Z \mid X_1X_2X_3}^{\otimes n}(Z^{n} | 0^n, 0^n, X_3^n(w_3))}{ \mathsf{M_2}  \mathsf{K_2}  \cdot \tilde{\Gamma}_{\ab{Z \mid X_3}}^{\ab{n}}(Z^n \mid X_3^n(w_3))} \Bigg ) \Bigg| (X_1^n(1,1), X_2^n(1,1), X_3^n(w_3), Z^n) \in \mathcal{B}_{\boldsymbol{\theta}}^{n} \Bigg] \cdot \Pr[A] \\
& \leq & \frac{  e^{\theta_0} }{\mathsf{M_1} \mathsf{M_2} \mathsf{K_1} \mathsf{K_2}} \mathop{\mathbb{E}}\left[ \frac{\Gamma^{\otimes n}_{Z \mid X_1X_2X_3} ( Z^n | 0^n, 0^n, X_3^n(w_3))
}{  \tilde{\Gamma}_{\ab{Z \mid X_3}}^n(Z^n \mid X_3^n(w_3))} \right] + \frac{  e^{\theta_1} }{\mathsf{M_1} \mathsf{K_1}} \mathop{\mathbb{E}}\left[ \frac{\Gamma^{\otimes n}_{Z \mid X_1X_2X_3} ( Z^n | 0^n, 0^n, X_3^n(w_3))}{  \tilde{\Gamma}_{\ab{Z \mid X_3}}^{\ab{n}}(Z^n \mid X_3^n(w_3))} \right] \nonumber \\
&&  +\frac{  e^{\theta_2} }{\mathsf{M_2} \mathsf{K_2}} \mathop{\mathbb{E}}\left[ \frac{\Gamma^{\otimes n}_{Z \mid X_1X_2X_3} ( Z^n | 0^n, 0^n, X_3^n(w_3))}{  \tilde{\Gamma}_{\ab{Z \mid X_3}}^{\ab{n}}(Z^n \mid X_3^n(w_3))} \right] \\
&\stackrel{(b)}{=}&   \frac{  e^{\theta_0} }{\mathsf{M_1} \mathsf{M_2} \mathsf{K_1} \mathsf{K_2}} \cdot \sum_{x_3^n} \sum_{z^n} P_{X_3 \mid T} ^{\otimes n} (x_3^n\mid t^n) \Gamma^{\otimes n}_{Z \mid X_1X_2X_3} ( z^n | 0^n, 0^n, x_3^n) \nonumber\\
&&+ \frac{  e^{\theta_1} }{\mathsf{M_1} \mathsf{K_1}} \cdot \sum_{x_3^n} \sum_{z^n} P_{X_3 \mid T} ^{\otimes n} (x_3^n\mid t^n) \Gamma^{\otimes n}_{Z \mid X_1X_2X_3} ( z^n | 0^n, 0^n, x_3^n) \nonumber \\
&&+ \frac{  e^{\theta_2} }{\mathsf{M_2} \mathsf{K_2}} \cdot \sum_{x_3^n} \sum_{z^n} P_{X_3 \mid T} ^{\otimes n} (x_3^n\mid t^n) \Gamma^{\otimes n}_{Z \mid X_1X_2X_3} ( z^n | 0^n, 0^n, x_3^n) \\
&\stackrel{(c)}{=}& \frac{  e^{\theta_0} }{\mathsf{M_1} \mathsf{M_2} \mathsf{K_1} \mathsf{K_2}} + \frac{  e^{\theta_1} }{\mathsf{M_1} \mathsf{K_1}} + \frac{  e^{\theta_2} }{\mathsf{M_2} \mathsf{K_2}}, \label{eq:l1} 
\end{IEEEeqnarray}
where $(a)$ holds by the definition of the set $\mathcal{B}_{\boldsymbol{\theta}}^n$ in \eqref{eq:def_b_tau_set_resolvability}; $(b)$ holds because $(X_3^n(w_3),Z^n) \sim \tilde{\Gamma}_{\ab{Z \mid X_3}}^{\ab{n}} P_{X_3|T}^{\otimes n}$; and $(c)$ holds because for all $t \in \mathcal{T}$ the term $P_{X_3 \mid T}(x_3 \mid t)\Gamma_{Z \mid X_1X_2X_3} ( z \mid 0, 0, x_3)$ denotes a valid probability distribution over $\mathcal{X}_3 \times \mathcal{Z}$ and hence sums to 1.

To bound the second summand in \eqref{eq:dde}, remark that by the definition of $\tilde{\Gamma}_{\ab{Z \mid X_3}}^{\ab{n}}$ in \eqref{eqn:def_Q_tilde_n}, for any pair $(x_3^n, z^n)$:
\begin{IEEEeqnarray}{rCl}
\tilde{\Gamma}_{\ab{Z \mid X_3}}^{\ab{n}}(z^n \mid x_3^n)
&=& \prod_{i=1}^n  \sum_{(x_{1,i}, x_{2,i}) \in \mathcal{X}_1 \times \mathcal{X}_2 }P_{X_{1,n}}(x_{1,i} \mid t_i) P_{X_{2,n}}(x_{2,i} \mid t_i)\Gamma_{Z \mid X_1X_2X_3}(z|x_{1,i}, x_{2,i}, x_{3,i})\label{eq:step1} \\
& =& \prod_{i=1}^n  \Big[ \rho_{1,t_i}  \rho_{2,t_i} \alpha_{n}^2 \Gamma_{Z \mid X_1X_2X_3}(z_i|1,1, x_{3,i}) + \rho_{1, t_i} \alpha_{n}(1- \rho_{2,t_i} \alpha_{n}) \Gamma_{Z \mid X_1X_2X_3}(z_i|1,0, x_{3,i}) \nonumber \\
&& \hspace{1cm}+ (1- \rho_{1,t_i} \alpha_{n}) \rho_{2,t_i} \alpha_{n} \Gamma_{Z \mid X_1X_2X_3}(z_i|0,1, x_{3,i}) \nonumber \\
&& \hspace{1cm}+ (1- \rho_{1,t_i} \alpha_{n})(1- \rho_{2,t_i} \alpha_{n}) \Gamma_{Z \mid X_1X_2X_3}(z_i|0,0, x_{3,i}) \Big] \\
& \geq& \prod_{i=1}^{n} (1- \rho_{1,t_i} \alpha_{n})(1- \rho_{2,t_i} \alpha_{n}) \Gamma_{Z \mid X_1X_2X_3}(z_i|0,0, x_{3,j}) \\ 
& \geq& \prod_{i=1}^{n} (1- \rho_{1,t_i} \alpha_{n})(1- \rho_{2,t_i} \alpha_{n}) \ab{\nabla_0}.
\end{IEEEeqnarray}
We can thus conclude that for any $(x_1^n, x_2^n, x_3^n, z^n) \ablast{\in  \mathcal{X}_1^n \times \mathcal{X}_2^n \times \mathcal{X}_3^n \times \mathcal{Z}^{n}}$ the following bound holds:
\begin{IEEEeqnarray}{rCl}
    && \log \left(1 + \frac{\Gamma_{Z \mid X_1X_2X_3}^{\otimes n}(z^{n} | x_1^n, x_2^n, x_3^n)}{\mathsf{M_1} \mathsf{M_2} \mathsf{K_1} \mathsf{K_2} \cdot  \tilde{\Gamma}_{\ab{Z \mid X_3}}^{\ab{n}}(z^{n} \mid x_3^n)  } + \frac{\tilde{\Gamma}_{\ab{Z \mid X_1X_3}}^{\ab{n}}(z^{n} | x_1^{n}, x_3^n)}{\mathsf{M_1}  \mathsf{K_1}  \cdot \tilde{\Gamma}_{\ab{Z \mid X_3}}^{\ab{n}}(z^n \mid x_3^n)} + \frac{ \tilde{\Gamma}_{\ab{Z \mid X_2X_3}}^{\ab{n}}(z^{n} | x_2^{n}, x_3^n)}{\mathsf{M_2}  \mathsf{K_2}  \cdot \tilde{\Gamma}_{\ab{Z \mid X_3}}^{\ab{n}}(z^n \mid x_3^n)} \right) \nonumber \\
    &=& \log \left(\frac{1}{\tilde{\Gamma}_{\ab{Z \mid X_3}}^{\ab{n}}(z^{n} \mid x_3^n)} \right) +  \log \Bigg(\tilde{\Gamma}_{\ab{Z \mid X_3}}^{\ab{n}}(z^{n} \mid x_3^n) + \frac{\Gamma_{Z \mid X_1X_2X_3}^{\otimes n}(z^{n} | x_1^n, x_2^n, x_3^n)}{\mathsf{M_1} \mathsf{M_2} \mathsf{K_1} \mathsf{K_2}} + \frac{ \tilde{\Gamma}_{\ab{Z \mid X_1X_3}}^{\ab{n}}(z^{n} | x_1^{n}, x_3^n)}{\mathsf{M_1} \mathsf{K_1} } \nonumber \\
    && \hspace{4cm} + \, \frac{ \tilde{\Gamma}_{\ab{Z \mid X_2X_3}}^{\ab{n}}(z^{n} | x_2^{n}, x_3^n)}{\mathsf{M_2}  \mathsf{K_2}  } \Bigg) \\
    &\stackrel{(a)}{\leq}& \log \left(\frac{1}{\prod_{i=1}^{n} (1- \rho_{1,t_i} \alpha_{n})(1- \rho_{2,t_i} \alpha_{n}) \ab{\nabla_0} }\right) + \log(4) \\
    &=& \sum_{i=1}^n \log \left(\frac{1}{(1- \rho_{1,t_i} \alpha_{n})(1- \rho_{2,t_i} \alpha_{n}) \ab{\nabla_0}}  \right) + \log(4) \\
    &\leq& \sum_{i=1}^n \log \left(\frac{1}{(1- \rho_{1,t_i} \alpha_{n})(1- \rho_{2,t_i} \alpha_{n}) \ab{\nabla_0}}  \right) + n \sum_{t \in \mathcal{T}} \boldsymbol{\pi}(t) \log(4) \\
    &=& \sum_{t \in \mathcal{T}} n \boldsymbol{\pi}(t) \log \left(\frac{4}{(1- \rho_{1,t} \alpha_{n})(1- \rho_{2,t} \alpha_{n}) \ab{\nabla_0}}  \right). \label{eq:first_log_term_resolvability_event_not_in_b_tau}
\end{IEEEeqnarray}
Therefore, \begin{IEEEeqnarray}{rCl} 
\lefteqn{ \mathbb{E}\Bigg[ \log \Bigg( 1 +  \frac{  \Gamma_{Z \mid X_1X_2X_3}^{\otimes n}(Z^{n} | X_{1}^{n}(1,1), X_{2}^{n}(1,1), X_3^n(w_3))}{\mathsf{M_1} \mathsf{M_2} \mathsf{K_1} \mathsf{K_2}  \cdot \tilde{\Gamma}_{\ab{Z \mid X_3}}^{\ab{n}}(Z^n \mid X_3^n(w_3))} + \frac{ \tilde{\Gamma}_{\ab{Z \mid X_1X_3}}^{\ab{n}}(Z^{n} | X_{1}^{n}(1,1), X_3^n(w_3))}{\mathsf{M_1}\mathsf{K_1} \cdot \tilde{\Gamma}_{\ab{Z \mid X_3}}^{\ab{n}}(Z^n \mid X_3^n(w_3))}} \quad  \nonumber \\
&& \hspace{2cm} + \frac{\tilde{\Gamma}_{\ab{Z \mid X_2X_3}}^{\ab{n}}(Z^{n} | X_{2}^{n}(1,1), X_3^n(w_3))}{\mathsf{M_2}\mathsf{K_2} \cdot \tilde{\Gamma}_{\ab{Z \mid X_3}}^{\ab{n}}(Z^n \mid X_3^n(w_3))} \Bigg ) \Bigg| (X_1^n(1,1), X_2^n(1,1), X_3^n(w_3), Z^n) \notin \mathcal{B}_{\boldsymbol{\theta}}^{n} \Bigg] \nonumber \\
&\leq& \sum_{t \in \mathcal{T}} n \boldsymbol{\pi}(t) \log \left(\frac{4}{(1- \rho_{1,t} \alpha_{n})(1- \rho_{2,t} \alpha_{n}) \ab{\nabla_0}}  \right).  \label{eq:l3}\IEEEeqnarraynumspace
\end{IEEEeqnarray}
We proceed to analyze the probability of the event $B$, which by the union bound can be \ablast{upper bounded as follows}:
\begin{align}
    \Pr \left[ B \right] &\leq \Pr \left[  \log \left( \frac{\Gamma_{Z\mid X_1X_2X_3}^{\otimes n}(Z^{n} \mid X_{1}^{n}(1,1), X_{2}^{n}(1,1), X_3^n(w_3)) }{\Gamma^{\otimes n}_{Z \mid X_1X_2X_3} ( Z^n | 0^n, 0^n, X_3^n(w_3))}  \right) \geq \theta_0 \right] + \Pr \left[  \log \left( \frac{{\tilde{\Gamma}}_{Z\mid X_1X_3}^{\ab{n}}(Z^{n} \mid X_1^{n}(1,1), X_3^n(w_3)) }{\Gamma^{\otimes n}_{Z \mid X_1X_2X_3} ( Z^n | 0^n, 0^n, X_3^n(w_3))}  \right) \geq \theta_1 \right] \nonumber \\
    &\hspace{1cm}+ \Pr \left[  \log \left( \frac{{\tilde{\Gamma}}_{Z\mid X_2X_3}^{\ab{n}}(Z^{n} \mid X_2^{n}(1,1), X_3^n(w_3)) }{\Gamma^{\otimes n}_{Z \mid X_1X_2X_3} ( Z^n | 0^n, 0^n, X_3^n(w_3))}  \right) \geq \theta_2 \right]\\
    &= \Pr \left[ \sum_{i=1}^{n} \log \left(\frac{\Gamma_{Z \mid X_1X_2X_3}(Z_i| X_{1,i}, X_{2,i}, X_{3,i})}{\Gamma_{Z \mid X_1X_2X_3}(Z_i| 0, 0, X_{3,i})} \right) \geq \theta_0 \right] + \Pr \left[ \sum_{i=1}^{n} \log \left(\frac{\Gamma_{Z \mid X_1X_3}^{{(t_i)}}(Z_i| X_{1,i}, X_{3,i})}{\Gamma_{Z \mid X_1X_2X_3}(Z_i| 0, 0, X_{3,i})} \right) \geq \theta_1 \right] \nonumber \\
    &\hspace{1cm}+ \Pr \left[ \sum_{i=1}^{n} \log \left(\frac{\Gamma_{Z \mid X_2X_3}^{{(t_i)}}(Z_i| X_{2,i}, X_{3,i})}{\Gamma_{Z \mid X_1X_2X_3}(Z_i| 0, 0, X_{3,i})} \right) \geq \theta_2 \right] \label{eqn:cr_4}
\end{align}
To show that the quantity in \eqref{eqn:cr_4} vanishes we will make  use of Bernstein's inequality. First we notice the expectations
\begin{IEEEeqnarray}{rCl}
\lefteqn{\sum_{\ab{i}=1}^{n} \mathop{\mathbb{E}} \left[\log \left( \frac{\Gamma_{Z\mid X_1X_2X_3}(Z_{\ab{i}} | X_{1,\ab{i}}, X_{2,\ab{i}}, X_{3,i})}{\Gamma_{Z \mid X_1X_2X_3}(Z_{\ab{i}} \mid  0, 0, X_{3,\ab{i}})} \right) \right] }\nonumber \\
    &=& \sum_{t \in \mathcal{T}} n \boldsymbol{\pi}(t) \mathbb{E} \left[\log \left( \frac{\Gamma_{Z\mid X_1X_2X_3}(Z \mid X_1, X_2, X_3)}{\Gamma_{Z \mid X_1X_2X_3}(Z \mid  0, 0, X_{3})}\right) \bigg| \, T=t \right] \label{eq:exp1} \\
  &\stackrel{(a)}{=}& n \sum_{t \in \mathcal{T}} \boldsymbol{\pi}(t) \mathbb{E}_{P_{X_3\mid T=t }}  \left[ \rho_{1,t} \rho_{2,t} \alpha_n^2 \mathbb{D}_Z^{(1,2)}(X_3) + \rho_{1,t} \alpha_n (1-\rho_{2,t} \alpha_n) \mathbb{D}_Z^{(1)}(X_3) + (1-\rho_{1,t} \alpha_n) \rho_{2,t} \alpha_n \mathbb{D}_Z^{(2)}(X_3) \ablast{\bigg| \, T=t} \right]  \IEEEeqnarraynumspace\\
&=& n \sum_{t \in \mathcal{T}} \boldsymbol{\pi}(t) \mathbb{E}_{P_{X_3\mid T=t  }} \left[ \rho_{1,t} \alpha_n \mathbb{D}_Z^{(1)}(X_3) + \rho_{2,t} \alpha_n \mathbb{D}_Z^{(2)}(X_3) \ablast{\bigg| \, T=t} \right]+ n\mathcal{O}(\alpha_n^2)  \\
&=& n \alpha_n \mathbb{E}_{\pi P_{X_3|T}}  \left[ \rho_{1,T} \mathbb{D}_Z^{(1)}(X_3) + \rho_{2,T} \mathbb{D}_Z^{(2)}(X_3)  \right] + n\mathcal{O}(\alpha_n^2) , \label{eqn:cr_5}
\end{IEEEeqnarray}
where the first expectation is taken with respect to the law $P_{X_{1,n}|T=t}P_{X_{2,n}|T=t}P_{X_{3}|T=t}\Gamma_{Z|X_1X_2X_3}$. Also,  Equality $(a)$ holds because when $(x_1,x_2) = (0,0)$, the log term is zero.

In a similar way, we have: 
\begin{IEEEeqnarray}{rCl}
\lefteqn{\sum_{\ab{i}=1}^{n} \mathop{\mathbb{E}} \left[\log \left( \frac{\Gamma_{Z\mid X_1X_3}(Z_{\ab{i}} | X_{1,\ab{i}}, X_{3,i})}{\Gamma_{Z \mid X_1X_2X_3}(Z_{\ab{i}} \mid  0, 0, X_{3,\ab{i}})} \right) \right] } \nonumber \\
&=& n \sum_{t \in \mathcal{T}} \boldsymbol{\pi}(t) \mathop{\mathbb{E}} \left[ \log \left( \frac{\sum_{x_2} P_{X_2 \mid T} (x_2 \mid t) \Gamma_{Z\mid X_1X_2X_3}(Z \mid X_1, x_2, X_3)}{\Gamma_{Z \mid X_1X_2X_3}(Z \mid 0, 0, X_3)} \right)  \bigg| \,T=t\right] \\
&=& n \sum_{t \in \mathcal{T}} \boldsymbol{\pi}(t) \mathop{\mathbb{E}} \left[ \log \left( \frac{\rho_{2,t} \alpha_n \Gamma_{Z\mid X_1X_2X_3}(Z \mid X_1, 1, X_3) + (1-\rho_{2,t} \alpha_n) \Gamma_{Z\mid X_1X_2X_3}(Z \mid X_1, 0, X_3)}{\Gamma_{Z \mid X_1X_2X_3}(Z \mid 0, 0, X_3)} \right)  \bigg| \, T=t\right]\\
&=& n \sum_{t \in \mathcal{T}} \boldsymbol{\pi}(t) \mathop{\mathbb{E}} \left[ \log \left(  \frac{ \Gamma_{Z\mid X_1X_2X_3}(Z \mid X_1, 0, X_3)}{\Gamma_{Z \mid X_1X_2X_3}(Z \mid 0, 0, X_3)} \cdot \left( 1+ \rho_{2,t} \alpha_n  \left(\frac{ \Gamma_{Z\mid X_1X_2X_3}(Z \mid X_1, 1, X_3) }{\Gamma_{Z \mid X_1X_2X_3}(Z \mid X_1, 0, X_3)} - 1\right) \right)\right) \bigg| \, T=t\right]\\
&=& n \sum_{t \in \mathcal{T}} \boldsymbol{\pi}(t) \mathop{\mathbb{E}} \left[ \log  \frac{ \Gamma_{Z\mid X_1X_2X_3}(Z \mid X_1, 0, X_3)}{\Gamma_{Z \mid X_1X_2X_3}(Z \mid 0, 0, X_3)} \bigg| \, T=t \right]  \nonumber\\
&& +n \sum_{t \in \mathcal{T}} \boldsymbol{\pi}(t) \ \mathbb{E}\left[ \log \left( 1+ \rho_{2,t} \alpha_n  \left(\frac{ \Gamma_{Z\mid X_1X_2X_3}(Z \mid X_1, 1, X_3) }{\Gamma_{Z \mid X_1X_2X_3}(Z \mid X_1, 0, X_3)} - 1\right) \right) \bigg| \, T=t\right]
\label{eq:ddded}
\end{IEEEeqnarray}
where the expectations  are with respect to the law  $P_{X_{1,n}|T=t}P_{X_{3}|T=t}\Gamma_{Z|X_1X_3}^{(t)}$. For the first summand, we have: 
\begin{IEEEeqnarray}{rCl} 
\lefteqn{ n \sum_{t \in \mathcal{T}} \boldsymbol{\pi}(t) \mathop{\mathbb{E}} \left[ \log  \frac{ \Gamma_{Z\mid X_1X_2X_3}(Z \mid X_1, 0, X_3)}{\Gamma_{Z \mid X_1X_2X_3}(Z \mid 0, 0, X_3)} \bigg| \, T=t \right] } \nonumber\\
&= & n \sum_{t \in \mathcal{T}} \boldsymbol{\pi}(t)  \rho_{1,t}\alpha_{n} (1- \rho_{2,t} \alpha_n)  \mathop{\mathbb{E}}_{P_{X_{3|T=t}}\Gamma_{Z|X_1=1,X_2=0, X_3}} \left[ \log \left( \frac{\Gamma_{Z\mid X_1X_2X_3}(Z \mid 1, 0, X_3)}{\Gamma_{Z \mid X_1X_2X_3}(Z \mid 0, 0, X_3)} \right )\bigg| \, T=t\  \right] + \mathcal{O}(\alpha_n^2)\\
& =&  \alpha_n n  {\mathbb{E}}_{{\pi} P_{X_{3|T}}}  \left[\rho_{1,T} \mathbb{D}_Z^{(1)}(X_3)   \right] + \mathcal{O}(\alpha_n^2), \label{eq:ddded2}
\end{IEEEeqnarray}
because the logarithmic term vanishes when $x_1=0$ and by using the definition of the \ablast{distribution} $\Gamma_{Z|X_1X_3}^{(t)}$. 
For the second summand, we use the upper bound  $\log(1+x) \leq x - \frac{x^2}{2}$ to obtain: 
\begin{IEEEeqnarray}{rCl} 
\lefteqn{n \sum_{t \in \mathcal{T}} \boldsymbol{\pi}(t) \ \mathbb{E}_{P_{X_{1,n}|T=t}P_{X_{3}|T=t}\Gamma_{Z|X_1X_3}^{(t)}} \left[ \log \left( 1+ \rho_{2,t} \alpha_n  \left(\frac{ \Gamma_{Z\mid X_1X_2X_3}(Z \mid X_1, 1, X_3) }{\Gamma_{Z \mid X_1X_2X_3}(Z \mid X_1, 0, X_3)} - 1\right) \right)  \bigg| \, T=t \right]} \quad \nonumber \\
&\leq &n \sum_{t \in \mathcal{T}} \boldsymbol{\pi}(t)  \rho_{2,t} \alpha_n   \mathbb{E}\left[ \frac{ \Gamma_{Z\mid X_1X_2X_3}(Z \mid X_1, 1, X_3) }{\Gamma_{Z \mid X_1X_2X_3}(Z \mid X_1, 0, X_3)} - 1 \bigg| \, T=t \right]
+ \mathcal{O}(\alpha_n^2)\\
& = & n \sum_{t \in \mathcal{T}} \boldsymbol{\pi}(t)  \rho_{2,t} \alpha_n   \mathbb{E}_{P_{X_{1,n}|T=t}}P_{X_{3,t}\Gamma_{Z|X_1,X_2=0,X_3}}\left[ \frac{ \Gamma_{Z\mid X_1X_2X_3}(Z \mid X_1, 1, X_3) }{\Gamma_{Z \mid X_1X_2X_3}(Z \mid X_1, 0, X_3)} - 1\bigg| \, T=t\  \right] + \mathcal{O}(\alpha_n^2) \\
& = & n \sum_{t \in \mathcal{T}} \boldsymbol{\pi}(t)  \rho_{2,t} \alpha_n  \mathbb{E}_{P_{X_{1,n}|T=t}}P_{X_{3,t}\Gamma_{Z|X_1,X_2=0,X_3}} [  \Gamma_{Z\mid X_1X_2X_3}(Z \mid X_1, 1, X_3) - 1\bigg| \, T=t\ ] 
+ \mathcal{O}(\alpha_n^2) \\
& = & + \mathcal{O}(\alpha_n^2) ,\label{eq:ddded3}
\end{IEEEeqnarray}
where the last equation holds because the expectation $ \mathbb{E}_{\Gamma_{Z|X_1=x_1,X_2=0,X_3}} [  \Gamma_{Z\mid X_1X_2X_3}(Z \mid x_1, 1, X_3)]=1$ for any value of $x_1$. 

Combining \eqref{eq:ddded}, \eqref{eq:ddded2}, and \eqref{eq:ddded3}, we obtain: 
\begin{IEEEeqnarray}{rCl}
\sum_{{i}=1}^{n} \mathop{\mathbb{E}} \left[\log \left( \frac{\Gamma_{Z\mid X_1X_3}(Z_{\ab{i}} | X_{1,\ab{i}}, X_{3,i})}{\Gamma_{Z \mid X_1X_2X_3}(Z_{\ab{i}} \mid  0, 0, X_{3,\ab{i}})} \right) \right] &\leq  & \alpha_n n  {\mathbb{E}}_{{\pi} P_{X_{3|T}}}  \left[\rho_{1,T} \mathbb{D}_Z^{(1)}(X_3)   \right] + \mathcal{O}(\alpha_n^2), 
\end{IEEEeqnarray}

Likewise,
\begin{IEEEeqnarray}{rCl}
\sum_{\ab{i}=1}^{n} \mathop{\mathbb{E}} \left[\log \left( \frac{\Gamma_{Z\mid X_2X_3}(Z_{\ab{i}} | X_{2,\ab{i}}, X_{3,i})}{\Gamma_{Z \mid X_1X_2X_3}(Z_{\ab{i}} \mid  0, 0, X_{3,\ab{i}})} \right) \right]  \leq \alpha_n n \mathbb{E}_{\pi P_{X_3|T}}  \left[ \rho_{2,T} \mathbb{D}_Z^{(2)}(X_3)\right] + \mathcal{O}(\alpha_n^2) . \label{eqn:cr_7}
\end{IEEEeqnarray}

Then, notice that the variances satisfy:
\begin{IEEEeqnarray}{rCl}
      \mathop{\mathbb{E}} \left[\log^2 \left( \frac{\Gamma_{Z\mid X_1X_2X_3}(Z \mid X_1, X_2, X_3)}{\Gamma_{Z \mid X_1X_2X_3}(Z \mid  0, 0, X_3)} \right)  \right] \overset{(a)}{=} \mathcal{O}(\alpha_n),
\end{IEEEeqnarray}
and,
\begin{IEEEeqnarray}{rCl}
      \mathop{\mathbb{E}} \left[\log^2 \left( \frac{\Gamma_{Z\mid X_1X_3}(Z \mid X_1,  X_3)}{\Gamma_{Z \mid X_1X_2X_3}(Z \mid  0, 0, X_3)} \right)  \right]  \overset{(b)}{\leq} & \mathcal{O}(\alpha_n),
\end{IEEEeqnarray}
and
\begin{IEEEeqnarray}{rCl}
      \mathop{\mathbb{E}} \left[\log^2 \left( \frac{\Gamma_{Z\mid X_2X_3}(Z \mid X_2,  X_3)}{\Gamma_{Z \mid X_1X_2X_3}(Z \mid  0, 0, X_3)} \right)  \right]  \overset{(c)}{\leq}& \mathcal{O}(\alpha_n),
\end{IEEEeqnarray}
where similarly to \eqref{eq:variance_log_squared_bernstein_step}, (a) follows because when $(x_1,x_2) = (0,0)$, the log term is zero, whereas (b) and (c) follows by \ablast{first splitting the log term and using the Taylor expansion and then using the same argument as for (a)}.
Since $\lim_{n \to \infty} \boldsymbol{\pi}(t) \to P_T(t)$ and $\lim_{n \to \infty} \alpha_n^2 = 0$, Bernstein's inequality allows us to conclude that with the choices
\begin{subequations}\label{eq:cr_8}
\begin{IEEEeqnarray}{rCl}
    \theta_0 &\triangleq& (1+\xi_4) \alpha_n n \mathbb{E}_{P_{TX_3}}  \left[ \rho_{1,T} \mathbb{D}_Z^{(1)}(X_3) + \rho_{2,T} \mathbb{D}_Z^{(2)}(X_3) \right],\\
    \theta_1 &\triangleq& (1+\xi_5) \alpha_n n \mathbb{E}_{P_{TX_3}}  \left[ \rho_{1,T} \mathbb{D}_Z^{(1)}(X_3) \right], \\
    \theta_2 &\triangleq& (1+\xi_6) \alpha_n n \mathbb{E}_{P_{TX_3}}  \left[ \rho_{2,T} \mathbb{D}_Z^{(2)}(X_3) \right],
\end{IEEEeqnarray}
\end{subequations}
for any $\xi_4 > 0$, $\xi_5 > 0$, $\xi_6 > 0$ there exists a constant $B_1 > 0$, $B_2 > 0$ and $B_3 > 0$ so that for sufficiently large blocklengths $n$:
\begin{subequations}\label{eqn:end_resolvability}
    \begin{IEEEeqnarray}{rCl}  
    \Pr \left[ B \right] \leq  e^{ - B_1 n \alpha_n} + e^{ - B_2 n \alpha_n} + e^{ - B_3 n \alpha_n}
       \end{IEEEeqnarray}
\end{subequations}
Plugging \eqref{eqn:end_resolvability} into \eqref{eqn:cr_4}  and further combining it with \eqref{eq:dde}, \eqref{eq:l1}, and \eqref{eq:l3}, we finally obtain
 \begin{IEEEeqnarray}{rCl}  
   \lefteqn{  \mathop{\mathbb{E}}_{ \{X_1^{n}w_1',s_1')\}_{w_1',s_1')}, \{X_2^{n}w_2',s_2')\}_{w_2',s_2')}, X_3^n(w_3)} \left[\mathbb{D}\left( \widehat{Q}_{\mathcal{C}, w_3}^n \Big\| \tilde{\Gamma}_{\ab{Z \mid X_3}}^{\ab{n}}(\cdot \mid X_3^n(w_3))\right) \right]  } \quad \nonumber \\
&\leq& \frac{  e^{\theta_0} }{\mathsf{M_1} \mathsf{M_2} \mathsf{K_1} \mathsf{K_2}} + \frac{  e^{\theta_1} }{\mathsf{M_1} \mathsf{K_1}} + \frac{  e^{\theta_2} }{\mathsf{M_2} \mathsf{K_2}} + n \mathbb{E}_{T}  \left[  \log \left(\frac{4}{(1- \rho_{1,T} \alpha_{n})(1- \rho_{2,T} \alpha_{n}) \ab{\nabla_0}}  \right) \right ] \cdot \left( e^{ - B_1 n \alpha_n} + e^{ - B_2 n \alpha_n} + e^{ - B_3 n \alpha_n} \right) \nonumber \\ \label{eq:exp_bound}
\end{IEEEeqnarray}

We notice that the second summand tends to 0 because $ne^{-n a}$ decays for any positive $a>0$. If moreover, 
\begin{subequations}
    \begin{IEEEeqnarray}{rCl}
         \ab{\limsup}_{n\to \infty} \left( \log \left( \mathsf{M_1}\mathsf{M_2}\mathsf{K_1}\mathsf{K_2}\right) - (1+\xi_4) \ab{\alpha_n n} \mathbb{E}_{P_{TX_3}}  \left[  \rho_{1,T} \mathbb{D}_Z^{(1)}(X_3) + \rho_{2,T} \mathbb{D}_Z^{(2)}(X_3) \right] \right)= - \infty , \IEEEeqnarraynumspace \label{eq:covertness} \\
         \ab{\limsup}_{n\to \infty} \left( \log \left( \mathsf{M_1}\mathsf{K_1}\right) - (1+\xi_5) \ab{\alpha_n n} \mathbb{E}_{P_{TX_3}}  \left[  \rho_{1,T} \mathbb{D}_Z^{(1)}(X_3) \right] \right)= - \infty,\IEEEeqnarraynumspace \label{eq:key_constraint_user_1}\\
         \ab{\limsup}_{n\to \infty} \left( \log \left( \mathsf{M_2}\mathsf{K_2}\right) - (1+\xi_6) \ab{\alpha_n n} \mathbb{E}_{P_{TX_3}}  \left[  \rho_{2,T} \mathbb{D}_Z^{(2)}(X_3) \right] \right)= - \infty, \IEEEeqnarraynumspace \label{eq:key_constraint_user_2} 
    \end{IEEEeqnarray}
\end{subequations}
then also the first summand of \eqref{eq:exp_bound} tends to 0 exponentially fast. 

Notice that for small values $\xi_4, \xi_5,\xi_6$ and large blocklengths $n$ Constraint \eqref{eq:covertness} is redundant in view of the per-user secret-key constraints \eqref{eq:key_constraint_user_1} and \eqref{eq:key_constraint_user_2}.\\

\textit{Analysis of  divergence $\mathbb{D}( \tilde{\Gamma}_{\ab{Z \mid X_3}}^{\ab{n}}(\cdot \mid x_3^n(w_3)) \| \Gamma^{\otimes n}_{Z \mid X_1X_2X_3} ( \cdot | 0^n, 0^n, x_3^n(w_3)))$:}

Recall that $\tilde{\Gamma}_{\ab{Z \mid X_3}}^{\ab{n}}$ and $\Gamma^{\otimes n}_{Z \mid X_1X_2X_3} ( \cdot | 0^n, 0^n, x_3^n(w_3))$ are both product distributions and thus
\begin{IEEEeqnarray}{rCl}
   \lefteqn{  \mathbb{D}( \tilde{\Gamma}_{\ab{Z \mid X_3}}^{\ab{n}}(\cdot \mid x_3^n(w_3)) \| \Gamma^{\otimes n}_{Z \mid X_1X_2X_3} ( \cdot | 0, 0, x_3^n(w_3)))}\nonumber \\
   & = &\sum_{t\in\mathcal{T} } \sum_{\substack{i \in\{1,\ldots, t\}\colon\\  t_i=t}  }\mathbb{D}( \Gamma_{Z \mid X_3}^{(t)}(z \mid x_3)  \| \Gamma_{Z \mid X_1X_2X_3} ( \cdot | 0, 0, x_{3,i}(w_3))) \label{eq:second} \\
      & = &\sum_{(t,x_3)\in\mathcal{T} } \sum_{\substack{i \in\{1,\ldots, t\}\colon\\  t_i=t\\ x_{3,i}(t)=x_3}  }\mathbb{D}( \Gamma_{Z \mid X_3}^{(t)}(z \mid x_3)  \| \Gamma_{Z \mid X_1X_2X_3} ( \cdot | 0^n, 0^n, x_{3}))\\
     &\stackrel{(a)}{=}&  \sum_{(x_3,t) \in \mathcal{X}_3 \times \mathcal{T}} n \lambda_{t}(x_3)  \left(1+ o(1) \right )  \cdot \frac{\left(\rho_{1,t} \alpha_n + \rho_{2,t} \alpha_n \right)^2 }{2} \chi^2(\rho_{1,t}, \rho_{2,t}, x_3) \nonumber \\
      &\stackrel{(b)}{=}&  \sum_{(x_3,t) \in \mathcal{X}_3 \times \mathcal{T}} \lambda_{t}(x_3)  \left(1+ o(1)\right )  \cdot \frac{(\rho_{1,t} + \rho_{2,t})^2 \cdot \omega_n^2}{2} \chi^2(\rho_{1,t}, \rho_{2,t}, x_3),\label{eq:alpha_bounds}
\end{IEEEeqnarray}
where in $(a)$ we used the definition
\begin{IEEEeqnarray}{rCl}
    \lambda_{t}(x_3) & \triangleq & \frac{1}{n} \sum_{i=1}^n \mathbbm{1} \{ x_{3,j}(w_3) = x_3, t_i = t\}
\end{IEEEeqnarray}
and we applied  Lemma~\ref{lemma_1_asymptotics} for each value of $t$ individually, also using the fact
\begin{equation}
 \chi^2(a, b, x_3)=  \chi^2\left(\frac{a}{a+b}, \frac{b}{a+b}, x_3\right);
 \end{equation}
and in $(b)$ we used the definition of $\alpha_n$. 

Since both $P_{e\ab{,0}}$ and $P_{e\ab{,1}}$ vanish as $n \to \infty$, and by the decoding rule in \eqref{eq:decoding_user_3}, we can conclude that the sequence of codes in Theorem \ref{main_theorem} satisfies for each $(x_3, t) \in \mathcal{X}_3 \times \mathcal{T}$
\begin{equation}
 \lim_{n\to \infty}   \left | \lambda_{t}(x_3) - P_T(t)P_{X_3 \mid T}(x_3 \mid t) \right | =0.
\end{equation}

\textit{Concluding the Divergence Proof:} 
With this observation, combining \eqref{eq:abs_diff} with \eqref{eq:exp_bound} and \eqref{eq:alpha_bounds}, under Condition \eqref{eq:exp_bound}, we obtain:
\begin{IEEEeqnarray}{rCl}
\frac{1}{\mathsf{M_3}} \sum_{w_3=1}^{\mathsf{M_3}} \delta_{n,w_3}& = &( 1+ o(1)) \frac{\omega_n^2 }{2} \mathbb{E}_{P_T} \left[ (\rho_{1,T} + \rho_{2,T})^2 \mathbb{E}_{P_{X_3 \mid T}} \left[ \chi^2(\rho_{1,T}, \rho_{2,T}, X_3) \right ] \right ]. 
\label{eq:b1}
\end{IEEEeqnarray}

\subsection{Concluding the Achievability  Proof:} 
By standard averaging arguments it can then be shown that there must exist at least one sequence of codebooks $\{\mathcal{C}_{n}\}_n$ for which the probabilities of error  under the two hypotheses  tend to 0 as $n\to \infty$ and with message and \ablast{secret-}key sizes as well as divergence satisfying
\begin{IEEEeqnarray}{rCl}
    \log  \mathsf{M}_1&=& (1- \xi_1) \alpha_n n \mathbb{E}_{P_{TX_3}}  \left[  \rho_{1,T} \mathbb{D}_Y^{(1)}(X_3) \right], \\
    \log \mathsf{M}_2 &= & (1- \xi_2) \alpha_n n \mathbb{E}_{P_{TX_3}}  \left[  \rho_{2,T} \mathbb{D}_Y^{(2)}(X_3) \right], \\
     \log \mathsf{M}_3 &= & (1- \xi_3) n I(X_3;Y|X_1=\ablast{0,} X_2=0,T),\\
    \log \left( \mathsf{M}_1\mathsf{K}_1\right) &= &(1+\xi_5) \alpha_n n \mathbb{E}_{P_{TX_3}}  \left[  \rho_{1,T} \mathbb{D}_Z^{(1)}(X_3) \right], \IEEEeqnarraynumspace\\
 \log \left( \mathsf{M}_2\mathsf{K_2}\right) &=& (1+\xi_6) \alpha_n n \mathbb{E}_{P_{TX_3}}  \left[  \rho_{2,T} \mathbb{D}_Z^{(2)}(X_3) \right], \\
\frac{1}{\mathsf{M_3}} \sum_{w_3=1}^{\mathsf{M_3}} \delta_{n,w_3} & = &( 1+ o(1)) \frac{\omega_n^2 }{2} \mathbb{E}_{P_T} \left[ (\rho_{1,T} + \rho_{2,T})^2 \mathbb{E}_{P_{X_3 \mid T}} \left[ \chi^2(\rho_{1,T}, \rho_{2,T}, X_3) \right ] \right ].
\end{IEEEeqnarray} 

\section{Modifications for the Generalized Scheme}\label{sec:modified}
For simplicity of notation we assume that $\cup_t\mathcal{L}_{\ablast{1,2}}\ablast{(t)}= \{1,\ldots, n_2\}$ and $\cup_t\mathcal{L}_{\ablast{1}}\ablast{(t)}= \{n_2+1,\ldots, n_1\}$. We denote the   first $n_2$ symbols of the corresponding codewords by $x_{1}^{n_2}(w_1, s_1)$,  $x_{2}^{n_2}(w_2, s_2)$, and  $x_{3}^{n_2}(w_3)$ and the following $n_1-n_2$ symbols of the corresponding codewords by $x_{1, n_2+1}^{n_1}(w_1, s_1)$ and  $x_{3,n_2+1}^{n_1}(w_3)$.

It is easy to observe that the divergence term $\delta_{n, w_3}$ now only depends on the first $n_1$ channel uses, as the terms corresponding to the last $n-n_1$ channel uses are zero. In analogy to \eqref{eq:resolvability_original} we can then obtain the bound (notice the new blocklength $n_1$ instead of $n$):
\begin{align}
    &\left | \mathbb{D}\left( \widehat{Q}_{\mathcal{C}, w_3}^{n_1} \Big\| \Gamma^{\otimes n_1}_{Z \mid X_1X_2X_3} ( \cdot | 0^{n_1}, 0^{n_1}, x_3^{n_1}(w_3))\right) -  \mathbb{D}\left( \tilde{\Gamma}_{{Z \mid X_3}}^{n_1} \Big\| \Gamma^{\otimes n_1}_{Z \mid X_1X_2X_3} ( \cdot | 0^{n_1}, 0^{n_1}, x_3^{n_1}(w_3))\right) \right | \nonumber \\
    &\leq \mathbb{D}\left( \widehat{Q}_{\mathcal{C}, w_3}^{n_1} \Big\| \tilde{\Gamma}_{{Z \mid X_3}}^{{n_1}}\right) + n_1 \log\left( \frac{1}{{\nabla_0}} \right)  \sqrt{\frac{1}{2}\mathbb{D}\left( \widehat{Q}_{\mathcal{C}, w_3}^{n_1} \Big\| \tilde{\Gamma}_{{Z \mid X_3}}^{{n_1}}\right)}, \label{eq:abs_diff}
\end{align} 
where $x_{3}^{n_1}(w_3)$ denotes the first $n_1$ symbols of codeword $x_3^n(w_3)$ and $\widehat{Q}_{\mathcal{C}, w_3}^{n_1}$ denotes the pmf of the warden's first $n_1$ output symbols. For the generalized scheme we have: 
\begin{IEEEeqnarray}{rCl}
\widehat{Q}_{\mathcal{C}, w_3}^{n_1}(z^{n_1}) &\triangleq&  \frac{1}{\mathsf{M}_1 \mathsf{M}_2\mathsf{K}_1\mathsf{K}_2} \sum_{(w_1,s_1)} \sum_{(w_2,s_2)}   \widehat{Q}_{\mathcal{C}, w_1, w_2, w_3, s_1, s_2}^{n_1}(z^{n_1}), 
\end{IEEEeqnarray}
where for any valid $(w_1, w_2, w_3, s_1, s_2) \ablast{\in \mathcal{M}_1 \times \mathcal{M}_2 \times \mathcal{M}_3 \times \mathcal{K}_1 \times \mathcal{K}_2}$, we have: 
\begin{IEEEeqnarray}{rCl}
 \widehat{Q}_{\mathcal{C}, w_1, w_2, w_3, s_1, s_2}^{n_1}(z^{n_1})& \triangleq& \Gamma^{\otimes n_2}_{Z \mid X_1X_2X_3} (z^{n_2}| x_1^{n_2}(w_1,s_1), x_2^{n_2}( w_2,s_2), x_3^{n_2}(w_3)) \nonumber \\
 && \cdot \tilde{\Gamma}^{(n_2\to n_1)}_{Z \mid X_1X_3} (z_{n_2+1}^{n_1}| x_{1,n_2+1}^{n_1}(w_1,s_1),x_{3,n_2+1}^{n_1}(w_3)),
 \end{IEEEeqnarray}
 for $z_{n_2+1}^{n_1}\triangleq (z_{n_2+1},\ldots, z_{n-1})$ and for $\tilde{\Gamma}^{(n_2\to n_1)}_{Z \mid X_1X_3} $  defined in analogy to $\tilde{\Gamma}^{n}_{Z \mid X_1X_3}$  in \eqref{eq:tG} but based on the sequence $t_{n_2+1}, \ldots, t_{n_1}$: 
  \begin{equation}
 \tilde{\Gamma}_{Z \mid X_1X_3}^{(n_2\to n_1)}(z_{n_2+1}^{n_1}| x_{1,n_2+1}^{n_1}(w_1,s_1),x_{3,n_2+1}^{n_1}(w_3)) \triangleq \prod_{i=n_2+1}^{n_1} \Gamma_{Z \mid X_1X_3}^{(t_i)}( z_i \mid x_{1,i},x_{3,i}).
  \end{equation}

Following similar steps as in \eqref{eq:second}--\eqref{eq:b1}, one can bound the second divergence on the left-hand side of \eqref{eq:abs_diff} as: 
\begin{equation}
\mathbb{D}\left( \tilde{\Gamma}_{Z \mid X_3}^{n_1} \Big\| \Gamma^{\otimes n_1}_{Z \mid X_1X_2X_3} ( \cdot | 0^{n_1}, 0^{n_1}, x_3^{n_1}(w_3))\right)  = ( 1+ o(1)) \frac{n_1}{n}  \cdot\frac{\omega_n^2 }{2} \mathbb{E}_{P_T} \left[ (\rho_{1,T} + \rho_{2,T})^2 \mathbb{E}_{P_{X_3 \mid T}} \left[ \chi^2(\rho_{1,T}, \rho_{2,T}, X_3) \right ] \right ]
\end{equation}

To bound the divergence term on the right-hand side of  \eqref{eq:abs_diff}, we slightly modify the steps in \eqref{eq:firsta}--\eqref{eq:l1}.  In particular, we write: 
\begin{IEEEeqnarray}{rCl}  
 \lefteqn{  \mathop{\mathbb{E}}_{ \mathcal{C}} 
 \left[\mathbb{D}\left( \widehat{Q}_{\mathcal{C}, w_3}^{n_1} \Big\| \tilde{\Gamma}_{{Z \mid X_3}}^{{n_1}}(\cdot \mid X_3^{n_1}(w_3))\right) \right]}\nonumber\\
&  = & \mathbb{E}\left[ 
\log \left(\sum_{(w_1,w_2,s_1,s_2)} \mathop{\mathbb{E}}_{\substack{\{ X_1^{n_1}(w_1,s_1) \} \setminus  X_1^{n_1}(1,1), \\ \{ X_2^{n_2}(w_2,s_2) \} \setminus  X_2^{n_2}(1,1)}}  \left[ \frac{ \widehat{Q}_{\mathcal{C}, w_1, w_2, w_3, s_1, s_2}^{n_1}(Z^{n_1})
}{\mathsf{M_1} \mathsf{M_2} \mathsf{K_1} \mathsf{K_2}  \cdot \tilde{\Gamma}_{Z \mid X_3}^{n_1}(Z^{n_1} \mid X_3^{n_1}(w_3))}  
\right ]
\right ) \right]
\\[1.2ex]
&\leq  &
\mathbb{E}\left[ \log \left(\sum_{\substack{(w_1, s_1)\neq (1,1) \\(w_2, s_2)\neq (1,1)} }    \mathop{\mathbb{E}}_{ \substack{X_1^{n_1}(w_1, s_1)\\  X_2^{n_2}(w_2, s_2)}} \left[ \frac{ \widehat{Q}_{\mathcal{C}, w_1, w_2, w_3, s_1, s_2}^{n_1}(Z^{n_1})}{\mathsf{M_1} \mathsf{M_2} \mathsf{K_1} \mathsf{K_2}  \cdot \tilde{\Gamma}_{{Z \mid X_3}}^{n_1}(Z^{n_1} \mid X_3^{n_1}(w_3))} \right]+  \frac{ \widehat{Q}_{\mathcal{C},1,1,w_3,1,1}^{n_1}(Z^{n_1})}{\mathsf{M_1} \mathsf{M_2} \mathsf{K_1} \mathsf{K_2}  \cdot \tilde{\Gamma}_{Z \mid X_3}^{{n_1}}(Z^{n_1} \mid X_3^{n_1}(w_3))}  \right. \right. \nonumber \\[1.2ex]
&& \hspace{1.9cm} + \sum_{(w_2,s_2)\neq (1,1)}  \mathop{\mathbb{E}}_{ X_2^{n_2}(w_2, s_2)} \left[ \frac{ \widehat{Q}_{\mathcal{C}, 1, w_2, w_3, 1, s_2}^{n_1}(Z^{n_1})}{\mathsf{M_1} \mathsf{M_2} \mathsf{K_1} \mathsf{K_2}  \cdot \tilde{\Gamma}_{{Z \mid X_3}}^{n_1}(Z^{n_1} \mid X_3^{n_1}(w_3))} \right] \nonumber \\
& &  \hspace{2.6cm}  + \left. \left.  \sum_{(w_1,s_1)\neq (1,1)}  \mathop{\mathbb{E}}_{ X_1^{n_1}(w_1, s_1)}  \left[ \frac{ \widehat{Q}_{\mathcal{C}, w_1, 1, w_3, s_1, 1}^{n_1}(Z^{n_1})}{\mathsf{M_1} \mathsf{M_2} \mathsf{K_1} \mathsf{K_2}  \cdot \tilde{\Gamma}_{{Z \mid X_3}}^{n_1}(Z^{n_1} \mid X_3^{n_1}(w_3))}\right] \right) \right]  \IEEEeqnarraynumspace\\[1.2ex]
& \stackrel{(a)}{=}  &
\mathbb{E}\left[ \log \left( \frac{(\mathsf{M_1} \mathsf{K_1}-1)(\mathsf{M_2}  \mathsf{K_2} -1)}{\mathsf{M_1} \mathsf{M_2} \mathsf{K_1} \mathsf{K_2} }  +  \frac{   \widehat{Q}_{\mathcal{C}, 1, 1, w_3, 1, 1}^{n_1}(Z^{n_1})}{\mathsf{M_1} \mathsf{M_2} \mathsf{K_1} \mathsf{K_2}  \cdot \tilde{\Gamma}_{{Z \mid X_3}}^{n_1}(Z^{n_1} \mid X_3^{n_1}(w_3))} \right. \right. \nonumber \\
&& \hspace{1.5cm}  \left. \left. + \frac{ (\mathsf{M_2}\mathsf{K_2}-1) \tilde{\Gamma}_{Z \mid X_1X_3}^{n_1}(Z^{n_1} | X_{1}^{n_1}(1,1), X_3^{n_1}(w_3))}{\mathsf{M_1} \mathsf{M_2} \mathsf{K_1} \mathsf{K_2}  \cdot\tilde{\Gamma}_{{Z \mid X_3}}^{n}(Z^{n_1} \mid X_3^{n_1}(w_3))}+  \frac{ (\mathsf{M_1}\mathsf{K_1}-1) \tilde{\Gamma}_{{Z \mid X_2X_3}}^{{n_2}}(Z^{n_2} | X_{2}^{n_2}(1,1), X_3^{n_2}(w_3))}{\mathsf{M_1} \mathsf{M_2} \mathsf{K_1} \mathsf{K_2}  \cdot\tilde{\Gamma}_{{Z \mid X_3}}^{n_2}(Z^{n_2} \mid X_3^{n_2}(w_3))}  \right ) \right]\nonumber \\
\\
&\leq &
\mathbb{E}\left[ \log \left( 1 +  \frac{   \widehat{Q}_{\mathcal{C}, 1, 1, w_3, 1, 1}^{n_1}(Z^{n_1}) }{\mathsf{M_1} \mathsf{M_2} \mathsf{K_1} \mathsf{K_2}  \cdot \tilde{\Gamma}_{{Z \mid X_3}}^{{n_1}}(Z^{n_1} \mid X_3^{n_1}(w_3))}  \right. \right. \nonumber \\
&& \hspace{1.5cm}  \left. \left. + \frac{ \tilde{\Gamma}_{Z \mid X_1X_3}^{n_1}(Z^{n_1} | X_{1}^{n_1}(1,1), X_3^{n_1}(w_3))}{\mathsf{M_1} \mathsf{K_1} \cdot\tilde{\Gamma}_{{Z \mid X_3}}^{n_1}(Z^{n_1} \mid X_3^{n_1}(w_3))}+  \frac{  \tilde{\Gamma}_{{Z \mid X_2X_3}}^{{n_2}}(Z^{n_2} | X_{2}^{n_2}(1,1), X_3^{n_2}(w_3))}{ \mathsf{M_2}  \mathsf{K_2}  \cdot\tilde{\Gamma}_{{Z \mid X_3}}^{n_2}(Z^{n_2} \mid X_3^{n_2}(w_3))}  \right ) \right]  \label{eq:cr_1_2}
\end{IEEEeqnarray}
where in $(a)$ we used that 
\begin{IEEEeqnarray}{rCl}  
  \mathop{\mathbb{E}}_{\substack{ X_1^{n_1}({w_1,s_1})  \\ X_2^{n_2}({w_2,s_2}) }}   \left[  \widehat{Q}_{\mathcal{C}, w_1, w_2, w_3, s_1, s_2}^{n_1}(z^{n_1}) \right]  &=& \tilde{\Gamma}_{Z \mid X_3}^{n_1}(z^{n_1} | X_3^{n_1}(w_3))
   \end{IEEEeqnarray}
and 
\begin{IEEEeqnarray}{rCl}  
     \mathop{\mathbb{E}}_{ X_1^{n_1}(w_1, s_1)} \left[  \widehat{Q}_{\mathcal{C},  w_1,1, w_3, s_1,1}^{n_1}(z^{n_2})\right]  &=& \tilde{\Gamma}_{Z \mid X_2X_3}^{n_2}(z^{n_2} | X_{2}^{n_2}(1,1), X_3^{n_2}(w_3)) \cdot \tilde{\Gamma}_{Z|X_3}^{n_2 \to n_1}(z_{n_2+1}^{n_1} | X_{3,n_2+1}^{n_1}(w_3))\\
     \mathop{\mathbb{E}}_{ X_2^{n_2}(w_2, s_2)} \left[  \widehat{Q}_{\mathcal{C}, 1, w_2, w_3, 1, s_2}^{n_1}(z^{n_1})\right]  &=& \tilde{\Gamma}_{Z \mid X_1X_3}^{n_1}(z^{n_1} | X_{1}^{n_1}(1,1), X_3^{n_1}(w_3))
\end{IEEEeqnarray}
and we simplified the last fraction by noting that 
\begin{equation}
\frac{ \tilde{\Gamma}_{Z \mid X_2X_3}^{n_2}(z^{n_2} | X_{2}^{n_2}(1,1), X_3^{n_2}(w_3)) \cdot \tilde{\Gamma}_{Z|X_3}^{n_2 \to n_1}(z_{n_2+1}^{n_1} | X_{3,n_2+1}^{n_1}(w_3))}{\tilde{\Gamma}_{{Z \mid X_3}}^{n_1}(Z^{n_1} \mid X_3^{n_1}(w_3))}  =\frac{  \tilde{\Gamma}_{{Z \mid X_2X_3}}^{{n_2}}(Z^{n_2} | X_{2}^{n_2}(1,1), X_3^{n_2}(w_3))} {\tilde{\Gamma}_{{Z \mid X_3}}^{n_2}(Z^{n_2} \mid X_3^{n_2}(w_3))}.
\end{equation}

Define now for any triple $\boldsymbol{\theta}= (\theta_0,\theta_1, \theta_2)$ the set:\footnote{Notice that $x_2^{n_2}$ is of length $n_2$ while the other sequences are of length $n_1$.}
\begin{IEEEeqnarray}{rrCl}  
\lefteqn{    \mathcal{B}_{\boldsymbol{\theta}}^{n_1} \triangleq \Bigg \{  (x_1^{n_1} , x_2^{n_2}, x_3^{n_1}, z^{n_1}) \in  \mathcal{X}_1^{n_1}\times \mathcal{X}_2^{n_2} \times \mathcal{X}_3^{n_1} \times \mathcal{Z}^{n_1} \colon } \hspace{2cm}
    \nonumber \\
    &  \log \left( \frac{\Gamma_{Z \mid X_1X_2X_3}^{\otimes n_2}(z^{n_2} \mid x_1^{n_2}, x_2^{n_2}, x_3^{n_2})\cdot \tilde{\Gamma}_{Z \mid X_1X_3}^{n_2\to n_1}(z_{n_2+1}^{n_1} \mid x_{1,n_2+1}^{n_1}, x_{3,n_2+1}^{n_1})  }{\Gamma_{Z \mid X_1X_2X_3}^{\otimes n_1}(z^{n_1} \mid 0^{n_1},0^{n_1}, x_3^{n_1})}  \right) &\leq& \theta_0, \nonumber \\
    & \log \left( \frac{\tilde{\Gamma}_{Z \mid X_1X_3}^{n_1}(z^{n_1} \mid x_1^{n_1}, x_3^{n_1}) }{\Gamma_{Z \mid X_1X_2X_3}^{\otimes n_1}(z^{n_1} \mid 0^{n_1}, 0^{n_1}, x_3^{n_1})}  \right)& \leq& \theta_1, \nonumber \\
    & \log \left( \frac{\tilde{\Gamma}_{Z \mid X_2X_3}^{n_2}(z^{n_2} \mid x_2^{n_2}, x_3^{n_2}) }{\Gamma_{Z \mid X_1X_2X_3}^{\otimes n_2}(z^{n_2} \mid 0^{n_2}, 0^{n_2}, x_3^{n_2})}  \right)& \leq &\theta_2 \Bigg \},
\end{IEEEeqnarray}
and denote by $A$ and $B$ the events $\{(X_1^{n_1}(1,1), X_2^{n_2}(1,1), X_3^{n_1}(w_3), Z^{n_1}) \in \mathcal{B}_{\boldsymbol{\theta}}^{n_1}\}$ and $\{(X_1^{n_1}(1,1), X_2^{n_2}(1,1), X_3^{n_1}(w_3), Z^{n_1}) \notin \mathcal{B}_{\boldsymbol{\theta}}^{n_1}\}$ respectively.
    
    We then continue by similar  steps to \eqref{eq:dde}--\eqref{eq:exp_bound}, 
 where the superscript $n$ to has to be changed to $n_1$ or $n_2$ accordingly, and the joint law $ \Gamma_{Z\mid X_1X_2X_3}^{\otimes n}$ to $\Gamma_{Z \mid X_1X_2X_3}^{\otimes n_2} \cdot \tilde{\Gamma}_{Z \mid X_1X_3}^{n_2-n_1}$.  In particular we apply Bernstein's inequality to analyze the probability of set $B$, and replace \eqref{eq:exp1} --\eqref{eqn:cr_7} by the following expresssions: 
 \begin{subequations}
 \begin{IEEEeqnarray}{rCl}  
 \lefteqn{   \mathbb{E} \left[  \log \left( \frac{\Gamma_{Z \mid X_1X_2X_3}^{\otimes n_2}(Z^{n_2} \mid X_1^{n_2}(1,1), X_2^{n_2}(1,1), X_3^{n_2}(w_3))\cdot \tilde{\Gamma}_{Z \mid X_1X_3}^{n_2\to n_1}(Z_{n_2+1}^{n_1} \mid X_{1,n_2+1}^{n_1}(1,1), X_{3,n_2+1}^{n_1}) (w_3) }{\Gamma_{Z \mid X_1X_2X_3}^{\otimes n_1}\ablast{(}Z^{n_1} \mid 0^{n_1},0^{n_1}, X_3^{n_1}(w_3)) } \right) \right] }\nonumber \quad \\
 &\leq &\alpha_n \left(    n_2 \mathbb{E}_{\pi P_{X_3|T}}  \left[ \rho_{1,T} \mathbb{D}_Z^{(1)}(X_3) + \rho_{2,T} \mathbb{D}_Z^{(2)}(X_3) \right]  +   (n_1-n_2) \mathbb{E}_{\pi P_{X_3|T}}  \left[ \rho_{1,T} \mathbb{D}_Z^{(1)}(X_3) ) \right]  \right)+ n \mathcal{O}(\alpha_n^2)
 \end{IEEEeqnarray}
 and 
  \begin{IEEEeqnarray}{rCl}  
  \mathbb{E} \left[   \log \left( \frac{\tilde{\Gamma}_{Z \mid X_1X_3}^{n_1}(Z^{n_1} \mid X_1^{n_1}(1,1), X_3^{n_1}(w_3)) }{\Gamma_{Z \mid X_1X_2X_3}^{\otimes n_1}(Z^{n_1} \mid 0^{n_1}, 0^{n_1}, X_3^{n_1}(w_3))}  \right) \right]&\leq& \alpha_n   n_1 \mathbb{E}_{\pi P_{X_3|T}}  \left[ \rho_{1,T} \mathbb{D}_Z^{(1)}(X_3)  \right]  + \mathcal{O}(\alpha_n^2)\\
 \mathbb{E} \left[   \log \left( \frac{\tilde{\Gamma}_{Z \mid X_2X_3}^{n_2}(Z^{n_2} \mid X_2^{n_2}(1,1), X_3^{n_2}(w_3)) }{\Gamma_{Z \mid X_1X_2X_3}^{\otimes n_2}(Z^{n_2} \mid 0^{n_2}, 0^{n_2}, X_3^{n_2}(w_3))}  \right) \right] &\leq &\alpha_n  n_2     \mathbb{E}_{\pi P_{X_3|T}}  \left[  \rho_{2,T} \mathbb{D}_Z^{(2)}(X_3)  \right] + \mathcal{O}(\alpha_n^2).
\end{IEEEeqnarray}
 \end{subequations}
 
 These steps  allow us to conclude that whenever
\begin{subequations}
    \begin{IEEEeqnarray}{rCl}
         {\limsup}_{n\to \infty} \left( \log \left( \mathsf{M_1}\mathsf{M_2}\mathsf{K_1}\mathsf{K_2}\right) - (1+\xi_4) {\alpha_{n}} \mathbb{E}_{P_{TX_3}}  \left[ n_1 \rho_{1,T} \mathbb{D}_Z^{(1)}(X_3) + n_2\rho_{2,T} \mathbb{D}_Z^{(2)}(X_3) \right] \right)= - \infty , \IEEEeqnarraynumspace \label{eq:covertness2} \\
         {\limsup}_{n\to \infty} \left( \log \left( \mathsf{M_1}\mathsf{K_1}\right) - (1+\xi_5) {\alpha_{n} n_1} \mathbb{E}_{P_{TX_3}}  \left[  \rho_{1,T} \mathbb{D}_Z^{(1)}(X_3) \right] \right)= - \infty,\IEEEeqnarraynumspace \label{eq:key_constraint_user_1_2}\\
         {\limsup}_{n\to \infty} \left( \log \left( \mathsf{M_2}\mathsf{K_2}\right) - (1+\xi_6) {\alpha_{n} n_2} \mathbb{E}_{P_{TX_3}}  \left[  \rho_{2,T} \mathbb{D}_Z^{(2)}(X_3) \right] \right)= - \infty, \IEEEeqnarraynumspace \label{eq:key_constraint_user_2_2} 
    \end{IEEEeqnarray}
\end{subequations}
then  the divergence on the right-hand side of \eqref{eq:abs_diff} tends to 0 exponentially fast, and thus the approximation
\begin{equation}
\mathbb{D}\left( \tilde{\Gamma}_{{Z \mid X_3}}^{n_1} \Big\| \Gamma^{\otimes n_1}_{Z \mid X_1X_2X_3} ( \cdot | 0^{n_1}, 0^{n_1}, x_3^{n_1}(w_3))\right)  = ( 1+ o(1))  \frac{n_1}{n}\frac{\omega_n^2 }{2} \mathbb{E}_{P_T} \left[ (\rho_{1,T} + \rho_{2,T})^2 \mathbb{E}_{P_{X_3 \mid T}} \left[ \chi^2(\rho_{1,T}, \rho_{2,T}, X_3) \right ] \right ]
\end{equation}
is exponentially tight.  We notice that in an asymptotic sense condition \eqref{eq:covertness2} is redundant in view of \eqref{eq:key_constraint_user_1_2} and \eqref{eq:key_constraint_user_2_2}. This concludes the resolvability analysis by considering that $n_1 \approx n \phi_1$ and $n_2 \approx n \phi_2$ and that $\alpha_n = \omega_n \sqrt{n}$.
 
 \section{Achievability Proof to Theorem~\ref{th:asymp_result}}\label{app:ach}
Start by noticing that without  loss in generality in Theorem~\ref{main_theorem} one can replace Constraint~\eqref{eq:th_1_log_m1} by   the difference between constraints  \eqref{eq:th_1_log_m1} and  \eqref{eq:th_1_log_k1_k2} and Constraint~\eqref{eq:th_1_log_m2} by the difference between constraints  \eqref{eq:th_1_log_m2} and \eqref{eq:th_1_log_k1}. Taking $n\to \infty$ with these new constraints proves achievability of the following quintuple $(r_1,r_2,R_3,k_1, k_2)$ 
 for arbitrary  pmfs $P_{TX_3}$,  nonnegative tuples $\{(\rho_{1, t}, \rho_{2,t})\}_{t\in\mathcal{T}}$, and  pairs $(\phi_1, \phi_2)\in[0,1]^2$: 
\begin{IEEEeqnarray}{rCl}
  r_{\ell} & = & \frac{\phi_\ell}{ \ablast{\sqrt{\max( \phi_{1} \ablast{;} \phi_2)}}} \sqrt{2} \frac{\mathbb{E}_{P_{TX_3}} \left[ \rho_{\ell,T} \D_{Y}^{(\ell)}(X_3) \right]}{\sqrt{\mathbb{E}_{P_{TX_3}} \left[ \left( \rho_{1,T} + \rho_{2,T} \right)^2  \cdot \chi^2(\rho_{1,T}, \rho_{2,T}, X_3) \right]}}, \qquad  \forall \ell \in \{1,2\},  \\[1ex]
 R_3 &=&  \mathbb{I}(X_3;Y \mid X_1=0,X_2=0,T), \\[1ex]
k_{\ell}  &=& \frac{\phi_\ell}{ \ablast{\sqrt{\max( \phi_{1}\ablast{;} \phi_2)}}}  \sqrt{2} \frac{\mathbb{E}_{P_{TX_3}}  \left[ \rho_{\ell,T} \left( \D_{Z}^{(\ell)}(X_3) - \D_{Y}^{(\ell)}(X_3) \right) \right]} {\sqrt{\mathbb{E}_{P_{TX_3}} \left[ \left( \rho_{1,T} + \rho_{2,T} \right)^2 \cdot \chi^2(\rho_{1,T}, \rho_{2,T}, X_3) \right]}}, \qquad  \forall \ell \in \{1,2\},
\end{IEEEeqnarray}
where we use the definition $0/0=0$.
Define $\beta_\ell \triangleq  \frac{\phi_\ell}{\sqrt{ \max( \phi_{1}\ablast{;} \phi_2)}}$ and notice that it lies in $[0,1]$. Notice further that the equalities on the \ablast{message} rates $r_1, r_2, R_3$ can be relaxed into $\leq$-inequalities because it is always possible to reduce the \ablast{message} rate by introducing dummy data-bits. 
Moreover, it is possible to relax the inequalities on the \ablast{secret-}key rates $k_1$ and $k_2$ into $\geq$- inequalities because it is always possible to ignore some of the \ablast{secret-}key bits. These latter observations establish achievability of the theorem.

\section{Converse Proof to Theorem~\ref{th:asymp_result}}\label{sec:proof2}
The converse proof relies on elements from the proofs of \ablast{\cite[Theorem~3]{bloch_first}, \cite[Theorem~1 and 2]{ligong_first} , \cite[Theorem~2]{ours_first} and \cite[Section~5.2.3]{hou_phdthesis}}. 

Consider a sequence of length-$n$ codes with vanishing probability of error $P_{e, \mathcal{H}} \to 0$ and vanishing covertness constraints $\delta_{n, w_3}\to 0$ as the blocklength $n\to \infty$. Consider now  a fixed blocklength $n$, and let $X_1^n, X_2^n, X_3^n$ be the random inputs generated under the chosen codes and $Y^n$ as well as $Z^n$ the corresponding outputs at the \ablast{legitimate} receiver and the warden \ab{under $\mathcal{H}=1$}. Define also a random variable $T$ to be uniform over the set  of channel uses $[|1,n|]$, independent of the inputs $X_1^n, X_2^n, X_3^n$ and the outputs $Y^n$ and $Z^n$.   With these definitions, since the three users have independent messages and keys, the joint pmf of the time-averaged inputs and outputs has the following form: 
\begin{equation}
    \label{eq:def_joint_proba_time_rv_converse_log_m1}
    P_{{X}_{1,T},{X}_{2,T},{X}_{3,T},{Y}_T,Z_T,T}(x_1,x_2,x_3,y,z,t) = P_T(t) P_{X_1 \mid T}(x_1 \mid t) P_{X_2 \mid T}(x_2 \mid t) P_{X_3 \mid T}(x_3 \mid t) \Gamma_{Y Z\mid X_1X_2X_3}(y,z \mid x_1, x_2, x_3).
\end{equation}

For our converse proof, we  also define $\alpha_{n, \ab{i}, \ell}$ as the probability of $X_{\ell,\ab{i}}$ equal  1: 
\begin{equation}\label{eq:def_alpha}
\alpha_{n,\ab{i},\ell} \triangleq \Pr[ X_{\ell,\ab{i}}=1], \quad \ab{i}\in \ab{\{1, \ldots, n\}}, \ab{ \; \ell \in\{1,2\},}
\end{equation}
and the derived \ab{positive} quantities
\begin{equation}\label{eq:def_rho}
\rho_{n,\ab{i},\ell} \triangleq \frac{  \alpha_{n,\ab{i},\ell}}{ \sum_{\ab{i}=1}^n\frac{  \alpha_{n,\ab{i},1} + \alpha_{n,\ab{i},2}}{n}} ,  \quad \ab{i}\in \ab{\{1, \ldots, n\}}, \; \ell \in\{1,2\}.
\end{equation}
We observe that by the uniform law of $T$:
\begin{equation}\label{eq:EPT}
\mathbb{E}_{P_T}\left[ \rho_{n,T,1}+  \rho_{n,T,2} \right]=\frac{1}{n} \sum_{\ab{i}=1}^n n \frac{  \alpha_{n,\ab{i},1}+ \alpha_{n,\ab{i},2}}{ \sum_{\ab{i}=1}^n  \alpha_{n,\ab{i},1} + \alpha_{n,\ab{i},2}} =1.\end{equation}
\subsection{Auxiliary Lemmas:} We will make use of the following lemma, which is an extension of \cite[Lemma~1]{bloch_first}. Recall the definitions in \eqref{eq:D_def}.
\begin{lemma}
\label{lemma_1_converse}
Let  $X_1, X_2$ be binary over $\{0,1\}$ and $T,X_3,Y$ over arbitrary finite alphabets $\mathcal{T}$, $\mathcal{X}_3$, and $\mathcal{Y}$, with joint pmf  of the form $P_T P_{X_{1} \mid T} P_{X_3 \mid T}\Gamma_{Y\mid X_1,X_2,X_3}$.
Then, for any $x_2\in\{0,1\}$ and $t\in\mathcal{T}$:
\begin{IEEEeqnarray}{rCl}
\mathbb{I} (X_1; Y \mid X_2=x_2, X_3, T=t) 
  &=&   P_{X_{1} \mid T=t}(1) \mathbb{E}_{P_{X_3 \mid T=t}} \left[ \ab{\mathbb{D}(\Gamma_{Y\mid X_1X_2X_3}(\cdot \mid 1, x_2, X_3) || \Gamma_{Y \mid X_1X_2X_3}(\cdot | 0, 0, X_3))} \right ] \nonumber \\
  &&\hspace{0.5cm} - \mathbb{E}_{P_{X_3 \mid T=t}} \left[\mathbb{D}(\Gamma_{Y\mid X_2X_3}(\cdot \mid x_2, X_3) || \Gamma_{Y \mid X_1X_2X_3}(\cdot | 0, \ab{0}, X_3)) \right]. \label{eq:bound1}
\end{IEEEeqnarray}
\ab{Similarly,} for any $x_1\in\{0,1\}$ and $t\in\mathcal{T}$:
\begin{IEEEeqnarray}{rCl}
\mathbb{I} (X_2; Y \mid X_1=x_1, X_3, T=t)  &=&     P_{X_{2} \mid T=t}(1) \mathbb{E}_{P_{X_3 \mid T=t}} \left[ \ab{\mathbb{D}(\Gamma_{Y\mid X_1X_2X_3}(\cdot \mid x_1, 1, X_3) || \Gamma_{Y \mid X_1X_2X_3}(\cdot | 0, 0, X_3))}\right ]  \nonumber \\
  &&\hspace{0.5cm} -\mathbb{E}_{P_{X_3 \mid T=t}} [\mathbb{D}(\Gamma_{Y\mid X_1X_3}(\cdot \mid x_1, X_3) || \Gamma_{Y \mid X_1X_2X_3}(\cdot | x_1, 0, X_3))]. \label{eq:bound2}
\end{IEEEeqnarray}
\end{lemma}
\begin{IEEEproof}
Follows by simple rewriting. Details omitted.
\end{IEEEproof}
The following lemma is a direct consequence of Lemma~\ref{lemma_1_converse} and the nonnegativity of \ablast{Kullback-Leibler} divergence.
\begin{lemma}
\label{lemma_11_converse}
Let $(T,X_1, X_2, X_3, Y)$ be as in Lemma~\ref{lemma_1_converse} where in addition we assume that  for any $t\in\mathcal{T}$:
\begin{equation}
\lim_{n\to \infty} P_{X_{1}|T=t}(1)= \lim_{n\to \infty}  P_{X_{2}|T=t}(1) =0.
\end{equation}
Then:
\begin{IEEEeqnarray}{rCl}
     \mathbb{I} (X_1; Y \mid X_2, X_3, T=t)  & \leq &  P_{X_{1}|T=t}(1) \left( \mathbb{E}_{P_{X_3 \mid T=t}} \left[ \mathbb{D}_{Y}^{(1)}(X_3) \right ] + o(1) \right)\label{eq:bound3} \\
     \mathbb{I} (X_2; Y \mid X_1, X_3, T=t)  & \leq &  P_{X_{2}|T=t}(1) \left( \mathbb{E}_{P_{X_3 \mid T=t}} \left[ \mathbb{D}_{Y}^{(2)}(X_3) \right ] + o(1)\right) \label{eq:bound4}
\end{IEEEeqnarray}
\end{lemma}
\begin{IEEEproof}
We can write the mutual information term as follows:
\begin{IEEEeqnarray}{rCl}
    \lefteqn{\mathbb{I} (X_1; Y \mid X_2, X_3, T=t )} \nonumber \\
    &=& P_{X_{2}|T=t}(0) \mathbb{I} (X_1; Y \mid X_2=0, X_3, T=t ) +  P_{X_{2}|T=t}(1)\mathbb{I} (X_1; Y \mid X_2=1, X_3, T=t )\\
    & \leq & \mathbb{I} (X_1; Y \mid X_2=0, X_3, T=t ) +P_{X_{2}|T=t}(1)\mathbb{I} (X_1; Y \mid X_2=1, X_3, T=t )\\
    & \overset{(a)}{\leq} &P_{X_{1}|T=t}(1) \mathbb{E}_{P_{X_3 \mid T=t}} \left[ \mathbb{D}_{Y}^{(1)}(X_3) \right ] + P_{X_{2}|T=t}(1)\cdot  P_{X_{1}|T=t}(1) \mathbb{E}_{P_{X_3 \mid T=t}} \ab{\left[\mathbb{D}(\Gamma_{Y\mid X_1X_2X_3}(\cdot \mid 1, 1, X_3) || \Gamma_{Y \mid X_1X_2X_3}(\cdot | 0, 0, X_3))\right]}  \nonumber \\
    & = &P_{X_{1}|T=t}(1)  \left( \mathbb{E}_{P_{X_3 \mid T=t}} \left[ \mathbb{D}_{Y}^{(1)}(X_3) \right] + o(1) \right),
\end{IEEEeqnarray}
where (a) holds by Lemma~\ref{lemma_1_converse}.

Similarly, one can show that
\begin{IEEEeqnarray}{rCl}
    \mathbb{I} (X_2; Y \mid X_1, X_3, T=t ) \leq P_{X_{2}|T=t}(1) \left(\mathbb{E}_{P_{X_3 \mid T=t}} \left[ \mathbb{D}_{Y}^{(2)}(X_3) \right ] + o(1)\right).
\end{IEEEeqnarray}
\end{IEEEproof}

\subsection{Lower bound on $\frac{1}{\mathsf{\mathsf{M_3}}}\sum_{w_3=1}^{\mathsf{\mathsf{M_3}}} \delta_{n,w_3} $:}

 Recalling also the definition of $  \widehat{Q}_{\mathcal{C}, w_3}^{n}(z^{n})$ in \eqref{eq:def_Q_C_w2}, we obtain for a specific code $\mathcal{C}$:
 \allowdisplaybreaks[4]
\begin{align}
    &\frac{1}{\mathsf{\mathsf{M_3}}}\sum_{w_3=1}^{\mathsf{\mathsf{M_3}}} \delta_{n,w_3} \nonumber \\
    &= \frac{1}{\mathsf{\mathsf{M_3}}}\sum_{w_3=1}^{\mathsf{\mathsf{M_3}}} \sum_{z^n} \widehat{Q}_{\mathcal{C},w_3}^n(z^n) \log \left( \frac{\widehat{Q}_{\mathcal{C},w_3}^{n}(z^n)}{\Gamma_{Z \mid X_1X_2X_3}^{\otimes n}(z^n| 0^n, 0^n, x_3^n(w_3))} \right)  \\
    & \overset{(a)}{\geq} \frac{1}{\mathsf{\mathsf{M_3}}}\sum_{w_3=1}^{\mathsf{\mathsf{M_3}}} \sum_{i=1}^n \sum_{z_i} \widehat{Q}_{\mathcal{C},w_3}^{(i)}(z_i)  \log \left( \frac{\widehat{Q}_{\mathcal{C},w_3}^{(i)}(z_i)}{\Gamma_{Z \mid X_1X_2X_3}(z_i | 0, 0, x_{3i}(w_3))} \right) \\
    &= \frac{1}{\mathsf{\mathsf{M_3}}}\sum_{w_3=1}^{\mathsf{\mathsf{M_3}}} \sum_{i=1}^n \mathbb{D} \left( \widehat{Q}_{\mathcal{C},w_3}^{(i)} \| \Gamma_{Z \mid X_1X_2X_3}(\cdot| 0, 0, x_{3i}(w_3)) \right) \\ 
    & \overset{(b)}{=} \frac{1}{\mathsf{\mathsf{M_3}}}\sum_{w_3=1}^{\mathsf{\mathsf{M_3}}} \sum_{i=1}^n \mathbb{D} \Big( \Big. \alpha_{n,i,1}\alpha_{n,i,2} \Gamma_{Z \mid X_1X_2X_3}(\cdot \mid 1, 1, x_{3i}(w_3)) + \alpha_{n,i,1}(1- \alpha_{n,i,2}) \Gamma_{Z \mid X_1X_2X_3}(\cdot \mid 1, 0, x_{3i}(w_3)) \nonumber \\
    & \qquad \qquad \qquad \qquad + (1- \alpha_{n,i,1}) \alpha_{n,i,2} \Gamma_{Z \mid X_1X_2X_3}(\cdot \mid 0, 1, x_{3i}(w_3)) + (1-\alpha_{n,i,1})(1- \alpha_{n,i,2}) \Gamma_{Z \mid X_1X_2X_3}(\cdot \mid 0, 0, x_{3i}(w_3)) \nonumber \\
    & \qquad \qquad \qquad \qquad \quad \| \Gamma_{Z \mid X_1X_2X_3}(\cdot | 0, 0, x_{3i}(w_3)) \Big. \Big),  \label{eq:divergence_upper_bound_converse_to_show_alpha_t_goes_to_zero}
\end{align}
where $(a)$ holds by the memoryless nature of the channel  and by defining $ \widehat{Q}_{\mathcal{C},w_3}^{(i)}(z_i)$ as the probability of the event $Z_i=z_i$ conditioned on $W_3=w_3$, and by writing out the expectations over the independent random variables $X_{1,i}$ and $X_{2,i}$.

Notice that since for all $w_3 \in \mathcal{M}_3$ we have that $\lim_{n \to \infty} \delta_{n,w_3} = 0$, by \eqref{eq:divergence_upper_bound_converse_to_show_alpha_t_goes_to_zero} we can conclude that
\begin{equation}
    \lim_{n \to \infty} \alpha_{n,i,\ab{\ell}} = 0, \quad \forall i \in \ab{\{1, \ldots,n \}}, \; \ab{\ell} \in \{1,2\}. \label{eq:alpha_n_i_k_go_zero_proof}
\end{equation}
Combining \eqref{eq:divergence_upper_bound_converse_to_show_alpha_t_goes_to_zero}, \eqref{eq:alpha_n_i_k_go_zero_proof} and Lemma~\ref{lemma_1_asymptotics}, we can conclude that 
\begin{align}
    &\frac{1}{\mathsf{\mathsf{M_3}}}\sum_{w_3=1}^{\mathsf{\mathsf{M_3}}} \delta_{n,w_3}  \geq   n \mathbb{E}_{P_{TX_3}} \left[(1+o(1)) \frac{\left(\alpha_{n,T,1} + \alpha_{n,T,2} \right )^2}{2} \chi_n^2(\rho_{n,T,1},\rho_{n,T,2}, X_3) \right], \label{eq:upper_bound_divergence_Q12_Q02_part_final}
\end{align}
where we define the time random variable $T$ to be uniform over $\{1, \ldots, n\}$ and independent of all other random variables.\\

\subsection{Upper bound on $\log(\mathsf{M_1})$:} 
Since the message $W_1$ is uniform over $\{1, \ldots, \mathsf{M_1}\}$ and independent of the local randomness $C_1,C_2$, we have 
\begin{align}
    \log(\mathsf{M_1}) & = \mathbb{H}(W_1) \\
    &= \mathbb{H}(W_1 \mid W_2, S_1, S_2, W_3) \\
    &= \mathbb{I}(W_1 ; Y^n \mid W_2, S_1, S_2, W_3) + \mathbb{H}(W_1 \mid Y^n, W_2,S_1, S_2,  C_1,C_2,W_3) \\
    & \overset{(a)}{\leq} \mathbb{I}(W_1; Y^n \mid W_2, S_1, S_2, C_1,C_2,X_2^n, X_3^n) + \mathbb{H}_b(P_{e,1}) + P_{e,1} \log(\mathsf{M_1}) \\
    & = \frac{1}{1-P_{e,1}}(\mathbb{I}(W_1 ; Y^n \mid W_2, S_1, S_2,  C_1,C_2,X_2^n, X_3^n) + \mathbb{H}_b(P_{e,1}) ) \\
    & \overset{(b)}{=} \frac{1}{1-P_{e,1}} \Bigg( \sum_{i=1}^n \mathbb{H}(Y_i \mid Y^{i-1},W_2, S_1, S_2,  C_1,C_2,X_2^n, X_3^n) \nonumber \\
    &\hspace{3cm}- \mathbb{H}(Y_i \mid Y^{i-1}, W_1, W_2, S_1, S_2,  C_1,C_2,X_1^n, X_2^n, X_3^n) + \mathbb{H}_b(P_{e,1}) \Bigg)\\
    & \overset{(c)}{\leq} \frac{1}{1-P_{e,1}} \Bigg( \sum_{i=1}^n \mathbb{H}(Y_i \mid X_{2i}, X_{3i})- \mathbb{H}(Y_i \mid X_{1i}, X_{2i}, X_{3i}) + \mathbb{H}_b(P_{e,1}) \Bigg)\\
    &= \frac{1}{1-P_{e,1}} \left( n \sum_{i=1}^n \frac{1}{n} \mathbb{I}(X_{1i} ; Y_i \mid X_{2i}, X_{3i}) + \mathbb{H}_b(P_{e,1}) \right)\\
    &\leq    \frac{1}{1-P_{e,1}}  (n \mathbb{I}(X_{1,T}; Y_T\mid X_{2,T}, X_{3,T}) + 1) \\
       & \overset{(d)}{\leq} \frac{1}{1-P_{e,1}}  n \left( \mathbb{E}_{P_{TX_{3,T}}}\left[ \alpha_{n, T, 1} \left( \D_Y^{(1)}(X_{3,T}) + o(1) \right) \right] + \frac{1}{n}\right). \label{eq:converse_bound_mutual_info_X1_Y_part_1}
\end{align}
Above sequence of (in)equalities are justified as follows:\\
(a) holds by Fano's inequality and because $X_2^n=\varphi_{2}^{(n)}(W_2, S_2,C_2)$ as well as $X_3^n=\varphi_3^{(n)}(W_3)$; \\
(b) holds by the chain rule of entropy and because $X_1^n=\varphi_{1}^{(n)}(W_1,S_1,C_1)$;\\
(c) holds respectively because conditioning {reduces} entropy and because conditioned on $X_{1,i}, X_{2,i}, X_{3,i}$ the output $Y_i$ is independent of all messages, keys, and randomness;\\
(d) is obtained by  applying  Lemma \ref{lemma_11_converse}  for each realization of $T$ and by upper-bounding the binary entropy by 1.
\subsection{Upper Bound on $\log(\mathsf{M_2})$}
The desired upper bound can be derived using similar steps to those for $\log(\mathsf{M_1})$.
\begin{IEEEeqnarray}{rCl}
\log(\mathsf{M_2}) &\leq& \frac{1}{1-P_{e,1}}n\left( \mathbb{E}_{P_{TX_3}} \left[ \alpha_{n,T,2}\left( \mathbb{D}_{Y}^{(2)}(X_3) + o(1)\right) \right]  + \frac{1}{n}\right) \label{eq:converse_bound_log_m2_final}
\end{IEEEeqnarray}
\subsection{Upper Bound on $\log(\mathsf{\mathsf{M_3}})/n$:}
Using standard steps, one can find the upper bound
\begin{equation}
\frac{1}{n} \log(\mathsf{M_3}) \leq\frac{1}{1-P_{e,1}} \mathbb{I}({X}_{3,T};{Y}_T| {X}_{1,T}=0, {X}_{2,T}=0, {T}). \label{eq:converse_ub_log_m3}
\end{equation}
\subsection{Upper bound on $\log (\mathsf{M_1})/ \sqrt{n \frac{1}{\mathsf{M_3}} \sum_{w_3=1}^{\mathsf{M_3}}\delta_{n,w_3}}$:}
By  \eqref{eq:converse_bound_mutual_info_X1_Y_part_1} and \eqref{eq:upper_bound_divergence_Q12_Q02_part_final}, we obtain the bound
\begin{IEEEeqnarray}{rCl}
\frac{\log(\mathsf{M_1})}{\sqrt{n \frac{1}{\mathsf{M_3}} \sum_{w_3=1}^{\mathsf{M_3}}\delta_{n,w_3}}} & \leq & \frac{\sqrt{2}}{1-P_{e,1}}\frac{n \left( \mathbb{E}_{P_{TX_3}} \left[ \alpha_{n,T,1} \left( \mathbb{D}_{Y}^{(1)}(X_3) + o(1) \right)\right]  + \frac{1}{n}\right)}{\sqrt{n\left(   n  (1+o(1))\mathbb{E}_{P_{TX_3}} \left[ \left(\alpha_{n,T,1} + \alpha_{n,T,2} \right )^2 \chi_n^2(\rho_{n,T,1},\rho_{n,T,2}, X_3) \right] \right)} }  \\
&\overset{(a)}{=}& \frac{\sqrt{2}}{1-P_{e,1}}\frac{ \mathbb{E}_{P_{TX_3}} \left[ \frac{ \alpha_{n,T,1}}{\mathbb{E}_{P_T}[\alpha_{n,T,1} + \alpha_{n,T,2}]} \left(\mathbb{D}_{Y}^{(1)}(X_3) + o(1) \right) \right]  + \frac{1}{n}}{\sqrt{ (1+o(1))  \mathbb{E}_{P_{TX_3}} \left[ \left(\frac{\alpha_{n,T,1} + \alpha_{n,T,2}}{\mathbb{E}_{P_T}[\alpha_{n,T,1} + \alpha_{n,T,2}]} \right)^2 \chi_n^2(\rho_{n,T,1},\rho_{n,T,2}, X_3) \right]} } \\
&\overset{(b)}{=}& \frac{\sqrt{2}}{1-P_{e,1}}\frac{ \mathbb{E}_{P_{TX_3}} \left[ \rho_{n,T,1} \left( \mathbb{D}_{Y}^{(1)}(X_3) + o(1) \right)\right]  + \frac{1}{n}}{\sqrt{(1+o(1)) \mathbb{E}_{P_{TX_3}} \left[ \left(\rho_{n,T,1} + \rho_{n,T,2} \right)^2 \chi_n^2(\rho_{n,T,1},\rho_{n,T,2}, X_3) \right]} } \label{eq:converse_ub_log_m1}
\end{IEEEeqnarray}
where (a) follows by normalization by $\mathbb{E}_{P_T}[\alpha_{n,T,1} + \alpha_{n,T,2}]$, and (b) by the definition of $\rho_{n,i,\ell}$ in \eqref{eq:def_rho} for all $(i,\ell) \in \{1, \ldots, n\} \times \{1, 2\}$.

Similarly, one can show that
\begin{IEEEeqnarray}{rCl}
\frac{\log(\mathsf{M_2})}{\sqrt{n \frac{1}{\mathsf{M_3}} \sum_{w_3=1}^{\mathsf{M_3}}\delta_{n,w_3}}} \leq  \frac{\sqrt{2}}{1-P_{e,1}}\frac{ \mathbb{E}_{P_{TX_3}} \left[ \rho_{n,T,2} \left(\mathbb{D}_{Y}^{(2)}(X_3) + o(1) \right) \right]  + \frac{1}{n}}{\sqrt{  (1+o(1))\mathbb{E}_{P_{TX_3}} \left[ \left(\rho_{n,T,1} + \rho_{n,T,2} \right)^2 \chi_n^2(\rho_{n,T,1},\rho_{n,T,2}, X_3) \right]} } \label{eq:converse_ub_log_m2}
  \end{IEEEeqnarray}

\subsection{\underline{Upper bound on $\frac{1}{\mathsf{M_3}} \sum_{w_3=1}^{\mathsf{M_3}}\delta_{n,w_3}$:}}
\label{sec:ub_avg_delta_n_w3_converse}
Notice that the new parameters $\rho_{n,T,\ell}$ are well defined because $\mathbb{E}_T[ \alpha_{n,T,1} + \alpha_{n,T,2}]$  characterizes the sum of the fractions of $1$-entries in the codebooks  $\{x_1^n(W_1,S_1, C_1)\}$  and $\{x_2^n(W_2,S_2,C_2)\}$,   and is thus non-zero because otherwise no communication is going on. 
Moreover, by Jensen's Inequality, $\mathbb{E}_T[\left(\rho_{n,T,1} + \rho_{n,T,2}\right)^2] \geq\left( \mathbb{E}_T[\rho_{n,T,1} + \rho_{n,T,2}] \right)^2 =1$.

It then follows  by Assumptions~\eqref{eq:channel_conditions}, that the right-hand sides of \eqref{eq:converse_ub_log_m3},  \eqref{eq:converse_ub_log_m1} and \eqref{eq:converse_ub_log_m2}  lie in  bounded intervals. Combining the inequalities in  \eqref{eq:converse_ub_log_m3},  \eqref{eq:converse_ub_log_m1} and \eqref{eq:converse_ub_log_m2}  with trivial positivity considerations, one can conclude that also the left-hand sides of these inequalities must lie in bounded intervals. Consequently there exists  subsequence of blocklengths  so that  both the left- and  ride-hand sides of \eqref{eq:converse_ub_log_m3}, \eqref{eq:converse_ub_log_m1} and \eqref{eq:converse_ub_log_m2} all converge. We shall restrict to such a \ablast{subsequence} of blocklengths $\{n_i\}_{i=1}^\infty$. 

Let then $\beta_1$ and $\beta_2$ be the two numbers in $[0,1]$ that satisfy
\begin{IEEEeqnarray}{rCl}
\lim_{i\to\infty} \frac{\log(\mathsf{M_\ell})}{\sqrt{n_i \frac{1}{\mathsf{M_3}} \sum_{w_3=1}^{\mathsf{M_3}}\delta_{n_i,w_3} }} & = & \sqrt{2}\beta_\ell \lim_{i\to \infty} \frac{ \mathbb{E}_{P_{TX_3}} \left[ \rho_{n_i,T,\ell} \mathbb{D}_{Y}^{(\ell)}(X_{3,T}) \right]  }{\sqrt{ \mathbb{E}_{P_{TX_3}} \left[ \left(\rho_{n_i,T,1} + \rho_{n_i,T,2} \right)^2 \chi_n^2(\rho_{n_i,T,1},\rho_{n_i,T,2}, X_{3,T}) \right]} },\quad \ell \in\{1,2\}\nonumber \\ \label{eq:upper_delta}\
\end{IEEEeqnarray}
where notice that the limit on the right-hand side coincides with the limits on the right-hand sides in \eqref{eq:converse_ub_log_m1} or \eqref{eq:converse_ub_log_m2}, respectively.

Assume for the moment that $\beta_1,\beta_2$ are strictly larger than 0, and thus we can divide by it. Then, \eqref{eq:upper_delta} combined with  \eqref{eq:converse_bound_mutual_info_X1_Y_part_1} and  \eqref{eq:converse_bound_log_m2_final}   implies  that for all blocklengths $n_i$: 
\begin{IEEEeqnarray}{rCl}
    \sqrt{n_i \frac{1}{\mathsf{M_3}} \sum_{w_3=1}^{\mathsf{M_3}}\delta_{n_i,w_3}}  &\leq &\frac{n_i}{\beta_\ell\sqrt{2}} \sqrt{ \mathbb{E}_{P_{TX_3}} \left[ \left(\rho_{n_i,T,1} + \rho_{n_i,T,2} \right)^2 \chi_n^2(\rho_{n_i,T,1},\rho_{n_i,T,2}, X_{3,T}) \right]} +o(1), \qquad \ell \in\{1,2\}. \IEEEeqnarraynumspace
    \label{eq:LB}
\end{IEEEeqnarray}
\subsection{\underline{Lower bound on $\log(\mathsf{M_1}\mathsf{M_2}\mathsf{K_1}\mathsf{K_2})$:}}
We start with the  lower bound
\begin{IEEEeqnarray}{rCl}
    \log(\mathsf{M_1}\mathsf{K_1})  &=& \mathbb{H}(W_1,S_1|C_1, C_2) \\
    &= & \mathbb{H}(W_1,S_1 \mid X_3^n, C_1, C_2) \\
    & \geq & \mathbb{I}(W_1,S_1 ; Z^n \mid X_3^n, C_1, C_2) \\
    & \overset{(a)}{\geq}& \mathbb{I}(X_1^n ; Z^n \mid X_3^n, C_1, C_2) \\
    &=& \mathbb{I}(X_1^n, X_2^n ; Z^n \mid X_3^n, C_1, C_2) - \mathbb{I}(X_2^n ; Z^n \mid X_1^n, X_3^n, C_1, C_2)\\
     & \overset{(b)}{\geq}& \mathbb{I}(X_1^n, X_2^n ; Z^n \mid X_3^n, C_1, C_2) - \mathbb{I}(X_2^n ; Z^n \mid X_1^n, X_3^n)  \label{eq:mutual}
\end{IEEEeqnarray}
where (a) holds because $X_1^n=x_1^n(W_1,S_1,C_1)$ is a function of $W_1$, $S_1$, and $C_1$ and $(b)$ holds because of the Markov chain $(C_1, C_2) \to (X_1^n, X_2^n, X_3^n) \to Z^n$ and because conditioning reduces entropy.

Likewise,
\begin{IEEEeqnarray}{rCl}
    \log(\mathsf{M_2}\mathsf{K_2})  & \geq& \mathbb{I}(X_1^n, X_2^n ; Z^n \mid X_3^n, C_1,C_2) - \mathbb{I}(X_1^n ; Z^n \mid X_1^n, X_3^n),\label{eq:mutual2}
\end{IEEEeqnarray}

We next focus on the first mutual-information term that is common to the RHS of \eqref{eq:mutual} and \eqref{eq:mutual2}. To this end, we define for each pair $(c_1, c_2)\in \mathcal{G}_1\times \mathcal{G}_2$ the warden's average output distribution conditioned on the local randomness $c_1$ and $c_2$ and on a non-covert message $w_3$, the distribution
\begin{IEEEeqnarray}{rCl}
\widehat{Q}_{\mathcal{C}, (c_1,c_2,w_3)}^{n}(z^{n}) &\triangleq&  \frac{1}{\mathsf{M}_1 \mathsf{M}_2\mathsf{K}_1\mathsf{K}_2 } \sum_{(w_1,s_1)} \sum_{(w_2,s_2)} \Gamma^{\otimes n}_{Z \mid X_1X_2X_3} (z^n| x_1^n(w_1,s_1, c_1), x_2^n( w_2,s_2,c_2), x_3^n(w_3))
\end{IEEEeqnarray}
and the divergence
\begin{equation}
\delta_{n,(c_1,c_2,w_3)}\triangleq \mathbb{D}\left(\widehat{Q}_{\mathcal{C}, (c_1,c_2, w_3)}^{n} \; \big\| \; \Gamma^{\otimes n}_{Z \mid X_1X_2X_3} ( \cdot | 0^n, 0^n, x_3^n(w_3))\right). 
\end{equation}
With this definition, we can write: 
\begin{IEEEeqnarray}{rCl}
\lefteqn{\mathbb{I}(X_1^n, X_2^n ; Z^n \mid X_3^n, C_1, C_2)} \\
&=& \mathbb{E}_{X_1^n, X_2^n, X_3^n, C_1, C_2} \left[ 
\mathbb{D}\left( \Gamma_{Z \mid X_1X_2X_3}^{\otimes n} (\cdot \mid X_1^n, X_2^n, X_3^n) \| \widehat{Q}_{\mathcal{C},(C_1,C_2,W_3)}^{n}\right) \right] \\
&=& \mathbb{E}_{X_1^n, X_2^n, X_3^n,C_1,C_2} \bigg[ \sum_{z^n} \Gamma_{Z \mid X_1X_2X_3}^{\otimes n} (z^n\mid X_1^n, X_2^n, X_3^n) \log \left(\frac{\Gamma_{Z \mid X_1X_2X_3}^{\otimes n} (z^n\mid X_1^n, X_2^n, X_3^n)}{\Gamma_{Z \mid X_1X_2X_3}^{\otimes n} (z^n \mid 0^n, 0^n, X_3^n)} \right) \nonumber \\
&& \hspace{3cm} -\sum_{z^n} \Gamma_{Z \mid X_1X_2X_3}^{\otimes n} (z^n\mid X_1^n, X_2^n, X_3^n) \log \left(\frac{\widehat{Q}_{\mathcal{C},(C_1,C_2,W_3)}^{n}(z^n)}{\Gamma_{Z \mid X_1X_2X_3}^{\otimes n} (z^n \mid 0^n, 0^n, X_3^n)} \right) \Bigg] \\
&\overset{(a)}{=}& \mathbb{E}_{X_1^n, X_2^n, X_3^n} \left[ \sum_{z^n} \Gamma_{Z \mid X_1X_2X_3}^{\otimes n} (z^n\mid X_1^n, X_2^n, X_3^n) \log \left(\frac{\Gamma_{Z \mid X_1X_2X_3}^{\otimes n} (z^n\mid X_1^n, X_2^n, X_3^n)}{\Gamma_{Z \mid X_1X_2X_3}^{\otimes n} (z^n \mid 0^n, 0^n, X_3^n)} \right) \right] \nonumber\\
&& - \mathbb{E}_{C_1,C_2,W_3} \left[ \delta_{n,(C_1,C_2,W_3)} \right] \\
&\overset{(b)}{\geq}& \sum_{i=1}^n \mathbb{E}_{X_{1,i}, X_{2,i}, X_{3,i}} \left[ \sum_{z_i} \Gamma_{Z \mid X_1X_2X_3}(z_i \mid X_{1,i}, X_{2,i}, X_{3,i})  \log \left(\frac{\Gamma_{Z \mid X_1X_2X_3}(z_i \mid X_{1i}, X_{2i}, X_{3i})}{\Gamma_{Z \mid X_1X_2X_3} (z_i \mid 0, 0, X_{3i})} \right) \right] - \mathbb{E}_{W_3} \left[ \delta_{n,W_3} \right]\IEEEeqnarraynumspace \\
&\overset{(c)}{=}& \sum_{i=1}^n \mathbb{E}_{X_{3i}} \left[ \left (\alpha_{n,i,1} \mathbb{D}_{Z}^{(1)}(X_{3i}) + \alpha_{n,i,2} \mathbb{D}_{Z}^{(2)}(X_{3,i})\right)(1+o(1)) \right] - \mathbb{E}_{W_3} \left[ \delta_{n,W_3} \right] \IEEEeqnarraynumspace \\
&\overset{(d)}{=}& n \mathbb{E}_{P_{TX_{3T}}} \left[\left( \alpha_{n,T,1} \mathbb{D}_{Z}^{(1)}(X_{3,T}) + \alpha_{n,T,2} \mathbb{D}_{Z}^{(2)}(X_{3,T})\right )(1+ o(1)) \right] - \mathbb{E}_{W_3} \left[ \delta_{n,W_3} \right], \label{eq:single_letterization_I_X_1_X2_Z_mid_X3}
\end{IEEEeqnarray}
where (a) holds by the definition of $\delta_{n,(c_1, c_2,w_3)}$ and by replacing the average over $X_3^n$ by the average over $W_3$; $(b)$ holds by  by convexity of the divergence; $(c)$ by writing out the expectations over the independent random variables $X_{1,i}$ and $X_{2,i}$ and by noting that for $X_{1,i}=X_{2,i}=0$ the term in the expectation evaluates to 0; $(d)$ holds because  $T$ is uniform over $\{1, \ldots, n\}$.

For the second mutual-information term on the RHS of \eqref{eq:mutual}, we have:
\begin{IEEEeqnarray}{rCl}
\mathbb{I}(X_1^n ; Z^n \mid X_2^n, X_3^n) &=& \mathbb{H}(Z^n \mid X_2^n, X_3^n) - \mathbb{H}(Z^n \mid X_1^n, X_2^n, X_3^n) \\
&=& \sum_{i=1}^n \mathbb{H}(Z_i \mid Z^{i-1}, X_2^n, X_3^n) - \mathbb{H}(Z_i \mid Z^{i-1},X_1^n, X_2^n, X_3^n) \\
&\overset{(a)}{=}& \sum_{i=1}^n \mathbb{H}(Z_i \mid X_2^n, X_3^n) - \mathbb{H}(Z_i \mid X_{1,i}, X_{2,i}, X_{3,i}) \\
&\overset{(b)}{\leq}& \sum_{i=1}^n \mathbb{H}(Z_i \mid X_{2i}, X_{3i}) - \mathbb{H}(Z_i \mid X_{1i}, X_{2i}, X_{3i}) \\
&=& n \sum_{i=1}^n \frac{1}{n} \mathbb{I}(X_{1i} ; Z_i \mid X_{2i}, X_{3i}) \\
&=& n \mathbb{I}({X}_{1,T} ; {Z}_T \mid {X}_{2,T}, {X}_{3,T}, T) \\
&\overset{(c)}{\leq}& n \mathbb{E}_{P_{TX_{3,T}}} \left[\alpha_{n,T,1} \mathbb{D}_{Z}^{(1)}(X_{3,T}) (1+ o(1)) \right], \label{eq:single_letterization_I_X1_Z_mid_X2_X3}
\end{IEEEeqnarray}
where $(a)$ holds  by the memoryless nature of the channel; $(b)$ because conditioning reduces entropy; and $(c)$ by applying Lemma~\ref{lemma_11_converse} to output $Z_T$ instead of $Y_T$.

In an analogous way, one can show that
\begin{IEEEeqnarray}{rCl}
\mathbb{I}(X_2^n ; Z^n \mid X_1^n, X_3^n) \leq n \mathbb{E}_{P_{TX_{3,T}}} \left[\alpha_{n,T,2} \mathbb{D}_{Z}^{(2)}(X_{3,T})(1+ o(1)) \right]. \label{eq:single_letterization_I_X2_Z_mid_X1_X3}
\end{IEEEeqnarray}
Combining \eqref{eq:single_letterization_I_X_1_X2_Z_mid_X3}, \eqref{eq:single_letterization_I_X1_Z_mid_X2_X3} and \eqref{eq:single_letterization_I_X2_Z_mid_X1_X3} with \eqref{eq:mutual} we can conclude that 
\begin{IEEEeqnarray}{rCl}
   \log(\mathsf{M_\ell}\mathsf{K}_\ell) &\geq&  n \left( \mathbb{E}_{P_{TX_3}} \left[\alpha_{n,T,\ell} \mathbb{D}_{Z}^{(\ell)}(X_3) (1+ o(1)) \right]- \frac{\mathbb{E}_{W_3}[\delta_{n,W_3}]}{n} \right), \quad \ell \in\{1,2\}. \label{eq:logM1_logK_upper_bound_part_final}
\end{IEEEeqnarray}

\subsection{\underline{Lower bound on $\frac{\log(\mathsf{M_2} \mathsf{K_2})}{\sqrt{n\frac{1}{\mathsf{M_3}}\sum_{w_3=1}^{\mathsf{M_3}} \delta_{n,w_3}}}$, $\frac{\log(\mathsf{M_2} \mathsf{K_2})}{\sqrt{n\frac{1}{\mathsf{M_3}}\sum_{w_3=1}^{\mathsf{M_3}} \delta_{n,w_3}}}$ and Discussion}}
\vspace{3mm}

By combining \eqref{eq:LB} with \eqref{eq:logM1_logK_upper_bound_part_final}, for $n\in\{n_i\}$ we obtain the bound for $\ell\in\{1,2\}$:
\begin{IEEEeqnarray}{rCl}
   \lefteqn{ \frac{\log(\mathsf{M_\ell} \mathsf{K_\ell})}{\sqrt{n\frac{1}{\mathsf{M_3}}\sum_{w_3=1}^{\mathsf{M_3}} \delta_{n,w_3}}} } \nonumber \\
   & \geq & \beta_\ell \left( \frac{ \mathbb{E}_{P_{TX_3}} \left [ \rho_{n,T,\ell} \mathbb{D}_{Z}^{(\ell)}(X_{3T}) \right]}{\sqrt{ \mathbb{E}_{P_{TX_3}} \left[ \left( \rho_{n,T,1} + \rho_{n,T,2} \right)^2 \chi_n^2(\rho_{n,T,1},\rho_{n,T,2}, X_3) \right]}} - \frac{\frac{\mathbb{E}_{W_3}[\delta_{n,W_3}]}{n \mathbb{E}_{P_T} \left[\alpha_{n,T,1} + \alpha_{n,T,2} \right]} }{\sqrt{ \mathbb{E}_{P_{TX_3}} \left[ \left( \rho_{n,T,1} + \rho_{n,T,2} \right)^2 \chi_n^2(\rho_{n,T,1},\rho_{n,T,2}, X_3) \right]}} \right). \nonumber \\ \label{eq:bound3}
\end{IEEEeqnarray}

Notice that the second term on the right-hand side of \eqref{eq:bound3} vanishes whenever $\log(\mathsf{M_1}) \to \infty$ or $\log(\mathsf{M}_2)  \to \infty$. In fact, we have
 \begin{IEEEeqnarray}{rCl}
 \lim_{n\to \infty} \mathbb{E}_{W_3}[\delta_{n,W_3}] &=& 0, \label{eq:limit1} \\
 \ab{\liminf}_{n \to \infty}  n \cdot \mathbb{E}_T[\alpha_{n,T,1} + \alpha_{n,T,2}]& > & 0, \label{eq:limit2}\\
  \ab{\liminf}_{n \to \infty} \mathbb{E}_{P_{TX_3}} \left[ \left( \rho_{n,T,1} + \rho_{n,T,2} \right)^2 \chi_n^2(\rho_{n,T,1},\rho_{n,T,2}, X_3) \right]  & > & 0.
  \label{eq:limit4}
 \end{IEEEeqnarray}  
Limit \eqref{eq:limit1} holds because $\delta_{n,w_3}\to 0$ for any $w_3 \in \mathcal{M}_3$; \\
Limit~\eqref{eq:limit2} holds  because when the left-hand side tends to 0 then the number of $1$-symbols in all the codewords tends to zero\ablast{,} in which case no  messages can be transmitted reliably; \\
Limit~\eqref{eq:limit4} holds because by  definition \eqref{eq:def_xi_distance} and assumptions~\eqref{eq:channel_conditions}  the random variable in the expectation is 0 only when  $\rho_{n,T,1}=\rho_{n,T,2}=0$, which cannot happen with probability 1 over  $T$ as this would contradict  \eqref{eq:EPT}.
   
\subsection{\underline{Limits and Cardinality Bound}}

Notice that so far we have assumed that $\beta_1, \beta_2 >0$. For $\beta_\ell \geq 0$, we simply note that: 
\begin{IEEEeqnarray}{rCl}
  \liminf_{n \to \infty} \frac{\log(\mathsf{M}_\ell\mathsf{K}_\ell)}{\sqrt{n \frac{1}{\mathsf{M_3}}\sum_{w_3=1}^{\mathsf{M_3}} \delta_{n,w_3}}} \geq 0. \end{IEEEeqnarray}

In view of this, and \ablast{collecting} all results in the previous subsections, we can conclude that there exists an increasing subsequences of blocklengths $\{n_i\}$  so that for \ablast{any $\ell \in\{1,2\}$ and some $\beta_\ell \in[0,1]$}:
\begin{subequations}\label{eq:bounds}
\begin{IEEEeqnarray}{rCl}
    \lim_{n_i \to \infty} \frac{\log(\mathsf{M}_\ell)}{\sqrt{n_i \frac{1}{\mathsf{M_3}}\sum_{w_3=1}^{\mathsf{M_3}} \delta_{n_i,w_3}}} =  \lim_{n_i \to \infty} \sqrt{2}\beta_\ell \frac{ \mathbb{E}_{P_{TX_3}} \left[ \rho_{n_i,T,\ell} \mathbb{D}_{Y}^{(\ell)}(X_3) \right]  }{\sqrt{ \mathbb{E}_{P_{TX_3}} \left[ \left( \rho_{n_i,1,T} + \rho_{n_i,2,T} \right)^2 \chi_{n_i}^2(\rho_{n_i,T,1},\rho_{n_i,T,2}, X_3) \right]} }, \label{eq:bounds1}
  \end{IEEEeqnarray}
\begin{equation}
\limsup_{n_i \to \infty} \frac{1}{n_i} \log \mathsf{M_3} \leq \lim_{n_i\to\infty} \mathbb{I}({X}_{3,T};{Y}_{T}| {X}_{1,T}=0, {X}_{2,T}=0, {T}),
\end{equation}
\begin{IEEEeqnarray}{rCl}
  \liminf_{n_i \to \infty} \frac{\log(\mathsf{M}_\ell\mathsf{K}_\ell)}{\sqrt{n_i \frac{1}{\mathsf{M_3}}\sum_{w_3=1}^{\mathsf{M_3}} \delta_{n_i,w_3}}} \geq \lim_{n_i \to \infty} \sqrt{2} \beta_\ell\frac{ \mathbb{E}_{P_{TX_3}} \left [ \rho_{n_i,T,\ell} \mathbb{D}_{Z}^{(\ell)}(X_3) \right]}{\sqrt{ \mathbb{E}_{P_{TX_3}} \left[ \left( \rho_{n_i,1,T} + \rho_{n_i,2,T} \right)^2 \chi_{n_i}^2(\rho_{n_i,T,1},\rho_{n_i,T,2}, X_3) \right]}}. \label{eq:bounds2}
\end{IEEEeqnarray}
\end{subequations}
Applying the Fenchel-Eggleston-Carathéodory theorem to vectors of the form
\begin{IEEEeqnarray}{rCl}
    \vect{v}= \begin{pmatrix}
    \rho_{n_i,t,1} \mathbb{E}_{P_{X_3 \mid T=t}} \left[ \mathbb{D}_{Y}^{(1)}(X_3) \right]\\
    \rho_{n_i,t,2} \mathbb{E}_{P_{X_3 \mid T=t}} \left[ \mathbb{D}_{Y}^{(2)}(X_3) \right]\\
    \mathbb{I}(X_3 ; Y \mid X_1=0, X_2=0, T=t )\\
    \left( \rho_{n_i,t,1} + \rho_{n_i,t,2} \right)^2 \mathbb{E}_{P_{X_3 \mid T=t}} \left[ \chi_{n_i}^2(\rho_{n_i,T,1},\rho_{n_i,T,2}, X_3) \right]\\
     \mathbb{E}_{P_{X_3 \mid T=t}} \left[ \rho_{n_i,t,1} \mathbb{D}_{Z}^{(1)}(X_3) \right] \\
     \mathbb{E}_{P_{X_3 \mid T=t}} \left[ \rho_{n_i,t,2} \mathbb{D}_{Z}^{(2)}(X_3) \right]
    \end{pmatrix}    ,
\end{IEEEeqnarray}
we conclude that for any blocklength $n_i$ there exists a modified distribution $\tilde{P}_T$ over an alphabet of size $|\mathcal{T}|=6$ so that the bounds \eqref{eq:bounds}  hold also if $P_{T}$ is replaced by this new pmf $\tilde{P}_T$. In the rest of the proof\ablast{,}  we can thus restrict to these modified distributions $\tilde{P}_T$ over $|\mathcal{T}|=6$. 

\subsection{\underline{The Limiting Distribution}}

To conclude the proof, we notice that by the Bolzano-Weierstrass theorem there exists an increasing subsequence $\{n_{i_k}\}$ of $\{n_i\}$ so that $\{P_{X_{3}|T}(\cdot|t)\}$ and $\{P_T(\cdot)\}$ converge on this subsequence. If also $\rho_{n_{i_k},t,1}$ and $\rho_{n_{i_k},t,2}$ converge for each value of $t\in \mathcal{T}\triangleq \{1,\ldots, 6\}$,  then by the continuity of the expressions, we obtain a converse result by considering the convergence points of  $\{P_{X_{3}|T}(\cdot|t)\}$, $\{P_T(\cdot)\}$, and  $\{(\rho_{1,t}, \rho_{2,t})\}_{t\in\mathcal{T}}$. In fact, we can conclude that 
\begin{IEEEeqnarray}{rCl}
    \lim_{k \to \infty} \frac{\log(\mathsf{M_\ell})}{\sqrt{n_{i_k} \frac{1}{\mathsf{M_3}}\sum_{w_3=1}^{\mathsf{M_3}} \delta_{n_{i_k},w_3}}} &=&  \sqrt{2} \beta_\ell\frac{ \mathbb{E}_{P_{TX_3}} \left[ \rho_{\ell,T} \mathbb{D}_{Y}^{(\ell)}(X_3) \right] }{\sqrt{ \mathbb{E}_{P_{TX_3}} \left[ \left( \rho_{1,T} + \rho_{2,T} \right)^2  \chi^2(\rho_{1,T},\rho_{2,T}, X_3) \right ]} }, \quad \ell \in \{1,2\}\label{eq:singlem}\\
\limsup_{k\to \infty} \frac{1}{n_{i_k}} \log \mathsf{M_3} &\leq& \mathbb{I}(X_3;Y| X_1=0, X_2=0, T), \\
\liminf_{k \to \infty} \frac{\log(\mathsf{M_\ell}\mathsf{K_\ell})}{\sqrt{n_{i_k} \frac{1}{\mathsf{M_3}}\sum_{w_3=1}^{\mathsf{M_3}} \delta_{n_{i_k},w_3}}} &\geq& \sqrt{2} \beta_\ell\frac{ \mathbb{E}_{P_{TX_3}} \left[ \rho_{\ell,T} \mathbb{D}_{Z}^{(\ell)}(X_3) \right]}{\sqrt{ \mathbb{E}_{P_{TX_3}} \left[ \left( \rho_{1,T} + \rho_{2,T} \right)^2   \chi^2(\rho_{1,T},\rho_{2,T}, X_3) \right]  } }, \quad \ell \in \{1,2\}.\label{eq:sum_last}
\end{IEEEeqnarray}
for some  pmf $P_{X_3T}$ over $\mathcal{X}_3 \times \mathcal{T}$ and some  set of positive pairs $\{(\rho_{1,t}, \rho_{2,t})\}_{t\in\mathcal{T}}$. 

If instead for some $t\in \mathcal{T}$ the sequence $\rho_{n_{i_k},t,\ell}$ diverges to $\infty$ we proceed as follows. We first notice that for each of these $t$-values the probability $P_{T}(t)\to 0$ as $n \to \infty$, because otherwise the expectation \eqref{eq:EPT} is violated, and one of the following three cases applies:
\begin{itemize}
\item[1.)] $P_{T}(t)\rho_{n_{i_k},t,\ell} \to 0$ and $P_{T}(t)\rho_{n_{i_k},t,\ell}^2 \to 0$;
\item[2.)] $P_{T}(t)\rho_{n_{i_k},t,\ell} \to 0$ and $\lim_{n_{i_k}\to \infty} P_{T}(t)\rho_{n_{i_k},t,\ell}^2=c$ for  some $c\in(0,\infty)$;
\item[3.)] $P_T(t) \rho_{n_{i_k},t,\ell}\in [0,1]$ and $P_{T}(t)\rho_{n_{i_k},t,\ell}^2 \to \infty$.
\end{itemize}
All $t$-values satisfying case 1.) can simply be ignored since they do not change the bounds. Whenever there exists a $t$-value in case 3.), then bounds \eqref{eq:bounds1} and \eqref{eq:bounds2} are 0 for both $\ell\in\{1,2\}$ and the result is trivial. In  case 2.) we can  modify the probabilities  $P_{T}(t)$ and the parameters $\rho_{n_{i_k},t,\ell}$ to  values in a bounded interval $[a,b]$ for $b>a >0$, while still approximating the bounds \eqref{eq:bounds} arbitrarily closely. We then fall back to the case where all sequences $\rho_{n_{i_k},t,\ell}$ converge, which we discussed above.

\subsection{Recovering the Bound on the \ablast{Secret-}Key Size}\label{app1} 
Above converse result  remains valid if we add the additional constraint that results when taking the difference between  \eqref{eq:sum_last} and \eqref{eq:singlem}: 
\begin{IEEEeqnarray}{rCl}
\liminf_{k \to \infty} \frac{\log(\mathsf{\mathsf{K_\ell})}}{\sqrt{n_{i_k} \frac{1}{\mathsf{M_3}}\sum_{w_3=1}^{\mathsf{M_3}} \delta_{n_{i_k},w_3}}} &\geq& \sqrt{2} \beta_\ell\frac{ \mathbb{E}_{P_{TX_3}} \left[ \rho_{\ell,T} \left( \mathbb{D}_{Z}^{(\ell)}(X_3) - \mathbb{D}_{Y}^{(\ell)}(X_3) \right)\right]}{\sqrt{ \mathbb{E}_{P_{TX_3}} \left[ \left( \rho_{1,T} + \rho_{2,T} \right)^2   \chi^2(\rho_{1,T},\rho_{2,T}, X_3) \right]  } }, \quad \ell \in \{1,2\}.\label{eq:single_key_last}
\end{IEEEeqnarray}
Relaxing  the sum-constraint \eqref{eq:sum_last} completely and further relaxing the equality in \eqref{eq:singlem} into an $\leq$-inequality establishes finally the desired converse proof to Theorem~\ref{th:asymp_result}.

 \section{Proof of Lemma~\ref{lem:convexity}}
\label{app:convexity}
Fix two pmfs $P_{TX_3}$ and $Q_{TX_3}$ as well as the tuples $(\rho_{1,t},\rho_{2,t})$ and $(\rho_{1,t}',\rho_{2,t}')$ in $\mathbb{R}_0^{+^{2}}$ for all $t \in \mathcal{T}$. Define the tuples
\begin{align}
    \bm{\rho}_t &\triangleq (\rho_{1,t},\rho_{2,t}), \\
    \bm{\rho}_t' &\triangleq (\rho_{1,t}',\rho_{2,t}'), \\
    \bm{\mu} &\triangleq (\bm{\rho}_1,\ldots,\bm{\rho}_6, \bm{\rho}_1',\ldots,\bm{\rho}_6'),
\end{align}
for all $t \in \mathcal{T}$.\\
Let $(r_1,r_2,r_3,k)$, $(r_1', r_2', r_3', k')$, and $(\tilde r_1,\tilde r_2,\tilde r_3,\tilde k)$ be  the tuples of messages and key rates given by the right-hand sides of \eqref{eq:asymp1}--\eqref{eq:asympkey} when evaluated for $P_{TX_3}$ and $(\bm{\rho}_1,\ldots,\bm{\rho}_6)$, for $Q_{TX_3}$ and $(\bm{\rho}_1',\ldots,\bm{\rho}_6')$, and for $R_{TX_3}$ and $\mu$.
We shall show that 
\begin{equation}
\lambda \begin{pmatrix} r_1 \\ r_2 \\ r_3 \\ k \end{pmatrix} + (1-\lambda) \begin{pmatrix} r_1' \\ r_2' \\ r_3' \\ k' \end{pmatrix} = \begin{pmatrix} \tilde{r}_1 \\ \tilde{r}_2 \\ \tilde{r}_3 \\\tilde{k} \end{pmatrix}, \quad \forall \lambda \in [0,1].
\end{equation}
The desired equality for the $\tilde{r}_3$-component is directly obtained by the linearity of conditional mutual information and because it does not depend on the $\bm{\rho}$-, $\bm{\rho}'$-, and $\bm{\mu}$-tuples.
To see the equality for the other three components, fix $\lambda \in [0,1]$, and set $\nu >0$ so that
\begin{equation}
\nu^2  \triangleq \frac{\mathbb{E}_{P_{TX_3}} \left[ \left(\rho_{1,T} + \rho_{2,T} \right)^2 \chi^2(\bm{\rho}_T, X_3) \right] } { \mathbb{E}_{Q_{TX_3}} \left[\left(\rho_{1,T}' + \rho_{2,T}' \right)^2 \chi^2(\bm{\rho}_T', X_3)\right ] }.
\end{equation}
For all $\ell \in \{1,2\}$, upon forming the new pmf $R_{TX_3}$ by choosing 
\begin{equation}
R_T(t)=\begin{cases} \lambda \cdot P_T(t)  & t\in \{1,\ldots,6\} \\
(1-\lambda ) \cdot Q_T(t-6) & t\in \{7,\ldots,12\}
 \end{cases}
\end{equation}
and 
\begin{equation}
R_{X_3|T}(x_3|t) = \begin{cases} P_{X_3|T}(x_3|t) & t\in \{1,\ldots,6\}\\
 Q_{X_3|T}(x_3|t-6) & t\in \{7,\ldots,12\},
\end{cases}
\end{equation}
and by defining the following tuples and constants
\begin{IEEEeqnarray}{rCl}
\bm{\tilde{\rho}}_t &\triangleq&  
    \begin{cases}
        \bm{\rho}_t, &  t\in \{1,\ldots,6\}\\
        \bm{\rho}_{t-6}'  , &  t\in \{7,\ldots,12\} \
    \end{cases} \\
\alpha_{\ell,t} &\triangleq&  
    \begin{cases}
        \rho_{\ell,t}, &  t\in \{1,\ldots,6\}\\
        \nu\cdot \rho_{\ell,t-6}'  , &  t\in \{7,\ldots,12\} \
    \end{cases} \\
\beta_t &\triangleq&  
    \begin{cases}
        \rho_{1,t}+\rho_{2,t}, &  t\in \{1,\ldots,6\}\\
        \nu \cdot \left(\rho_{1,t-6}' + \rho_{2,t-6}' \right)  , &  t\in \{7,\ldots,12\} \
    \end{cases}
\end{IEEEeqnarray}
One can notice that for any function $f\colon \mathcal{X}_3\to\mathbb{R}$ we have
\begin{IEEEeqnarray}{rCl}
\lefteqn{\lambda  \frac{ \mathbb{E}_{P_{X_3T}} \left[ \rho_{\ell,T}  f(X_3) \right ]} { \sqrt{ \mathbb{E}_{P_{X_3T}} \left[ \left(\rho_{1,T} + \rho_{2,T} \right)^2 \chi^2(\bm{\rho}_T, X_3)\right ] }} +(1-\lambda )  \frac{ \mathbb{E}_{Q_{X_3T}} \left[ \rho_{\ell,T}' f(X_3) \right ]} { \sqrt{ \mathbb{E}_{Q_{X_3T}} \left[ \left(\rho_{1,T}' + \rho_{2,T}' \right)^2 \chi^2(\bm{\rho}_T', X_3)\right ] }}} \nonumber \\
&= &\lambda  \frac{ \mathbb{E}_{P_{X_3T}} \left[ \rho_{\ell,T}  f(X_3) \right ]} { \sqrt{ \mathbb{E}_{P_{X_3T}} \left[ \left(\rho_{1,T} + \rho_{2,T} \right)^2 \chi^2(\bm{\rho}_T, X_3)\right ] }} +(1-\lambda )  \frac{ \mathbb{E}_{Q_{X_3T}} \left[  \nu \rho_{\ell,T}' f(X_3) \right ]} { \sqrt{ \mathbb{E}_{Q_{X_3T}} \left[ \nu^2 \left(\rho_{1,T}' + \rho_{2,T}' \right)^2 \chi^2(\bm{\rho}_T', X_3)\right ] }} \\
& \stackrel{(a)}{=} &    \frac{\lambda \mathbb{E}_{P_{X_3T}} \left[ \rho_{\ell,T} f(X_3) \right ] +(1-\lambda )  \ \mathbb{E}_{Q_{X_3T}} \left[ \nu \rho_{\ell,T}'  f(X_3) \right ]}{ \sqrt{ \lambda \mathbb{E}_{P_{X_3T}} \left[ \left(\rho_{1,T} + \rho_{2,T} \right)^2 \chi^2(\bm{\rho}_T, X_3) \right ] +(1-\lambda)\mathbb{E}_{Q_{X_3T}} \left[\nu^2  \left(\rho_{1,T}' + \rho_{2,T}' \right)^2 \chi^2(\bm{\rho}_T', X_3)\right ] }}  \\
& =&     \frac{ \mathbb{E}_{R_{X_3T}} \left[ \alpha_{\ell,T} f(X_3) \right ]}{ \sqrt{ \mathbb{E}_{R_{X_3T}} \left[ \beta_T^2 \chi^2(\bm{\tilde{\rho}}_T, X_3)\right ] }}  \label{eq:convexity_rate_region_part_one}
\end{IEEEeqnarray}
\begin{IEEEeqnarray}{rCl}
\lefteqn{\mathbb{E}_{Q_{X_3T}} \left[\nu^2  \left(\rho_{1,T}' + \rho_{2,T}' \right)^2 \chi^2(\bm{\rho}_T', X_3) \right ]} \nonumber \\
& \stackrel{(b)}{=}&  \mathbb{E}_{P_{X_3T}} \left[\left(\rho_{1,T} + \rho_{2,T} \right)^2 \chi^2(\ablast{\bm{\rho}_T}, X_3)\right ] \\
 & = & \lambda \mathbb{E}_{P_{X_3T}} \left[\left(\rho_{1,T} + \rho_{2,T} \right)^2 \chi^2(\ablast{\bm{\rho}_T}, X_3)\right ] + (1-\lambda) \mathbb{E}_{Q_{X_3T}} \!\left[\nu^2  \left(\rho_{1,T}' + \rho_{2,T}' \right)^2 \chi^2(\bm{\rho}_T', X_3)\right]\!, \label{eq:convexity_rate_region_part_two}\IEEEeqnarraynumspace
\end{IEEEeqnarray}
where in $(b)$ we used the definition of $\nu$. This concludes the proof of the Lemma.
\end{appendices}

\printbibliography

\end{document}